\documentclass[numbers,preprint,12pt]{elsarticle}

\usepackage{amsmath,amssymb,amsfonts}
\usepackage{algorithmic}
\usepackage{textcomp}
\usepackage[dvipsnames]{xcolor}
\def\BibTeX{{\rm B\kern-.05em{\sc i\kern-.025em b}\kern-.08em
    T\kern-.1667em\lower.7ex\hbox{E}\kern-.125emX}}
    
\usepackage{listings}
\usepackage{physics}
\usepackage{algorithm}
\usepackage{adjustbox}
\usepackage{subcaption}
\usepackage{easyReview}
\usepackage{dirtytalk}
\usepackage{subcaption}

\usepackage{tcolorbox}

\usepackage[spaces,hyphens]{xurl}
\usepackage[section]{placeins}
\usepackage[hidelinks]{hyperref} \usepackage{cleveref}

\usepackage{latexsym,tabularx,multirow,multicol,xspace,tikz}
\usepackage{makecell}
\usepackage{booktabs}
\usepackage{ltxtable}
\usepackage{longtable}
\usepackage{colortbl}
\usepackage{arydshln}
\usepackage{rotating} 

\usepackage{personalcommand}
\usepackage{comment}

\newcommand{\expComp}[1]{\textbf{#1} UML Components\xspace}
\newcommand{\expNode}[1]{\textbf{#1} UML Nodes\xspace}
\newcommand{\expUC}[1]{\textbf{#1} UML Use Cases\xspace}
\newcommand{\expTime}{\textit{200}\xspace}
\newcommand{\expSolutions}{\textit{70,000}\xspace}

\journal{Information and Software Technology}

\begin{document}

\begin{frontmatter}

\title{Many-Objective Optimization of Non-Functional Attributes based on Refactoring of Software Models}

\author[inst1]{Vittorio Cortellessa\orcidlink{0000-0002-4507-464X}}\ead{vittorio.cortellessa@univaq.it}
\author[inst1]{Daniele Di Pompeo\orcidlink{0000-0003-2041-7375}}\ead{daniele.dipompeo@univaq.it}
\author[inst1]{Vincenzo Stoico\orcidlink{0000-0002-3681-372X}}\ead{vincenzo.stoico@graduate.univaq.it}

\affiliation[inst1]{organization={University of L'Aquila},country={Italy}}

\author[inst2]{Michele Tucci\orcidlink{0000-0002-0329-1101}}

\affiliation[inst2]{organization={Charles University},country={Czech Republic}}
\ead{tucci@d3s.mff.cuni.cz}

\begin{abstract}

\textbf{Context:} Software quality estimation is a challenging and time-consuming activity, and models are crucial to face the complexity of such activity on modern software applications. In this context, software refactoring is a crucial activity within development life-cycles where requirements and functionalities rapidly evolve.

\textbf{Objective:} One main challenge is that the improvement of distinctive quality attributes may require contrasting refactoring actions on software, as for trade-off between performance and reliability (or other non-functional attributes). In such cases, multi-objective optimization can provide the designer with a wider view on these trade-offs and, consequently, can lead to identify suitable refactoring actions that take into account independent or even competing objectives.

\textbf{Method:} In this paper, we present an approach that exploits the \nsga as the genetic algorithm to search optimal Pareto frontiers for software refactoring while considering many objectives. We consider performance and reliability variations of a model alternative with respect to an initial model, the amount of performance antipatterns detected on the model alternative, and the architectural distance, which quantifies the effort to obtain a model alternative from the initial one.

\textbf{Results:} We applied our approach on two case studies: a Train Ticket Booking Service, and CoCoME.
We observed that our approach is able to improve  performance (by up to 42\%) while preserving or even improving the reliability (by up to 32\%) of generated model alternatives.
We also observed that there exists an order of preference of refactoring actions among model alternatives.

\textbf{Conclusion:} Based on our analysis, we can state that performance antipatterns confirmed their ability to improve performance of a subject model in the context of many-objective optimization. In addition, the metric that we adopted for the architectural distance seems to be suitable for estimating the refactoring effort.

\end{abstract}

\begin{highlights}
\item Many-objective optimization of non-functional properties, such as performance, reliability, and performance antipatterns.
\item The role of performance antipatterns on many-objective optimization problem
\item Optimization of refactoring driven by meta-heuristics
\item Automation in model refactoring activity
\end{highlights}


\end{frontmatter}


\section{Introduction}\label{sec:introduction}

Software refactoring~\citep{fowler2018refactoring} can be triggered by different causes, such as the introduction of additional requirements, the adaptation to new execution contexts, or the degradation of non-functional properties. The identification of optimal refactoring actions is a non-trivial task, mostly due to the large space of solutions, while there is still lack of automated support to this task.
Search-based techniques have been involved in such a context~\cite{Bavota:2014kr,Kessentini:2012cb,Mariani:2017jd,Ouni:2015db,Ouni:2017db,Ramirez:2018uz,Ray:2014ip}, and they have proven to suit within the non-functional analysis due to the quantifiable nature of non-functional attributes~\cite{DBLP:conf/mompes/AletiBGM09,Aleti:2013gp,Martens:2010bn}. Among the search-based techniques, those related to multi-objective optimization have been recently applied to model refactoring optimization problems~\cite{CORTELLESSA2021106568,NI2021106565}.
A common aspect of multi-objective optimization approaches applied to model-based software refactoring problems is that they search among design alternatives (\eg through architectural tactics~\cite{Koziolek:2011cg,NI2021106565}).

In this paper, we present an approach based on a many-objective evolutionary algorithm (\ie \nsga~\citep{Deb:2002ut}) that searches sequences of refactoring actions, to be applied on models, leading to the optimization of four objectives: i) performance variation (analyzed through Layered Queueing Networks~\citep{DBLP:journals/tse/NeilsonWPM95}), ii) reliability (analyzed through a closed-form model~\citep{CortellessaSC02}), iii) number of performance antipatterns (automatically detected~\citep{DBLP:journals/infsof/ArcelliCP18}) and iv) architectural distance~\citep{Arcelli:2018vo}. 
A performance antipattern is a bad design decision that might lead to a performance degradation~\citep{DBLP:conf/wosp/SmithW00,DBLP:conf/cmg/SmithW01a}.\footnote{We provide more detail in \Cref{sec:background:pas}.}
In particular, we analyze the composition of model alternatives generated through the application of refactoring actions to the initial model, and we analyze the contribution of the architectural distance to the generation of Pareto frontiers.
Furthermore, we study the impact of performance antipatterns on the quality of refactoring solutions. 
Since it has been shown that removing performance antipatterns leads to systems that show better performance than the ones affected by them~\cite{DBLP:journals/infsof/ArcelliCP18,DBLP:conf/cmg/SmithW01a,Smith:2003wv}, we aim at studying if this result persists in the context of many-objective optimization, where performance improvement is not the only objective. 

Our approach applies to UML models augmented by MARTE \cite{MARTE} and DAM \cite{BernardiMP11} profiles that allow to embed performance and reliability properties. However, UML does not provide native support for performance analysis, thus we introduce a model-to-model transformation that generates Layered Queueing Networks (LQN) from annotated UML models. The solution of LQN models feeds the performance variation objective.

Here, we consider refactoring actions that are designed to improve performance in most cases. Since such actions may also have an impact on other non-functional properties, we introduce the reliability among the optimization objectives to study whether satisfactory levels of performance and reliability can be kept at the same time.
In order to quantify the reliability objective, we adopt an existing model for component-based software systems~\cite{CortellessaSC02} that can be generated from UML models.

We also minimize the distance between the initial UML model and the ones resulting from applying refactoring actions. Indeed, without an objective that minimizes such distance, the proposed solutions could be impractical because they could require to completely disassemble and re-assemble the initial UML model.

In a recent work~\cite{SEAA2021}, we extended the approach in~\cite{Arcelli:2018vo,CORTELLESSA2021106568}, by investigating UML models optimization, thus widening the scope of eligible models. 
In this paper, we extensively apply the approach to two case studies from the literature: Train Ticket Booking Service~\cite{DBLP:conf/staf/Pompeo0CE19,DBLP:journals/tse/ZhouPXSJLD21}, and CoCoME~\cite{Herold2008}.
We analyze the sensitivity of the search process to configuration variations. We refine the cost model of refactoring actions, introduced in~\cite{SEAA2021}, and we investigate how it contributes to the generation of Pareto frontiers. Also, we analyze the characteristics of computed Pareto frontiers in order to extract common properties for both case studies.

This study answers the following research questions:
\begin{itemize}
    \item \emph{RQ1}: To what extent do experimental configurations affect quality of Pareto frontiers?
    \begin{itemize}
        \item \emph{RQ1.1}: Does antipattern detection contribute to find better solutions compared to the case where antipatterns are not considered at all? 
        \item \emph{RQ1.2}: Does the probabilistic nature of fuzzy antipatterns detection help to include higher quality solutions in Pareto frontiers with respect to deterministic one? 
        \item \emph{RQ1.3}: To what extent does the architectural distance contribute to find better alternatives?
    \end{itemize}
    \item \emph{RQ2}: Is it possible to increase reliability without performance degradation?
    \item \emph{RQ3}: What type of refactoring actions are more likely to lead to better solutions? 
\end{itemize}
The experimentation lasted approximately \expTime hours and generated more than \expSolutions model alternatives.

Generally, multi-objective optimization is beneficial when the solution space is so large that an exhaustive search is impractical. 
Hence, due to the search of the solution space, multi-objective optimization requires a lot of time and resources.

Our results show that, by considering the reduction of performance antipatterns as an objective, we are able to obtain model alternatives that show better performance and, in the majority of cases, better reliability as well. We also find that a more sophisticated architectural distance objective estimation helps the optimization process to generate model alternatives showing better quality indicators. Also, we strengthen the idea that performance antipatterns are promising proxies of performance degradation of software models. 
Finally, to encourage reproducibility, we publicly share the implementation of the approach~\footnote{\url{https://github.com/SEALABQualityGroup/EASIER}}, as well as the data gathered during the experimentation~\footnote{\url{https://github.com/SEALABQualityGroup/2022-ist-replication-package}}.

The structure of the paper is the following: \secref{sec:background} introduces basic concepts, \secref{sec:approach} describes the approach, \secref{sec:case-study} describes the two involved case studies, and \secref{sec:settings} details used configurations, in \secref{sec:results} we evaluate our approach and discuss the results, threats to validity are described in \secref{sec:t2v}, \secref{sec:related} reports related work, and \secref{sec:conclusion} concludes the paper.
 \section{Background}\label{sec:background}
We identify four competing objectives of our evolutionary approach as follows: \perfq is a performance quality indicator that quantifies the performance improvement/detriment between an initial model and one obtained by applying the refactoring actions of a solution (\secref{sec:background:perfq}); \reliability is a measure of the reliability of the software model  (\secref{sec:background:reliability}); \pas is a metric that quantifies the amount of performance antipattern occurrences while considering the intrinsic uncertainty rising from thresholds used by the detection mechanism (\secref{sec:background:pas}); \achanges represents the distance between an initial model and one obtained by applying the refactoring actions of a solution (\secref{sec:background:distance}).

We employ the Non-dominated Sorting Algorithm II (\nsga) as our genetic algorithm~\cite{Deb:2002ut}, since it is extensively used in the software engineering community, \eg~\cite{Koziolek:2011cg,Mansoor:2015bm}. 
\nsga randomly creates an initial population of model alternatives, and it used to create the offspring population by applying the \emph{Crossover} with probability $P_{crossover}$, and the \emph{Mutation} with probability $P_{Mutation}$ operators. The union of the initial and the offspring populations is sorted by the \emph{Non-dominated sorting} operator, which identifies different Pareto frontiers with respect to considered objectives. Finally, the \emph{Crowding distance} operator cuts off the worse half of the sorted union population. Hence, the remaining model alternatives become the initial population for the next step.

\subsection{Performance Quality Indicator (\perfq)}\label{sec:background:perfq}

\perfq quantifies the performance improvement/detriment between two models, and it is defined as follows:
\[perfQ(M)=\frac{1}{c}\sum\limits_{j=1}^{c} p_j\cdot \frac{F_j-I_j}{F_j+I_j}\]
where $M$ is a model obtained by applying a refactoring solution to the initial model, $F_j$ is the value of a performance index in $M$, and $I_j$ is the value of the same index on the initial model. $p\in\{-1,1\}$ is a multiplying factor that holds: i) $1$ if the $j$--th index has to be maximized (i.e., the higher the value, the better the performance), like the throughput; ii) $-1$ if the $j$--th index has to be minimized (i.e., the smaller the value, the better the performance), like the response time. 

Notice that, for performance measures representing utilization, $p$ also holds $1$ but we deﬁne a \emph{utilization correction factor} $\Delta_j$ to be added to each j–th term above, as defined in~\cite{Arcelli:2018vo}. The utilization correction factor penalizes refactoring actions that push the utilization too close to 1, i.e., its maximum value.
Finally, the global \perfq is computed as the average across the number $c$ of performance indices considered in the performance analysis.

As mentioned in the introduction, in order to obtain performance indices of a UML model, the analysis has been conducted on Layered Queueing Networks (LQNs)~\citep{DBLP:journals/tse/NeilsonWPM95}\footnote{\url{http://www.sce.carleton.ca/rads/lqns/LQNSUserMan-jan13.pdf}} that are obtained through a model transformation approach from UML to LQN, which we have introduced in~\citep{SEAA2021}. 
We chose Layered Queueing Networks as our performance model notation because it is extensively used in the literature and it allows a more explicit representation of software and hardware components (and their interactions) than the one of conventional Queueing Networks~\citep{Koziolek:2011cg,DBLP:conf/qest/LiAZCP17,NI2021106565}.

\subsection{Reliability model}\label{sec:background:reliability}

The reliability model that we adopt here to quantify the \reliability objective is based on the model introduced in~\cite{CortellessaSC02}. The mean failure probability $\theta_S$ of a software system $S$ is defined by the following equation:
\[ \theta_S = 1 - \sum\limits_{j=1}^K p_j \left( \prod\limits_{i=1}^N (1 - \theta_i)^{InvNr_{ij}} \cdot \prod\limits_{l=1}^L (1 - \psi_{l})^{MsgSize(l,j)} \right) \]
This model takes into account failure probabilities of components ($\theta_i$) and communication links ($\psi_{l}$), as well as the probability of a scenario to be executed ($p_j$). Such probabilities are combined to obtain the overall reliability on demand of the system ($\theta_S$), which represents how often the system is not expected to fail when its scenarios are invoked.

The model is considered to be composed of $N$ components and $L$ communication links, whereas its behavior is made of $K$ scenarios. The probability ($p_j$) of a scenario $j$ to be executed is multiplied by an expression that describes the probability that no component or link fails during the execution of the scenario. This expression is composed of two terms: $\prod_{i=1}^N (1 - \theta_i)^{InvNr_{ij}}$, which is the probability of the involved components not to fail raised to the power of their number of invocations in the scenario (denoted by $InvNr_{ij}$), and $\prod_{l=1}^L (1 - \psi_{l})^{MsgSize(l,j)}$, which is the probability of the involved links not to fail raised to the power of the size of messages traversing them in the scenario (denoted by $MsgSize(l,j)$). 

\subsection{Performance Antipatterns}\label{sec:background:pas}

A performance antipattern describes bad design practices that might lead to performance degradation in a system. Smith and Williams have introduced the concepts of performance antipatterns in~\citep{DBLP:conf/cmg/SmithW01a,DBLP:books/sp/03/SmithW03}. These textual descriptions were later translated into a first-order logic (FOL) equations~\citep{DBLP:journals/sosym/CortellessaMT14}.

\begin{table*}[htbp]
    \centering
   \begin{tabular}{p{.28\textwidth}p{.7\textwidth}}
    \toprule
        Performance antipattern & Description \\
    \midrule
        Pipe and Filter              & Occurs when the slowest filter in a ``pipe and filter'' causes the system to have unacceptable throughput. \\
	\midrule
        Blob                         & Occurs when a single component either i) performs the greatest part of the work of a software system or ii) holds the greatest part of the data of the software system. Either manifestation results in excessive message traffic that may degrade performance. \\
	\midrule
        Concurrent Processing System & Occurs when processing cannot make use of available processors. \\
	\midrule
        Extensive Processing         & Occurs when extensive processing in general impedes overall response time.\\ 
	\midrule
        Empty Semi-Truck             & Occurs when an excessive number of requests is required to perform a task. It may be due to inefficient use of available bandwidth, an inefficient interface, or both. \\
	\midrule
        Tower of Babel               & Occurs when processes use different data formats and they spend too much time in convert them to an internal format. \\
    \bottomrule
    \end{tabular}
	\caption{Detectable performance antipatterns in our approach. Left column lists performance antipattern names, while right column lists performance antipattern descriptions~\cite{Smith:2003wv}.}
    \label{tab:supported-pas}
\end{table*}
 
A performance antipattern FOL is a combination of multiple literals, where each one represents a system aspect (\eg the number of connections among components). 
These literals must be compared to thresholds in order to reveal the occurrence of a performance antipattern. The identification of such thresholds is a non-trivial task, and using deterministic values may result in an excessively strict detection where the smallest change in the value of a literal determines the occurrence of the antipattern. For these reasons, we employ a fuzzy detection~\cite{DBLP:conf/fase/ArcelliCT15}, which assigns to each performance antipattern a probability to be an antipattern. An example of a performance antipattern fuzzy detection is the following:
\[1 - \frac{UB(literal) - literal}{UB(literal) - LB(literal)}\]
The upper (UB) and the lower (LB) bounds, in the above equation, are the maximum and minimum values of the $literal$ computed on the entire system. 
Instead of detecting a performance antipattern in a deterministic way, such thresholds lead to assign probabilities to antipattern occurrences. 
In this study, we detect the performance antipatterns listed in \Cref{tab:supported-pas}.

\subsection{Architectural distance}\label{sec:background:distance}

The architectural distance, that we express here as \achanges, represents the distance of the model obtained by applying refactoring actions from the initial one~\citep{Arcelli:2018vo}. 
On one side, a \emph{baseline refactoring factor (\brf)} is associated to each refactoring action in our portfolio, and it expresses the refactoring effort to be spent when applying the action.
On the other side, an \emph{architectural weight (AW)} is associated to each model element on the basis of the number of connections to other elements in the model.
Hence, we quantify the effort needed to perform a refactoring as the product between the \emph{baseline refactoring factor} of an action and the \emph{architectural weight} of the model element on which that action is applied. \achanges is obtained by summing the efforts of all refactoring actions contained in a solution.

Furthermore, \brf and AW can assume any positive value (\ie zero is a non-admitted value because it would lead the optimizer to always select only actions by that type).

As an example, let us assume that a refactoring sequence is made up of two refactoring actions: A1 with $\brf(A1)=1.23$, and A2 with $\brf(A2)=2.3$. For each refactoring action, the algorithm randomly selects a target element in the model. For instance, let those target elements be: E1 with $AW(E1)=1.43$, and E2 with $AW(E2)=1.32$. The resulting \achanges of A1 and A2 would be:
\[\achanges(A1,A2) = 1.23 \cdot 1.43 + 2.3 \cdot 1.32 \]
Details about the \emph{baseline refactoring factor} for each considered refactoring action are provided in \secref{sec:approach:brf}.
 \section{Approach}\label{sec:approach}

\begin{figure}
      \centering
      \includegraphics[width=0.8\linewidth]{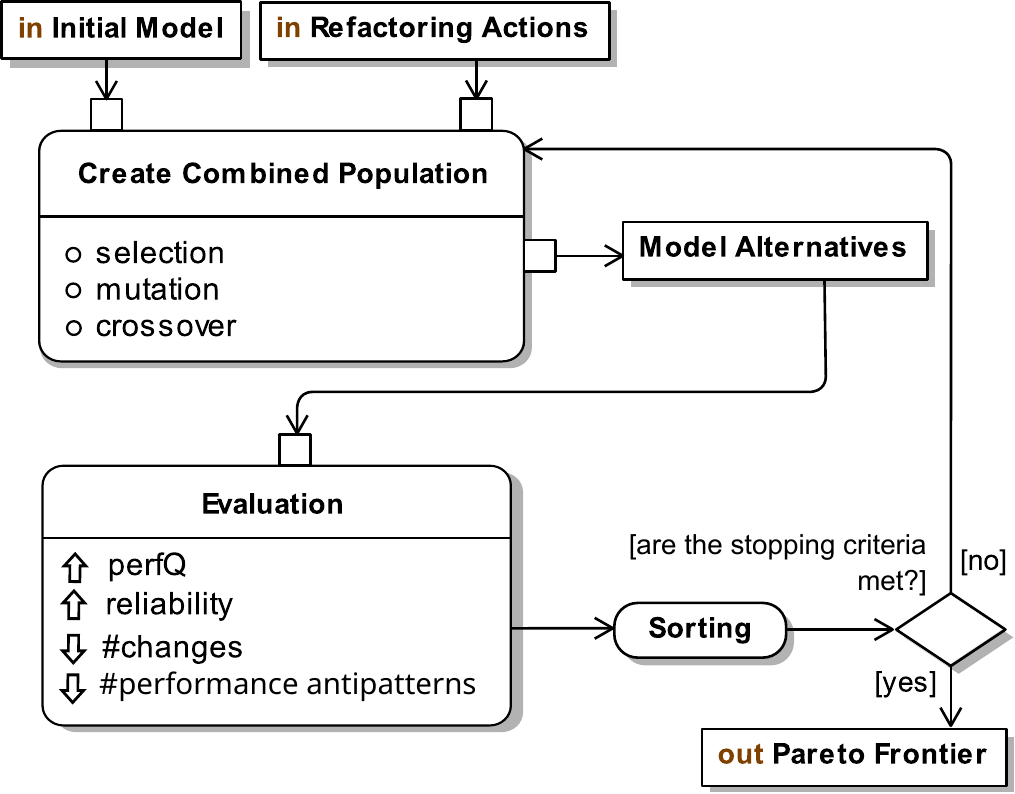}
      \caption{Our multi-objective evolutionary approach}
      \label{fig:process}
    \end{figure}

\figref{fig:process} depicts the process we present in this paper.
The process uses a UML model and a set of refactoring actions as input. The \emph{Initial Model} and the \emph{Refactoring Actions} are involved within the \emph{Create Combined Population} step, where mating operations (\ie selection, mutation, and crossover) are put in place to create \emph{Model Alternatives}. The mating operations randomly apply the refactoring actions, which generate alternatives functionally equivalent to the initial model.
Therefore, the \emph{Evaluation} step is applied to each model alternative. Subsequently, the model alternatives are ranked (\emph{Sorting} step) according to four objectives: \perfq, \reliability, \achanges, and \pas. 
The optimal model alternatives (\ie non-dominated alternatives) become the input of the next iteration. The process continues until the stopping criteria are met. Finally, the process generates a \emph{Pareto Frontier}, which contains all non-dominated model alternatives. 

\subsection{Assumptions on UML models}
\label{ref:assumptions}

In our approach, we consider UML models including three views, namely \emph{static, dynamic} and \emph{deployment} views. The static view is modeled by a UML Component diagram in which static connections among components are represented by interface realizations and their usages. The dynamic view is described by UML Use Case and Sequence diagrams. A Use Case diagram defines user scenarios, while a Sequence diagram describes the behavior inside a single scenario through component operations (as defined in their interfaces) and interactions among them. A Deployment diagram is used to model platform information and map Components to Deployment Nodes. As mentioned before, we use an augmented UML notation by embedding two existing profiles, namely MARTE~\cite{MARTE} that expresses performance concepts, and DAM~\cite{BernardiMP11} that expresses reliability concepts.

Although our assumptions on UML models seem to require an upfront modeling phase, the accuracy of results is affected by the quality of model and annotations.
We mitigate the modeling effort through the usage of UML.
In fact, a plethora of UML modeling tools is available, each equipped with  entry-level or advanced capabilities that differently help software models design.\footnote{\url{https://en.wikipedia.org/wiki/List_of_Unified_Modeling_Language_tools}}

\subsection{The Refactoring Engine}\label{sec:approach:refactoring}

The automated refactoring of UML models is a key point when evolutionary algorithms are employed in order to optimize some model attributes.
For the sake of full automation of our approach, we have implemented a refactoring engine that applies refactoring actions on UML software models~\citep{DBLP:conf/wcre/ArcelliCP19}.

Each solution that our evolutionary algorithm produces is a sequence of refactoring actions that, once applied to an initial model, leads to a model alternative that shows different non-functional properties. 
Since our refactoring actions are combined during the evolutionary approach, we exploit the feasibility engine that verifies in advance whether a sequence of refactoring actions is feasible or not~\cite{DBLP:conf/kbse/ArcelliCP18}. 

Our refactoring actions are equipped with pre- and post-condition.
While the pre-condition represents the model state for enabling the action, the post-condition represents the model state when the action has been applied.
The approach extracts a refactoring action and adds it to the sequence.
As soon as the action is selected, it randomly extracts a model element (\ie the target element).
Thus, the refactoring engine checks the feasibility of the (partial) sequence of refactoring actions.When the latest added action makes the sequence unfeasible, the engine discards the action and replaces it with a new one.
The engine reduces a sequence of refactoring actions to a single refactoring action, which includes all the changes (see Equation~\eqref{eq:refFeasibility}).

For example, considering two refactoring actions ($M_{i}$, and $M_{j}$), then the global pre-condition is obtained by logical ANDing the first action pre-condition (${^{P_r}}M_{i}$) and all the parts of $M_{j}$ pre-condition that are not yet verified by $M_{i}$ post-conditions ($ M_{j}^{P_r}\ / \ M_i{^{P_o}} $) (see Equations~\eqref{eq:refPrecond}). Since the status of the model after a refactoring is synthesized by its post-condition, we can discard the parts of a subsequent refactoring pre-condition that, by construction, are already verified by its post-condition.
The global post-condition is obtained by logical ANDing all post-conditions within the sequence ($ M_{i}^{P_o} \wedge M_{j}^{P_o} $) (see Equation~\eqref{eq:refPost}).
\begin{subequations}
\begin{align}
    {^{P_r}}M_{i}^{P_o} \wedge {^{P_r}}M_{j}^{P_o} \longmapsto  {^{P_r}}M^{P_o} \label{eq:refFeasibility}\\
	{^{P_r}}M_{i} \wedge M_{j}^{P_r}\ / \ M_i{^{P_o}} \longmapsto {^{P_r}}M \label{eq:refPrecond}\\
	M_{i}^{P_o} \wedge M_{j}^{P_o} \longmapsto M^{P_o} \label{eq:refPost}
\end{align}
\end{subequations}
Our feasibility engine also allows to reduce the number of invalid refactoring sequences, thus reducing the computational time.

\subsubsection{Refactoring Action portfolio}

\begin{figure}[ht]
                \begin{subfigure}[b]{.46\textwidth}
                        \includegraphics[width=\textwidth]{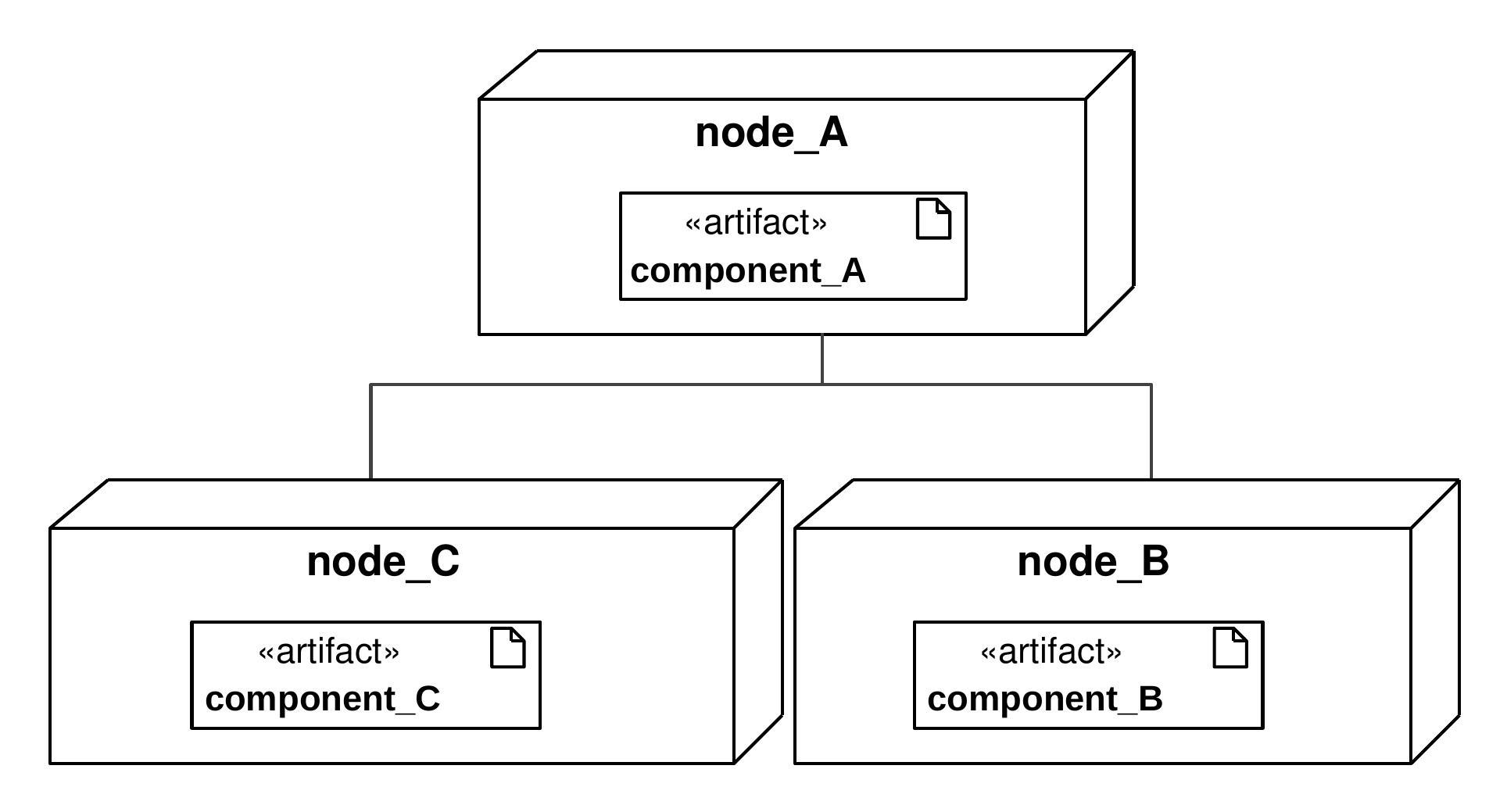}
                        \caption{Initial}\label{fig:clon-deploy}
                \end{subfigure}
                \begin{subfigure}[b]{.46\textwidth}
                        \includegraphics[width=.9\textwidth]{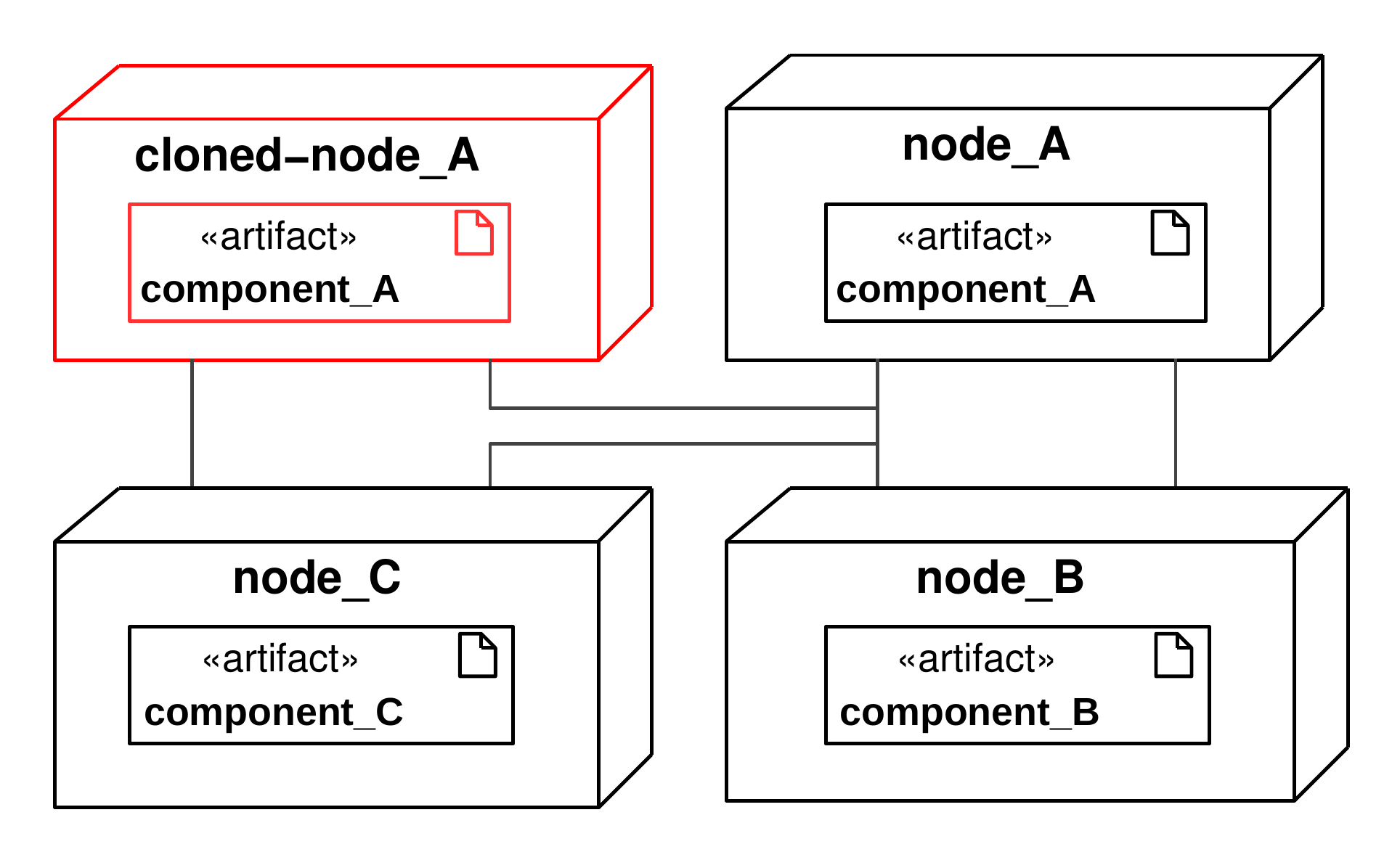}
                        \caption{Refactored}\label{fig:ref-clon-deploy}
                \end{subfigure}
		\caption{The \textit{Clon} refactoring action example on \emph{node\_A} through a UML Software Model}
                \label{fig:ref-clon-uml-diagrams}
        \end{figure}

\begin{figure}[ht]
                \begin{subfigure}[b]{.46\textwidth}
                \includegraphics[width=\textwidth]{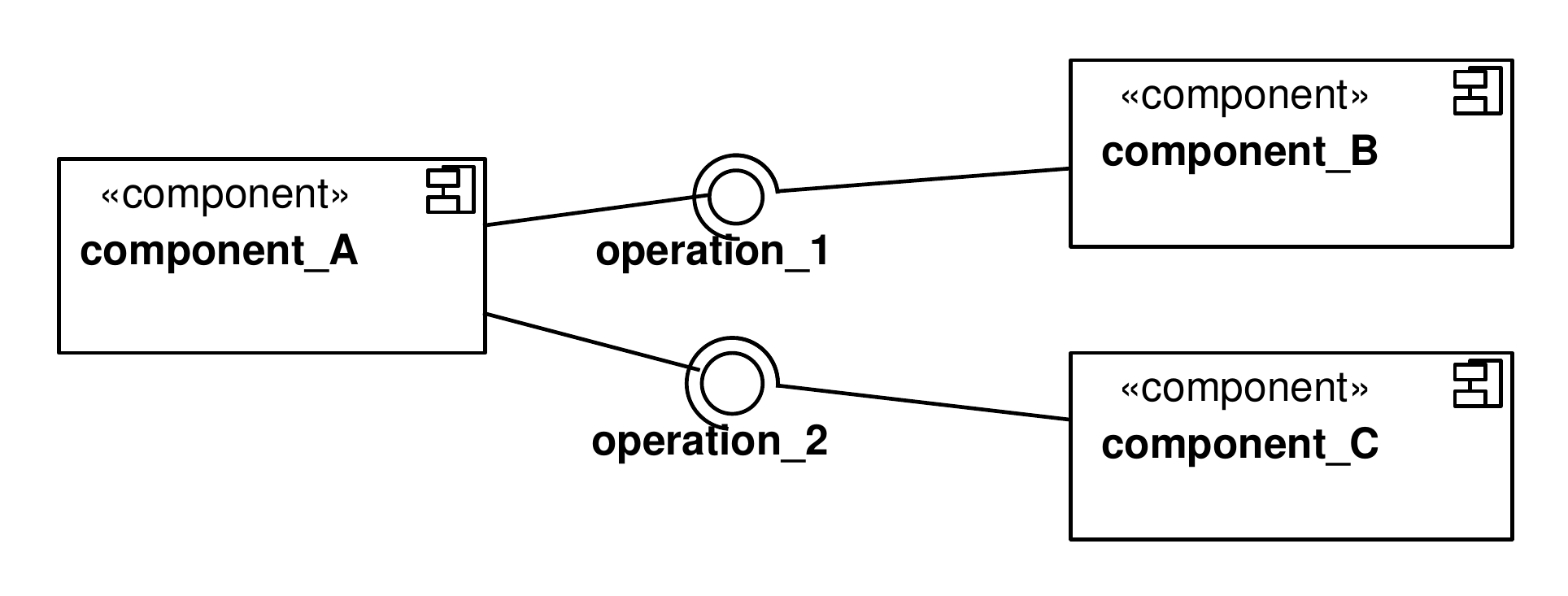}
                        \caption{Initial}\label{fig:mo2n-comp}
                \end{subfigure}
                \begin{subfigure}[b]{.46\textwidth}
                        \includegraphics[width=\textwidth]{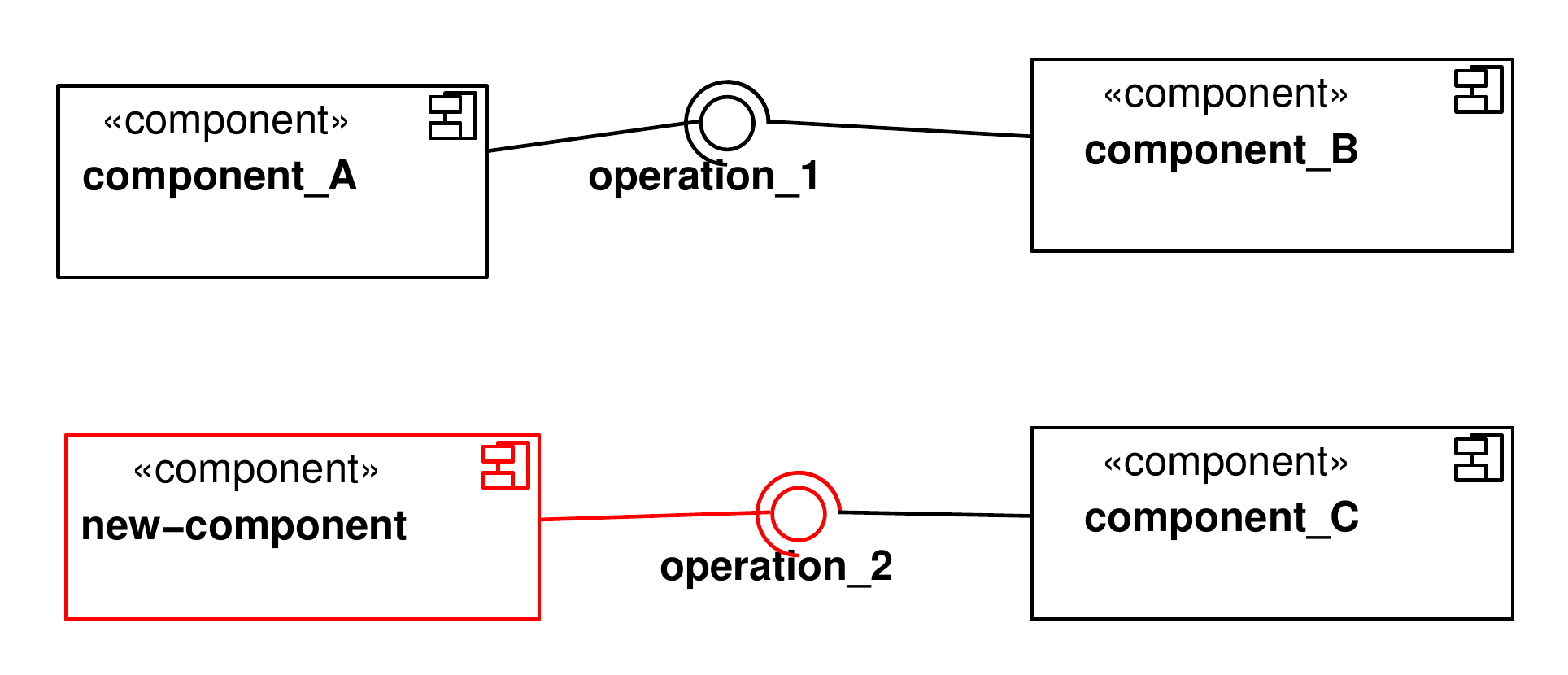}
                        \caption{Refactored}\label{fig:ref-mo2n-comp}
                \end{subfigure}
                \begin{subfigure}[b]{.46\textwidth}
                        \includegraphics[width=\textwidth]{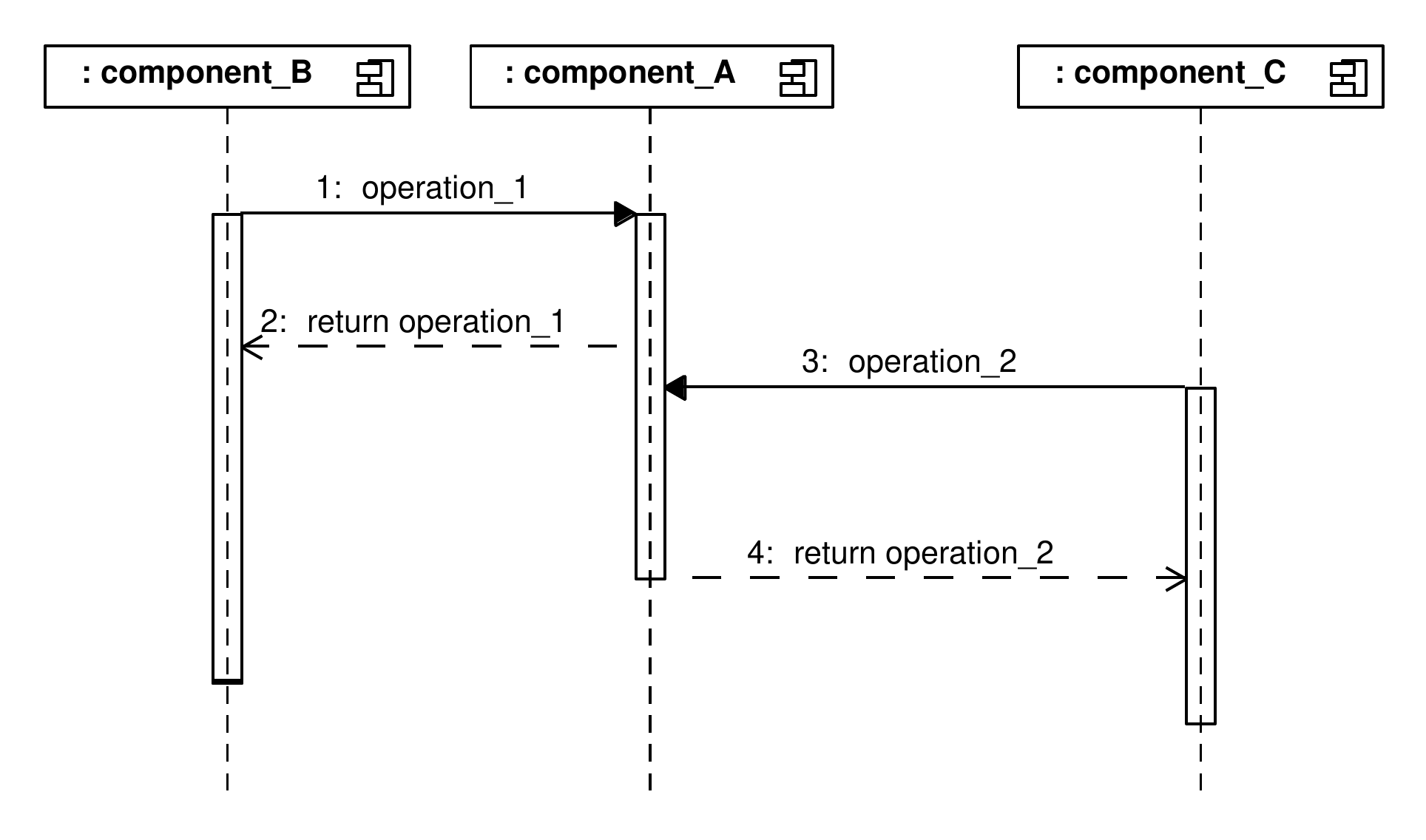}
                        \caption{Initial}\label{fig:mo2n-dynamic}
                \end{subfigure}
                \begin{subfigure}[b]{.46\textwidth}
                        \includegraphics[width=1.25\textwidth]{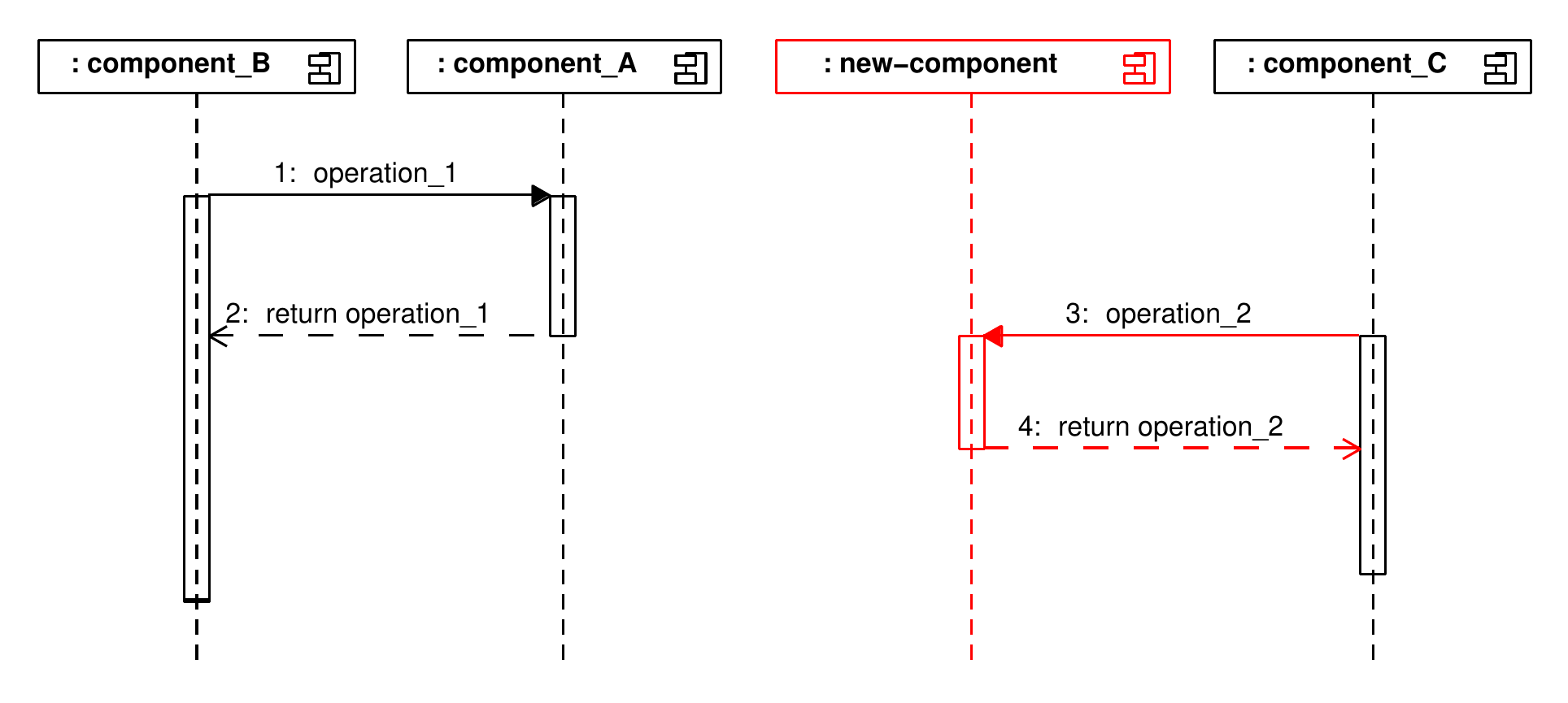}
                        \caption{Refactored}\label{fig:ref-mo2n-dynamic}
                \end{subfigure}
                \begin{subfigure}[b]{.46\textwidth}
                        \includegraphics[width=\textwidth]{img/ref-actions/deployment}
                        \caption{Initial}\label{fig:mo2n-deploy}
                \end{subfigure}
                \begin{subfigure}[b]{.46\textwidth}
                        \includegraphics[width=\textwidth]{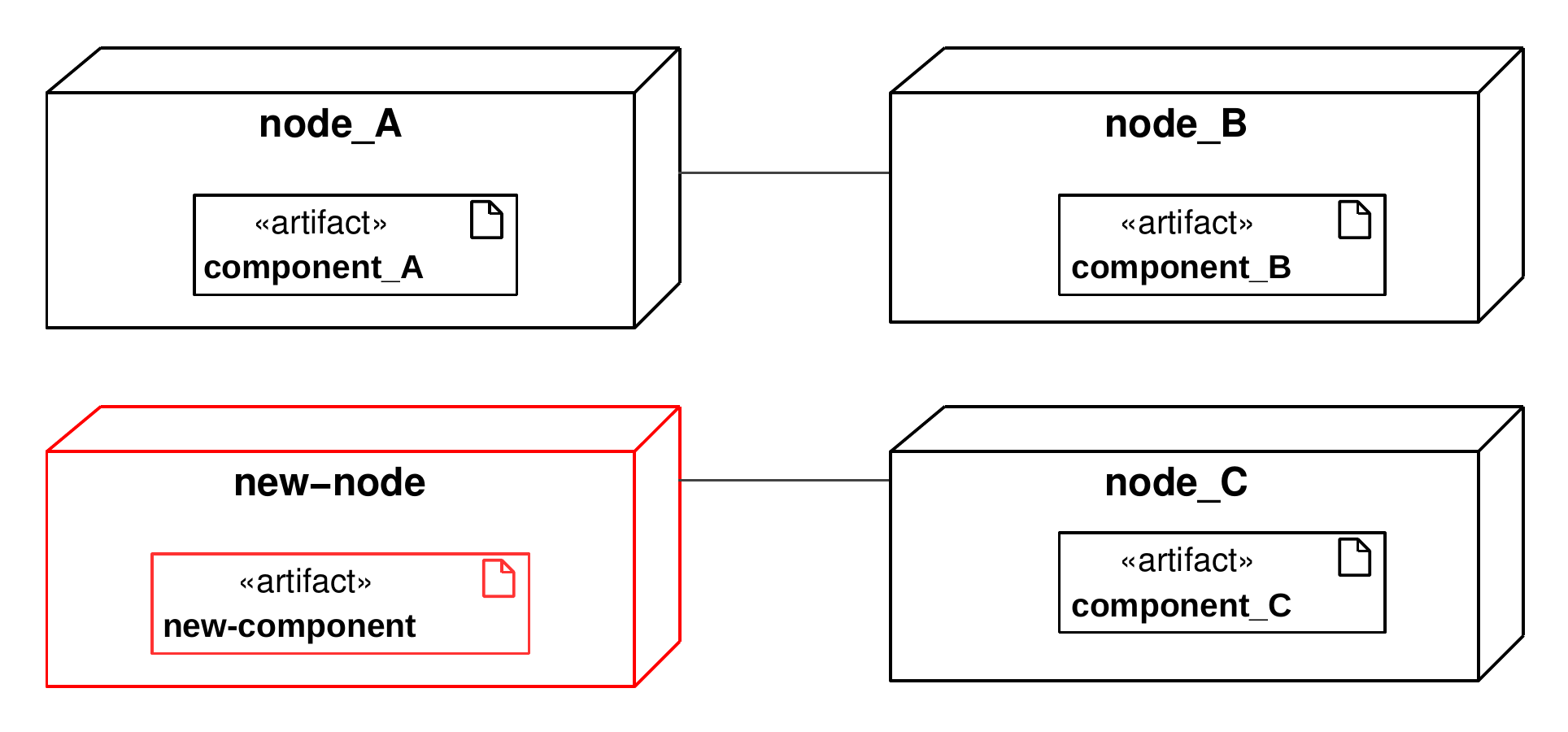}
                        \caption{Refactored}\label{fig:ref-mo2n-deploy}
                \end{subfigure}
                \caption{The \textit{MO2N} refactoring action example on \emph{operation\_2} through a UML Software Model}
                \label{fig:ref-mo2n-uml-diagrams}
        \end{figure}

\begin{figure}[ht]
                \begin{subfigure}[b]{.46\textwidth}
                       \includegraphics[width=\textwidth]{img/ref-actions/staticview}
                       \caption{Initial}\label{fig:mo2c-comp}
                \end{subfigure}
                \begin{subfigure}[b]{.46\textwidth}
                        \includegraphics[width=\textwidth]{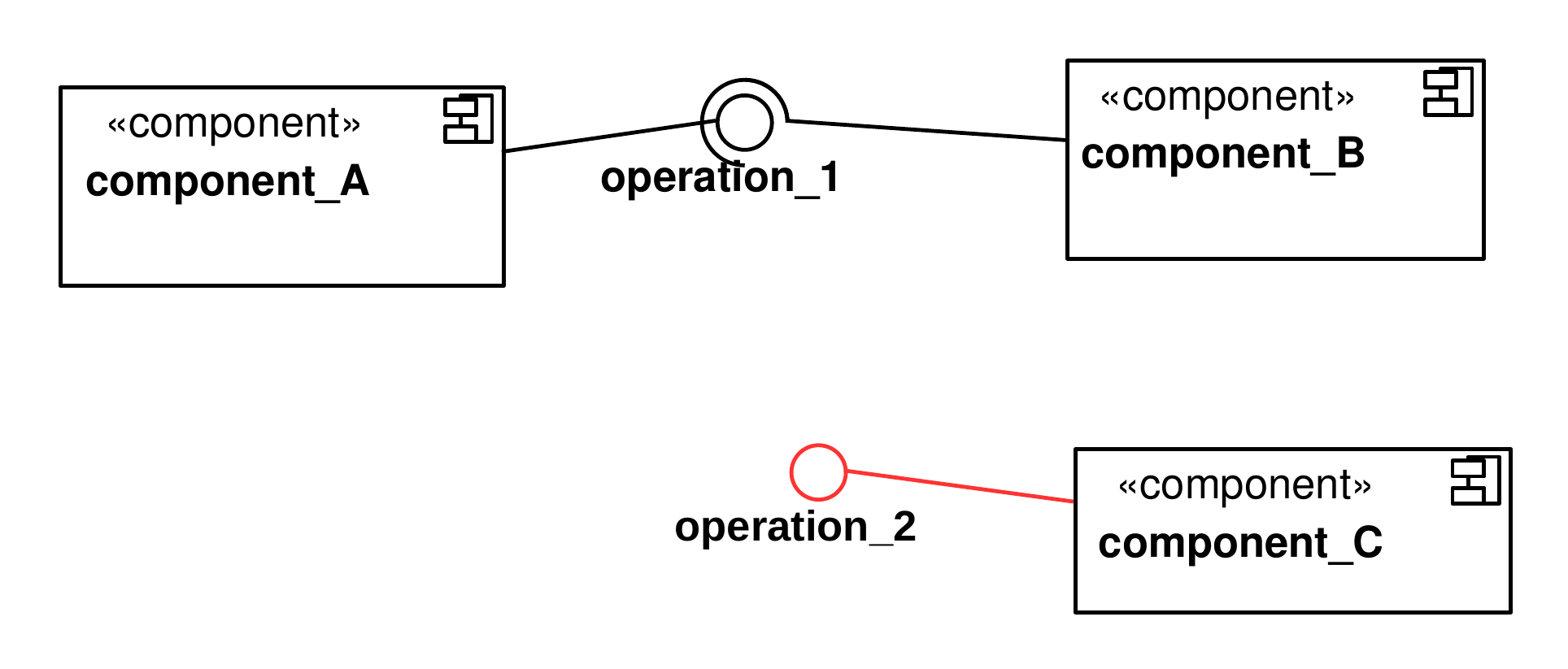}
                        \caption{Refactored}\label{fig:ref-mo2c-comp}
                \end{subfigure}
                \begin{subfigure}[b]{.46\textwidth}
                        \includegraphics[width=\textwidth]{img/ref-actions/dynamic}
                        \caption{Initial}\label{fig:mo2c-dynamic}
                \end{subfigure}
                \begin{subfigure}[b]{.46\textwidth}
                        \includegraphics[width=1.07\textwidth]{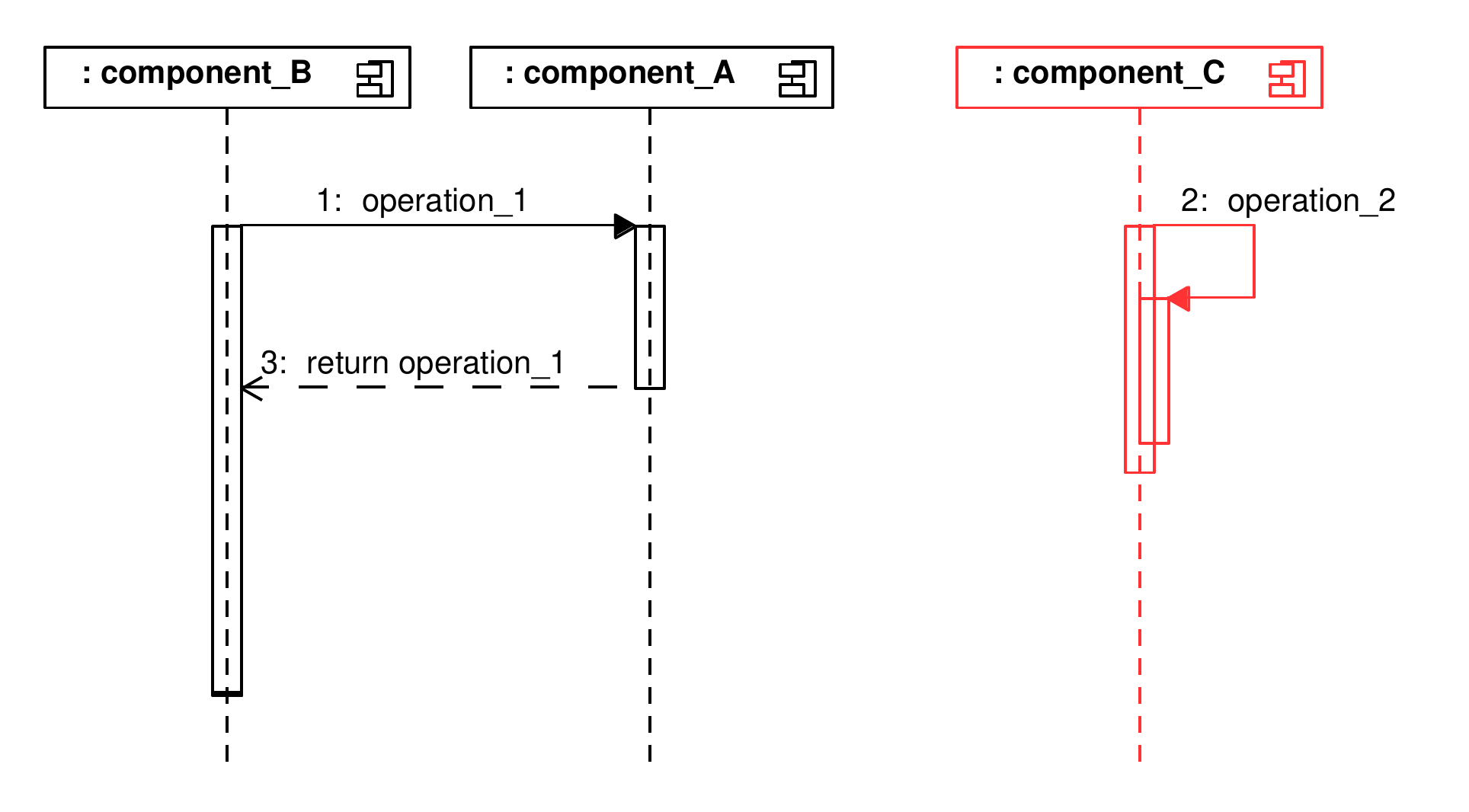}
                        \caption{Refactored}\label{fig:ref-mo2c-dynamic}
                \end{subfigure}
		\caption{The \textit{MO2C} refactoring action example on \emph{operation\_2} and \emph{component\_C} through a UML Software Model}
                \label{fig:ref-mo2c-uml-diagrams}
   \end{figure}

\begin{figure}[ht]
                \begin{subfigure}[b]{.46\textwidth}
                       \includegraphics[width=\textwidth]{img/ref-actions/deployment}
                       \caption{Initial}\label{fig:rede-comp}
                \end{subfigure}
                \begin{subfigure}[b]{.46\textwidth}
                        \includegraphics[width=\textwidth]{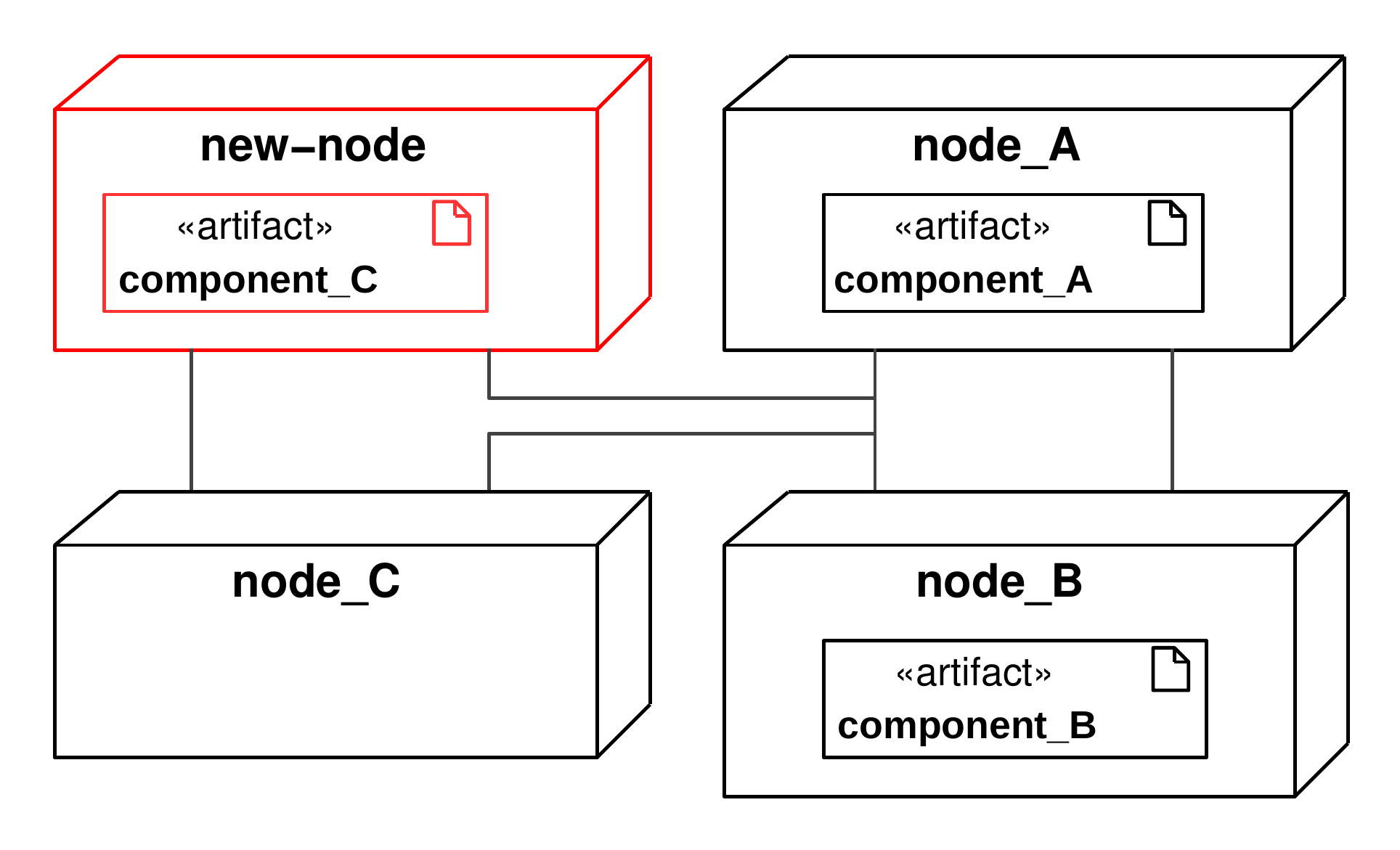}
                        \caption{Refactored}\label{fig:ref-rede-comp}
                \end{subfigure}
                \caption{The \textit{ReDe} refactoring action example on \emph{component\_C} through a UML Software Model}
		\label{fig:ref-rede-uml-diagrams}
        \end{figure}

\Cref{fig:ref-clon-uml-diagrams} through \Cref{fig:ref-rede-uml-diagrams} show a graphic representation of each refactoring action.
Each figure's left side shows the original model (\eg static view in \Cref{fig:mo2n-comp}, dynamic view in \Cref{fig:mo2n-dynamic}, and deployment view in \Cref{fig:mo2n-deploy}), while the refactored version is shown on the right side (\eg static view in \Cref{fig:ref-mo2n-comp}, dynamic view in \Cref{fig:ref-mo2n-dynamic}, and deployment view in \Cref{fig:ref-mo2n-deploy}).
The red highlights indicate changes.

\paragraph{Clone a Node (Clon)}
This action is aimed at introducing a replica of a Node. Adding a replica means that every deployed artifact and every connection of the original Node has to be in turn cloned. Stereotypes and their tagged values are cloned as well. The rationale of this action is to introduce a replica of a platform device with the aim of reducing its utilization.

\paragraph{Move an Operation to a new Component deployed on a new Node (MO2N)}
This action is in charge of randomly selecting an operation and moving it to a new Component. All the elements related to the moving operation (\eg links) will move as well. 
Since we adopt a multi-view model, and coherence among views has to be preserved, this action has to synchronize dynamic and deployment views. A lifeline for the newly created Component is added in the dynamic view, and messages related to the moved operation are forwarded to it. In the deployment view, instead, a new Node, a new artifact, and related links are created. The rationale of this action is to lighten the load of the original Component and Node.

\paragraph{Move an Operation to a Component (MO2C)}
This action is in charge of randomly selecting and transferring an Operation to an arbitrary existing target Component. The action consequently modifies each UML Use Case in which the Operation is involved. Sequence Diagrams are also updated to include a new lifeline representing the Component owning the Operation, but also to re-assign the messages invoking the operation to the newly created lifeline. The rationale of this action is quite similar to the previous refactoring action, but without adding a new UML Node to the model. 

\paragraph{Deploy a Component on a new Node (ReDe)}
This action simply modifies the deployment view by redeploying a Component to a newly created Node.
In order to be consistent with the initial model, the new Node is connected with all other ones directly connected to the Node on which the target Component was originally deployed. The rationale of this action is to lighten the load of the original UML Node by transferring the load of the moving Component to a new UML Node. 

\subsection{Baseline Refactoring Factor}\label{sec:approach:brf}
As described in \secref{sec:background:distance}, we measure the architectural distance by summing the products of baseline refactoring factor (\brf) and architectural weight (AW) for each refactoring action $a_i(el_j)$ within a sequence ($\mathbb{A}$). 
\[\achanges(\mathbb{A}) = \sum_{a_i(el_j) \in \mathbb{A}} \brf(a_i) \times AW(el_j)\]

    \begin{table}
\centering
    \begin{tabular}{llrr}
    \toprule
        Action & \brf & \ttbs & \ccm \\
    \midrule
MO2N & 1.80  & 70   & $\approx 4.8 \times 10^{3}$ \\
MO2C                   & 1.64  & $\approx 1.5 \times 10^{6}$  &	$\approx 1.3 \times 10^{8}$ \\
ReDe                & 1.45  & $\approx 3 \times 10^{2}$      & $\approx 7 \times 10^{2}$ \\
Clon                                  & 1.23  & $\approx 3 \times 10^{2}$      &	70 \\
        \midrule
       \multicolumn{2}{c}{$\Omega$} &	$9.45 \times 10^{12}$   &	$3.05 \times 10^{16}$ \\
    \bottomrule
    \end{tabular}
    \caption{A detailed size of the solution space ($\Omega$) computation.}
    \label{tab:solution_space_size}
\end{table}
AW is the weight of the target of the refactoring action, \brf is the intrinsic cost that one should pay in order to apply the specific action on a model element. There are different ways to compute the effort for implementing software artefacts or maintaining them (\eg COCOMO-II~\cite{boehm2009software}, and CoBRA~\cite{trendowicz2013software}). Nevertheless, we consider the cost in terms of the effort that one should spend on the model to complete a refactoring action, and we assign \brf values on the basis of our past experience in manual refactoring. We have not used a cost estimator model, such as CoBRA, because it requires to collect business information that is not available for non-industrial case studies. \tabref{tab:solution_space_size} lists the \brf values used in this study.
It is worth remarking that, in our optimization problem, the ratio among \brf values is more important than how each single value has been extracted.

\subsection{Computing reliability on UML models}

The reliability parameters of the model introduced in \secref{sec:background:reliability} are annotated on UML models by means of the MARTE-DAM profile.
The probability of executing a scenario ($p_j$) is specified by annotating UML Use Cases with the \emph{GaScenario} stereotype. This stereotype has a tag named \emph{root} that is a reference to the first \emph{GaStep} in a sequence. We use the \emph{GaScenario.root} tag to point to the triggering UML Message of a Sequence Diagram and the \emph{GaStep.prob} to set the execution probability.
Failure probabilities of components ($\theta_i$) are defined by applying the \emph{DaComponent} stereotype on each UML Component and by setting, in the \emph{failure} tag, a \emph{DaFailure} element with the failure probability specified in the \emph{occurrenceProb} tag.
Analogously, failure probabilities of links ($\psi_{l}$) are defined in the \emph{failure.occurrenceProb} tag of the \emph{DaConnector} stereotype that we apply on UML CommunicationPath elements. Such elements represent the connection links between UML Nodes in a Deployment Diagram.
Sequence Diagrams are traversed to obtain the number of invocations of a component $i$ in a scenario $j$ (denoted by $InvNr_{ij}$ in our reliability model), but also to compute the total size of messages passing over a link $l$ in a scenario $j$ (denoted by $MsgSize(l,j)$). The size of a single UML Message is annotated using the \emph{GaStep.msgSize} tag. The Java implementation of the reliability model is available online.\footnote{\url{https://github.com/SEALABQualityGroup/uml-reliability}}

\subsection{Pareto Frontier Quality Indicators}\label{sec:approach:qi}

We compare the performance of the \nsga while varying the configuration eligible values listed in \tabref{tab:config_params}. We used well-established quality indicators also provided in the JMetal framework~\cite{DBLP:conf/gecco/NebroDV15}. We use quality indicators to quantify the difference among computed Pareto frontiers (\computedP) with respect to the reference Pareto frontier (\referenceP)~\cite{Ali_Arcaini_Pradhan_Safdar_Yue_2020}. Therefore, we can declare which configuration outperform the others.

\begin{table}
	\centering
	\begin{tabular}{lll}
		\toprule
		{} & Configuration & Eligible values  \\
		\midrule
		\multirow{3}{*}{\shortstack{Experiment\\settings}} & Baseline Refactoring Factor & no, yes  \\
		& Performance Antipattern fuzziness &  0.55, 0.80, 0.95 \\
		& Case Study & \ttbs, \ccm \\
		\midrule
		\multirow{6}{*}{\nsga} & Number of genetic evolutions &  72, 82, 102  \\
		& Population Size & 16 \\
		& Number of independent runs & 3 \\
		& Selection operator & Binary Tournament Selection \\
		& $P_{crossover}$ & 0.80 \\
		& Crossover Operator & Single Point \\
		& $P_{mutation}$ & 0.20 \\
		& Mutation Operator & Simple Mutation \\ 
		\bottomrule
	\end{tabular}
	\caption{\label{tab:config_params} Eligible configuration values.}
\end{table} 

In the following, we recall some characteristics for each quality indicator.

\paragraph{GSPREAD} The Generalized SPREAD is a quality indicator to be minimized, and it measures the spread of solution within \computedP~\cite{DBLP:conf/cec/ZhouJZST06}. It is computed as follows:
\[GSPREAD(PF^c) = \frac{\sum_{i=1}^{m} d(e_i,PF^c) + \sum_{s \in PF^c} \abs{id(s, PF^c) - \bar{id}}}{\sum_{i=0}^{m} d(e_i, PF^c) + \abs{PF^c}* \bar{id}}\]
where $e_i$ is the optimal value for the objective $f_i$, \ie $(e_1, \dots, e_m)$ is the extreme solution in \referenceP, $id(s, PF^c) = d(s, PF^c \backslash \{s\})$ is the minimal distance of a solution $s$ from the solutions in \computedP, and $\bar{id}$ is the mean value of $id(s, PF^c)$ across the solutions $s$ in \referenceP.

\medskip 
\paragraph{IGD$^+$} The Inverse Generational Distance plus is a quality indicator to be minimized. It measures the distance from a solution in \referenceP to the nearest solutions in \computedP~\cite{DBLP:conf/cec/IshibuchiMN16}. It is computed as follows:
\[IGD^+(PF^c) = \frac{\sqrt{\sum_{s \in PF^{ref}} d(s, PF^c)^2}}{\abs{PF^{ref}}}\]

\medskip 
\paragraph{Hypervolume} The Hypervolume indicator is to be maximized and it measures the volume of the solution space $\Omega$ covered by \computedP~\cite{DBLP:journals/tec/ZitzlerT99}. It is computed as follows:
\[HV(PF^c) = volume(\cup_{s_i \in PF^c} hc(s_i))\]
where $s_i$ is a solution within the \computedP, $hc(s_i)$ is the hypercube having $s_i$ and $w$ as diagonal points. The variable $w$ is the reference point computed using the worst objective function values among all the possible solutions in \computedP.

\medskip
\paragraph{EPSILON} The EPSILON quality indicator measures the smallest distance that each solution within \computedP should be translated so that \computedP dominates \referenceP~\cite{DBLP:journals/tec/ZitzlerTLFF03}. EPSILON is a quality indicator to be  minimized, and it uses the notation of epsilon-dominance $\succ_{\epsilon}$. It is computed as follows:
\[EP(PF^c) = inf\{ \epsilon \in \mathbb{R} | (\forall{x \in PF^{ref}, \exists y \in PF^c : y \succ_{\epsilon} x })\}\]

In our study, we have computed a \referenceP for each case study by extracting every non-dominated solutions across each \computedP, \ie one for each configuration. Hence, the quality indicators in \tabref{tab:best-qi-ttbs} and \tabref{tab:best-qi-ccm} have been computed with respect to the \referenceP for the \ttbs, and \ccm case study respectively.
 \section{Case Studies}\label{sec:case-study}

In this section, we apply our approach to the Train Ticket Booking Service (\ttbs) case study~\cite{DBLP:conf/staf/Pompeo0CE19,DBLP:journals/tse/ZhouPXSJLD21}, and to the well-established model case study \ccm, whose UML model has been derived by the specification in~\cite{Herold2008}. 

\subsection{Train Ticket Booking Service}

Train Ticket Booking Service (\ttbs) is a web-based booking application, whose architecture is based on the microservice paradigm. The system is made up of 40 microservices, and it provides different scenarios through users that can perform realistic operations, \eg book a ticket or watch trip information like intermediate stops. The application employs a docker container for each microservice, and connections among them are managed by a central pivot container.

Our UML model of \ttbs is available online.\footnote{\url{https://github.com/SEALABQualityGroup/2022-ist-replication-package/tree/main/case-studies/train-ticket}}
The static view is made of \expComp{11}, where each component represents a microservice. In the deployment view, we consider \expNode{11}, each one representing a docker container.

Among all \ttbs scenarios shown in~\cite{DBLP:conf/staf/Pompeo0CE19}, in this paper we have considered \expUC{3}, namely \emph{login}, \emph{update user details} and \emph{rebook}. We selected these three scenarios because they commonly represent performance-critical ones in a ticketing booking service. Each scenario is described by a UML Sequence Diagram. Furthermore, the model comprises two user categories: simple and admin users. The simple user category can perform the login and the rebook scenarios, while the admin category can perform the login and the update user details scenarios.

\subsection{\ccm}

The component-based system engineering domain has always been characterized by a plethora of standards for implementing, documenting, and deploying components.  These standards are well-known as component models. Before the birth of the common component modeling example (\ccm) \cite{Herold2008}, it was hard for researchers to compare different component models. \ccm is a case study that acts as a single specification to be implemented using different component models.

\ccm describes a Trading System containing several stores. A store might have one or more cash desks for processing goodies. A cash desk is equipped with all the tools needed to serve a customer (e.g., a Cash Box, Printer, Bar Code Scanner). \ccm covers possible scenarios performed at a cash desk (e.g., scanning products, paying by credit card, generating reports, or ordering new goodies). A set of cash desks forms a cash desk line. The latter is connected to the store server for registering cash desk line activities. Instead, a set of stores is organized in an enterprise having its server for monitoring stores operations. 

\ccm describes 8 scenarios involving more than 20 components. We have modeled this case study using UML and following the structure described in \secref{ref:assumptions}. 
From the \ccm original specification, we analyzed different operational profiles, \ie scenarios triggered by different actors (such as Customer, Cashier, StoreManager, StockManager), and we excluded those related to marginal parts of the system, such as scenarios of the \emph{EnterpriseManager} actor.
Thus, we selected \expUC{3}, \expComp{13}, and \expNode{8} from the \ccm specification. 
Beside this, we focused on three scenarios, namely: UC1 that describes the arrival of a customer at the checkout, identification, and sale of a product; UC4 that represents how products are registered in the store database upon their arrival; UC5 that represents the possibility of generating a report of store activities.

\bigskip

We computed the size of the solution space ($\Omega$) as the Cartesian product of the combination of refactoring actions $C_{n,k}=\binom{n}{k}$ where $n$ is the number of target model elements, and $k$ is the length of the chromosome (\ie \xspace the length of the sequence of refactoring actions, which is 4 in our case), and we summarize data in \tabref{tab:solution_space_size}. We remark that a manual investigation of the solution space is unfeasible due to its size. Hence, the evolutionary search is helpful for looking for model alternatives showing better quality than the initial one.
\Cref{tab:case_study_params} summarizes the case study characteristics.

\begin{table}
    \centering
    \begin{tabular}{llllc}
    \toprule
        Case Study  & UML Node & UML Component & UML Message & $\Omega$ \\
    \midrule
         \ttbs & 11 & 11 & 8  & $1.20 \times 10^{13}$ \\
         \ccm                       & 8  & 13 & 20 & $3.26 \times 10^{16}$ \\
    \bottomrule
    \end{tabular}
    \caption{Number of UML elements in our Case Studies, and the size of the relative solution space ($\Omega$).}
    \label{tab:case_study_params}
\end{table}

\section{Experimental setup}\label{sec:settings}

A configuration is defined by the combination of parameters related to the genetic algorithm, and the ones related to the specific optimization model. The eligible configuration values in our approach are listed in \tabref{tab:config_params}. In order to investigate which configuration produces better Pareto frontiers, we have executed multiple tuning runs to find a set of optimal configurations.

In order to set the parameters related to the genetic algorithm, we have performed a tuning phase with the intent of increasing the quality of the Pareto frontiers. In particular, we have set the length of refactoring sequences to four actions, which represents a good approximation of the number of refactoring actions usually applied by a designer in a single session.
We have set the $P_{crossover}$ and $P_{mutation}$ probabilities to 0.8 and 0.2, respectively, following common configurations~\cite{DBLP:journals/ese/ArcuriF13}. The higher the values of these two probabilities, the greater the chance of generating an unfeasible sequence of refactoring actions, which in turn causes a longer simulation time due to a higher number of discarded sequences. For example, the $P_{crossover}$ increase could cause a lot of permutation among sequences, and it might lead to wrong or unfeasible sequences of refactoring actions.

The initial population size might drive the genetic algorithm in local minima, and thus result in stagnant solutions.
In general, a densely populated initial population minimizes the probability of stagnant solutions in local minima. However, the generation of a crowded initial population is computational demanding and, in case of rare local minima, the computational cost represents a clear slowdown for the evolutionary approach~\cite{Arcuri_Fraser_2011}. For that reason, we set the population size to \textbf{16} elements (i.e., 16 different UML model alternatives), which did not show stagnant issues in our tuning phase. Furthermore, we will investigate in a future work the impact of denser populations in our analysis, in terms of computational time and quality of the computed Pareto frontiers (\computedP).
In addition, multiple runs have been executed for each configuration in order to reduce the randomness of the genetic algorithm. 

We considered three fuzziness thresholds, \ie \{0.55, 0.80, 0.95\}, to study the impact of performance antipatterns on computed Pareto frontiers. Since we are considering a fuzzy detection of performance antipatterns, we should use values greater than 50\% to reduce the probability of false positives, but less than 100\% to not fall in a case of performance antipatterns deterministic detection. Therefore, we decided to use those three fuzziness values to analyze the uncertainty of a fuzzy performance antipatterns detection.

With regard to parameters related to refactoring actions, we ran the experiment twice, one by excluding \brf, and one by including it. For the latter, we set \brf of each refactoring action as reported in \tabref{tab:solution_space_size}. As we said in \secref{sec:approach:brf}, we did not employ a complex cost model for baseline refactoring factor values. However, we remark that we are interested in the ratio between \brf values rather than in their specific values, and we will deeply investigate the impact of other values on future work.

Our experimental settings on \ttbs and \ccm case studies have generated \expSolutions model alternatives and have taken \expTime hours of computation. We performed our experiments on a server equipped with two Intel Xeon E5-2650 v3 CPUs at 2.30GHz, 40 cores and 80GB of RAM.

\section{Results and discussion}\label{sec:results}

Results presented in this section are aimed at answering the aforementioned three research questions.

\subsection{RQ1}\label{sec:results:rq1}

\RQ{RQ1}{To what extent do experimental configurations affect quality of Pareto frontiers?}

\noindent RQ1 focuses on the contribution of experimental configurations to the quality of the computed Pareto frontiers (\computedP). 

In \tabref{tab:best-qi-ttbs} and \tabref{tab:best-qi-ccm} it is possible to observe the configurations that result in better Pareto frontiers. 
Generally, quality indicators are obtained with respect to the optimal reference Pareto frontier (\referenceP), and each one has its ideal value (\eg $HV=1$, $IDG^+=0$). Moreover, values in tables have been sorted in ascending order when the best quality indicator is the lowest one, and in descending order otherwise. 
Since we did not have the optimal \referenceP for our case studies, we computed, for each case study, the quality indicators with respect to a \referenceP that contains every non-dominated solution across all \computedP.
Once quality indicators have been obtained and sorted, we identify which \maxeval and \probpas have generated better indicators. Finally, we also report data about \brf. 

At a glance, we can see that in most cases for both case studies, $\maxeval=72$ and lower fuzziness generates better quality indicators, whereas \brf has a different impact on the two case studies. 

In the following, we split \emph{RQ1} into three sub-questions, each one related to a specific experimental configuration attribute. \emph{RQ1.1} analyzes the influence of performance antipatterns on \computedP. \emph{RQ1.2} investigates whether the fuzziness of performance antipattern detection helps to find better \computedP. \emph{RQ1.3} studies the contribution of \brf to the quality of \computedP.

\begin{table}
    \centering
    \footnotesize
    \begin{tabular}{lcccr}
\toprule
\brf & \maxeval & \probpas & q\_indicator &     value \\
\midrule
yes &       72 &       95 &          HV &  0.329645 \\
yes &       82 &       95 &          HV &  0.304931 \\
yes &       82 &       95 &          HV &  0.267898 \\
yes &       72 &       80 &          HV &  0.266588 \\
yes &       82 &       55 &          HV &  0.254973 \\
\midrule
yes &       72 &       95 &        IGD$^+$ &  0.135226 \\
yes &       82 &       95 &        IGD$^+$ &  0.149903 \\
yes &       72 &       55 &        IGD$^+$ &  0.157150 \\
yes &       82 &       95 &        IGD$^+$ &  0.167142 \\
yes &       72 &       80 &        IGD$^+$ &  0.173162 \\
\midrule
yes &       72 &       95 &          EP &  0.295681 \\
yes &       82 &       95 &          EP &  0.296014 \\
yes &       82 &       95 &          EP &  0.316964 \\
yes &       72 &       55 &          EP &  0.316964 \\
yes &       82 &       55 &          EP &  0.323661 \\
\midrule
yes &      102 &       55 &     GSPREAD &  0.125487 \\
yes &      102 &       95 &     GSPREAD &  0.127085 \\
yes &      102 &       80 &     GSPREAD &  0.144666 \\
yes &      102 &       55 &     GSPREAD &  0.148802 \\
yes &       72 &       55 &     GSPREAD &  0.203504 \\
\bottomrule
\end{tabular}
    \caption{Best five of each quality indicator for the Train Ticket Booking Service case study while varying the performance antipattern fuzziness and the genetic algorithm evolutions.}
    \label{tab:best-qi-ttbs}
\end{table}

\begin{table}
    \centering
    \footnotesize
    \begin{tabular}{lcccr}
\toprule
\brf & \maxeval & \probpas & q\_indicator &     value \\
\midrule
no &       72 &       95 &          HV &  0.360432 \\
no &       82 &       95 &          HV &  0.359415 \\
no &      102 &       95 &          HV &  0.342563 \\
no &       72 &       55 &          HV &  0.326384 \\
no &       82 &       95 &          HV &  0.305201 \\
\midrule
no &       72 &       95 &        IGD+ &  0.091767 \\
no &       82 &       95 &        IGD+ &  0.105173 \\
no &      102 &       95 &        IGD+ &  0.106406 \\
no &       82 &       95 &        IGD+ &  0.132800 \\
no &       72 &       55 &        IGD+ &  0.135904 \\
\midrule
no &       82 &       95 &          EP &  0.250000 \\
no &       72 &       55 &          EP &  0.250000 \\
no &       72 &       95 &          EP &  0.250000 \\
no &       82 &       95 &          EP &  0.313857 \\
yes &       72 &       95 &          EP &  0.333333 \\
\midrule
no &       82 &       55 &     GSPREAD &  0.145989 \\
yes &      102 &       55 &     GSPREAD &  0.193488 \\
yes &      102 &       95 &     GSPREAD &  0.196790 \\
no &      102 &       55 &     GSPREAD &  0.200320 \\
no &      102 &       80 &     GSPREAD &  0.203431 \\
\bottomrule
\end{tabular}
    \caption{Best five of each quality indicator for the CoCoME case study while varying the performance antipattern fuzziness and the genetic algorithm evolutions.}
    \label{tab:best-qi-ccm}
\end{table}

\subsubsection{RQ1.1}

\RQ{RQ1.1}{Does antipattern detection contribute to find better solutions compared to the case where antipatterns are not considered at all?}

\noindent In order to answer this research question, we have conducted an additional experimentation for every problem configuration, where we have removed performance antipattern occurrences from the fitness function, thus reducing the optimization to the remaining three objectives.

\begin{figure}[!htbp]
 \centering
\includegraphics[width=.98\linewidth]{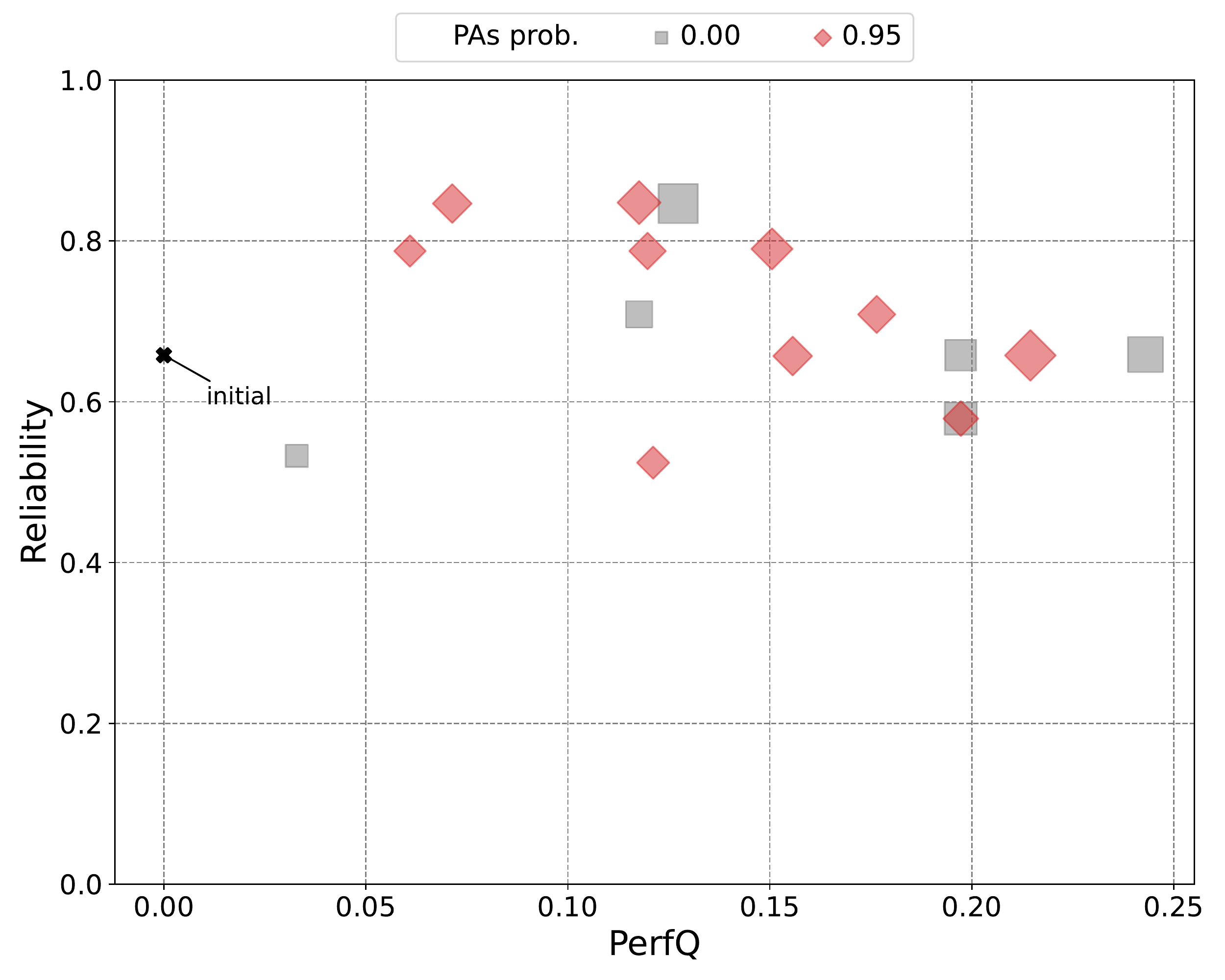}
\caption{\label{fig:ttbs-2dscatter-0-95}The scatter plot of \computedP of \ttbs with \texttt{72} genetic evolutions while considering, and excluding performance antipatterns in the optimization process (\ie $\probpas=0.95$, and $\probpas=0$)}
\end{figure}

\paragraph{Train Ticket Booking Service}
\figref{fig:ttbs-2dscatter-0-95} depicts the Pareto frontiers of \texttt{72} genetic evolutions while considering the lowest fuzziness (\ie $\probpas=0.95$) and no performance antipatterns (\ie $\probpas=0$). We can see that frontiers with performance antipatterns are generally more densely populated than the case where $\probpas=0$. Also, performance antipatterns help finding model alternatives showing lower \achanges than the ones found when they have been ignored. Although $\probpas=0$ generates the highest value of \perfq (\ie $\perfq=0.24$), there are more solutions in the topmost part of the plot when performance antipatterns drive the search process. From our analysis, it emerges that $\probpas=0.95$ produces better frontiers among those with performance antipatterns. Therefore, we can state that, for the \ttbs case study, the lower fuzziness the better the quality of frontiers in terms of \perfq, \reliability, and \achanges. 

\begin{figure}[!htbp]
 \centering
  \includegraphics[width=.98\linewidth]{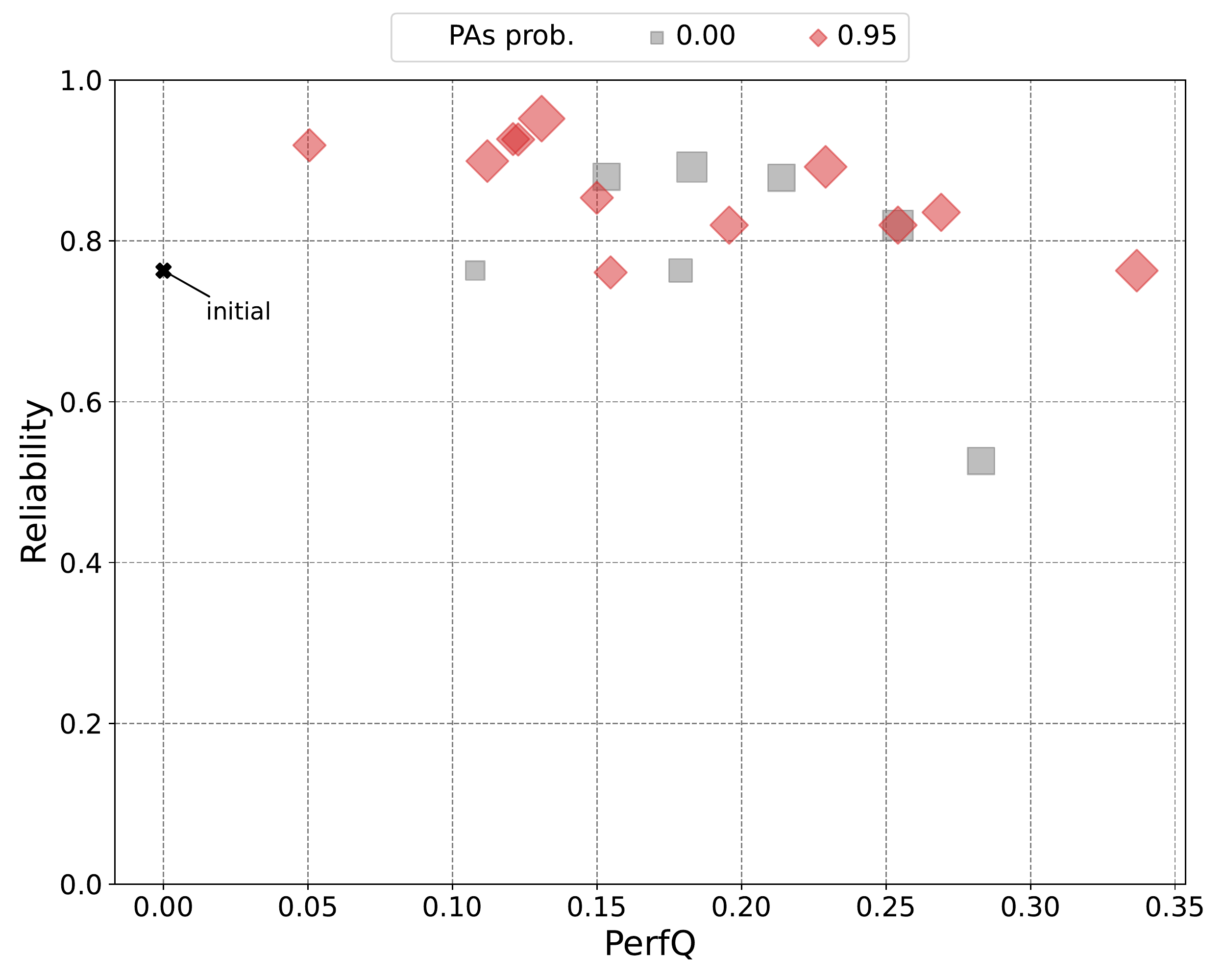}
  \caption{\label{fig:ccm-2dscatter-0-95}The scatter plot of \computedP of \ccm $72$ genetic evolutions while considering, and excluding performance antipatterns in the optimization process (\ie $\probpas=0.95$, and $\probpas=0$).}
\end{figure}

\paragraph{CoCoME}
\figref{fig:ccm-2dscatter-0-95} depicts the Pareto frontiers with \texttt{72} genetic evolutions while considering the lowest fuzziness (\ie $\probpas=0.95$) and no performance antipatterns (\ie $\probpas=0$). Most of the solutions lay in the topmost part of the plot, thus meaning that \computedP shows better \perfq and \reliability of the initial solution (see the black cross in the figure). Frontiers generated by performance antipatterns are more densely populated than those without performance antipatterns. Thus, the reduction of the number of performance antipatterns occurrences, if it is included among the objectives, helps the process finding more alternative models showing higher \perfq and \reliability with lower \achanges.

\paragraph{Discussion}

Based on our analysis, the reduction of performance antipatterns helps the optimization problem to generate alternatives showing better performance and reliability in most of the cases. 
The \ccm case study has mainly shown a light search for better reliability, likely due to the high reliability value of the initial model. 

\medskip

\answerRQ{On the basis of our experimentation, we can state that the consideration of performance antipattern occurrences in the optimization process leads to better solutions than the ones found when ignoring them.}

\subsubsection{RQ1.2}

\RQ{RQ1.2}{Does the probabilistic nature of fuzzy antipatterns detection help to include higher quality solutions in Pareto frontiers with respect to the deterministic one?}

\begin{figure}
 \centering
 \begin{subfigure}{.48\textwidth}
  \includegraphics[width=\textwidth]{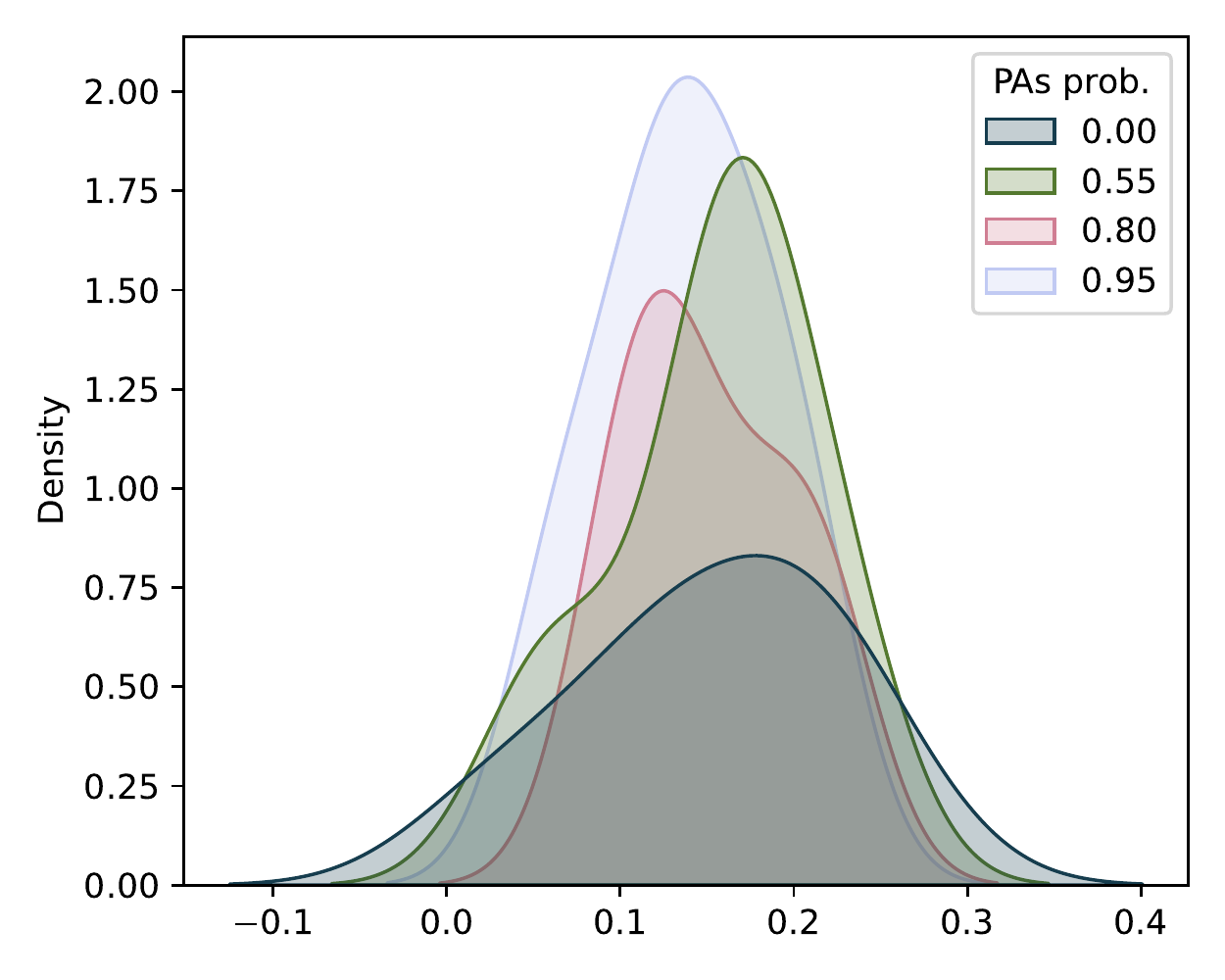}
  \caption{\label{fig:ttbs-perfq-kde}\perfq}
  \end{subfigure}
  \begin{subfigure}{.48\textwidth}
  \includegraphics[width=\textwidth]{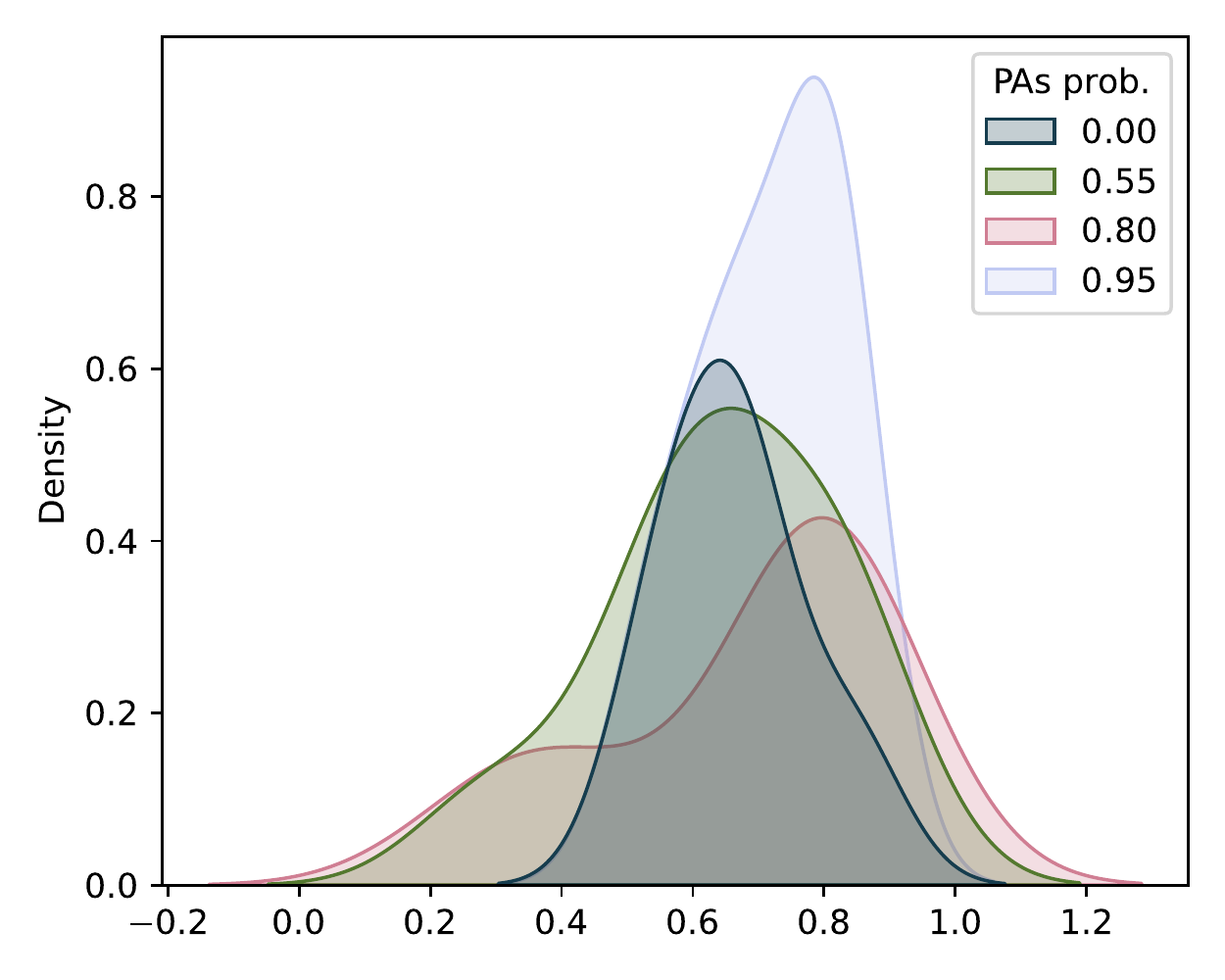}
 \caption{\label{fig:ttbs-rel-kde}\reliability}
\end{subfigure}
  \begin{subfigure}{.48\textwidth}
  \includegraphics[width=\textwidth]{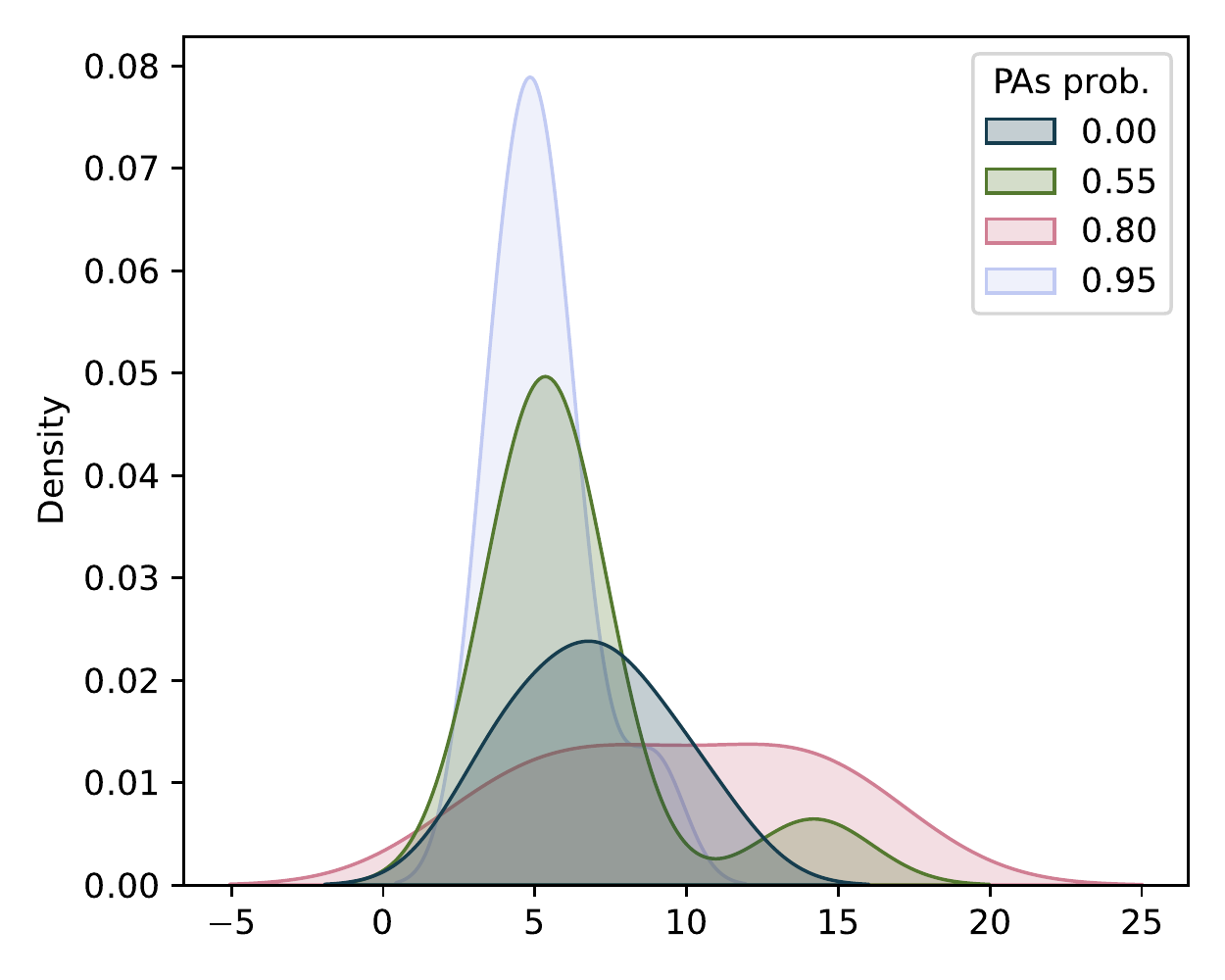}
 \caption{\label{fig:ttbs-changes-kde}\achanges}
\end{subfigure}
\begin{subfigure}{.48\textwidth}
  \includegraphics[width=\textwidth]{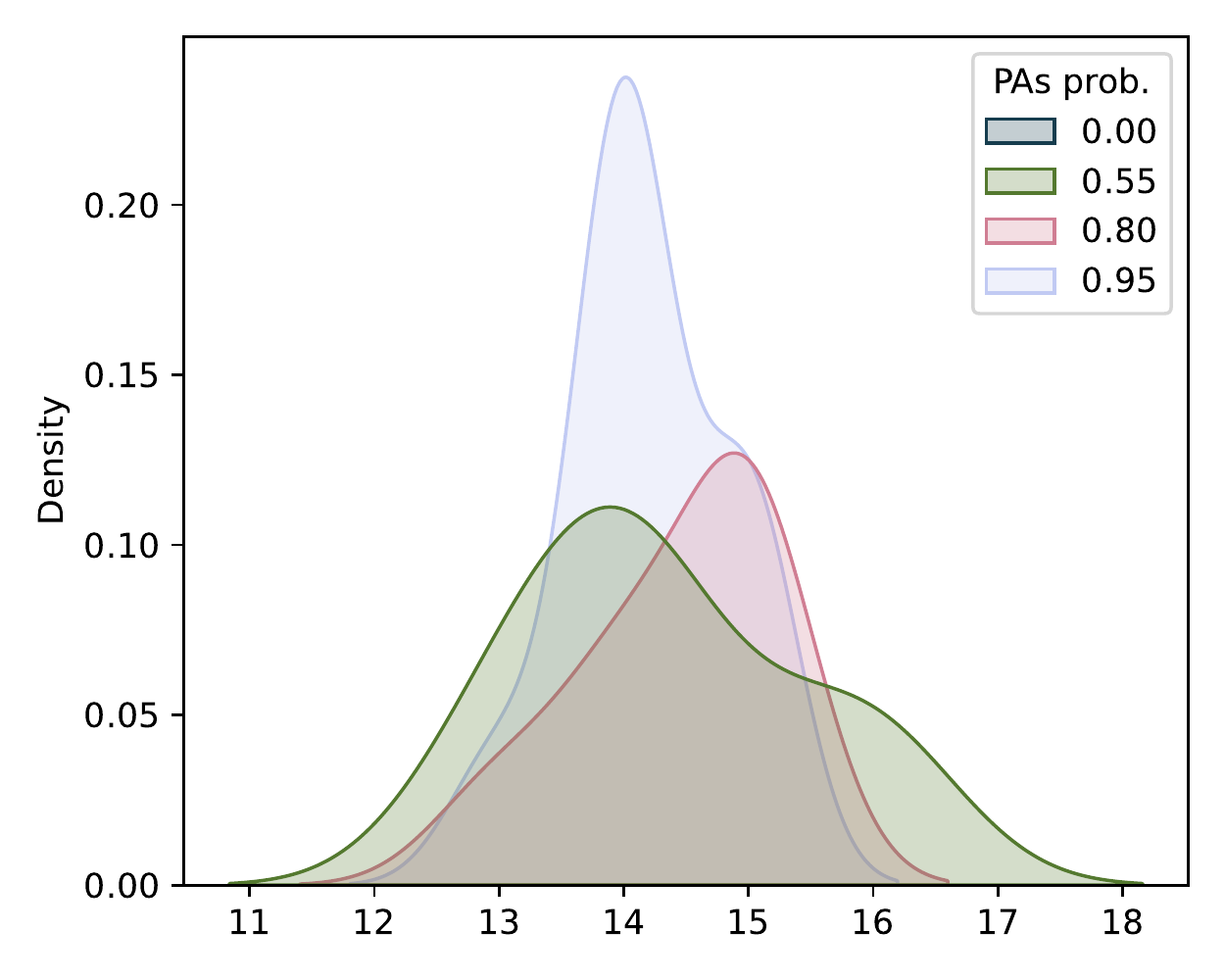}
 \caption{\label{fig:ttbs-pas-kde}\pas}
\end{subfigure}
\caption{\label{fig:ttbs-kdes}The KDE plots of the Train Ticket Booking Service case study while varying the Performance Antipattern fuzziness probabilities. The $\probpas=0.00$ means performance antipatterns were ignored as objectives. Each plot is referring to the objective in the label.}
\end{figure}

\begin{figure}
 \centering
 \begin{subfigure}{.48\linewidth}
  \includegraphics[width=\textwidth]{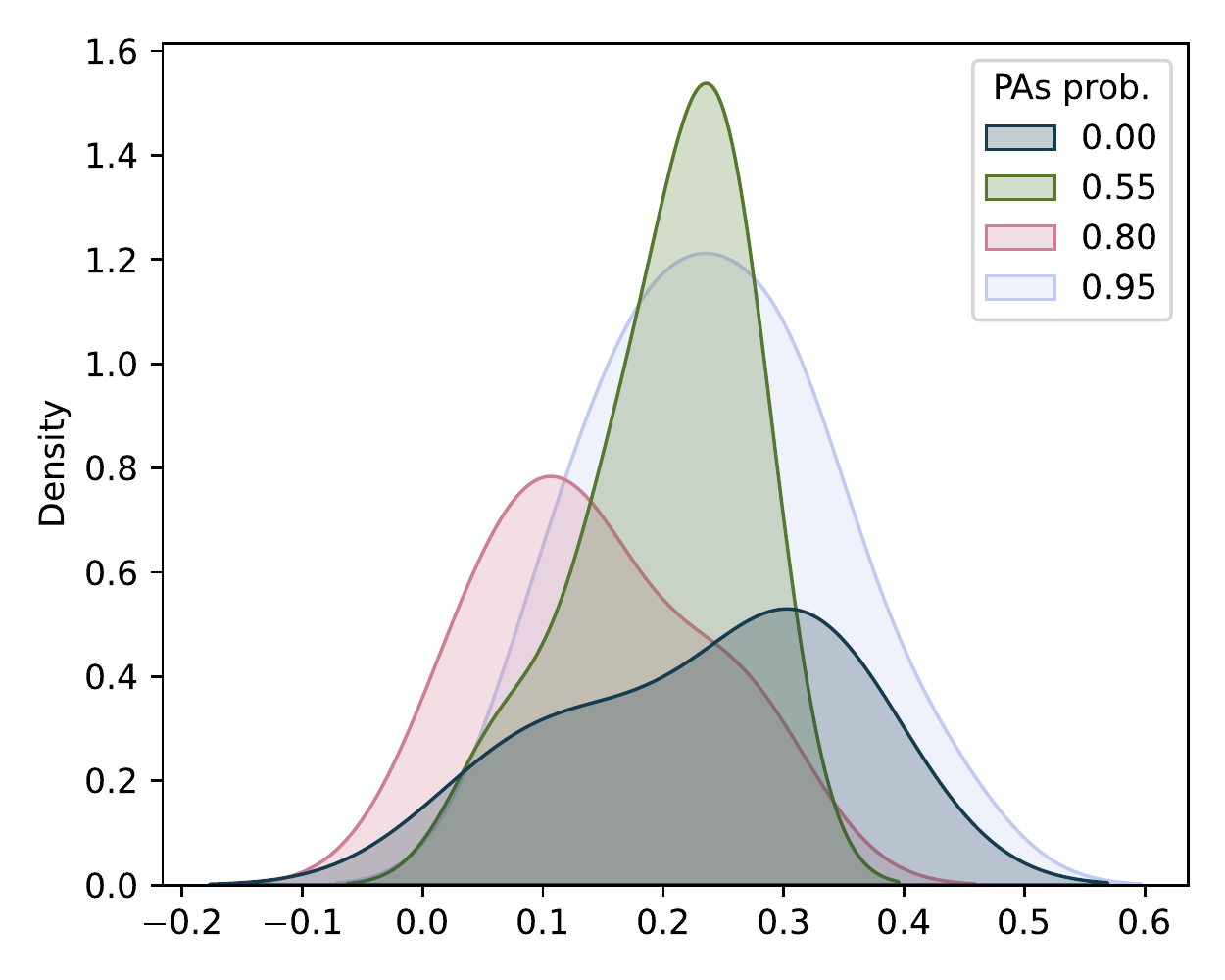}
  \caption{\label{fig:ccm-perfq-kde}\perfq}
 \end{subfigure}
 \begin{subfigure}{.48\linewidth}
  \includegraphics[width=\textwidth]{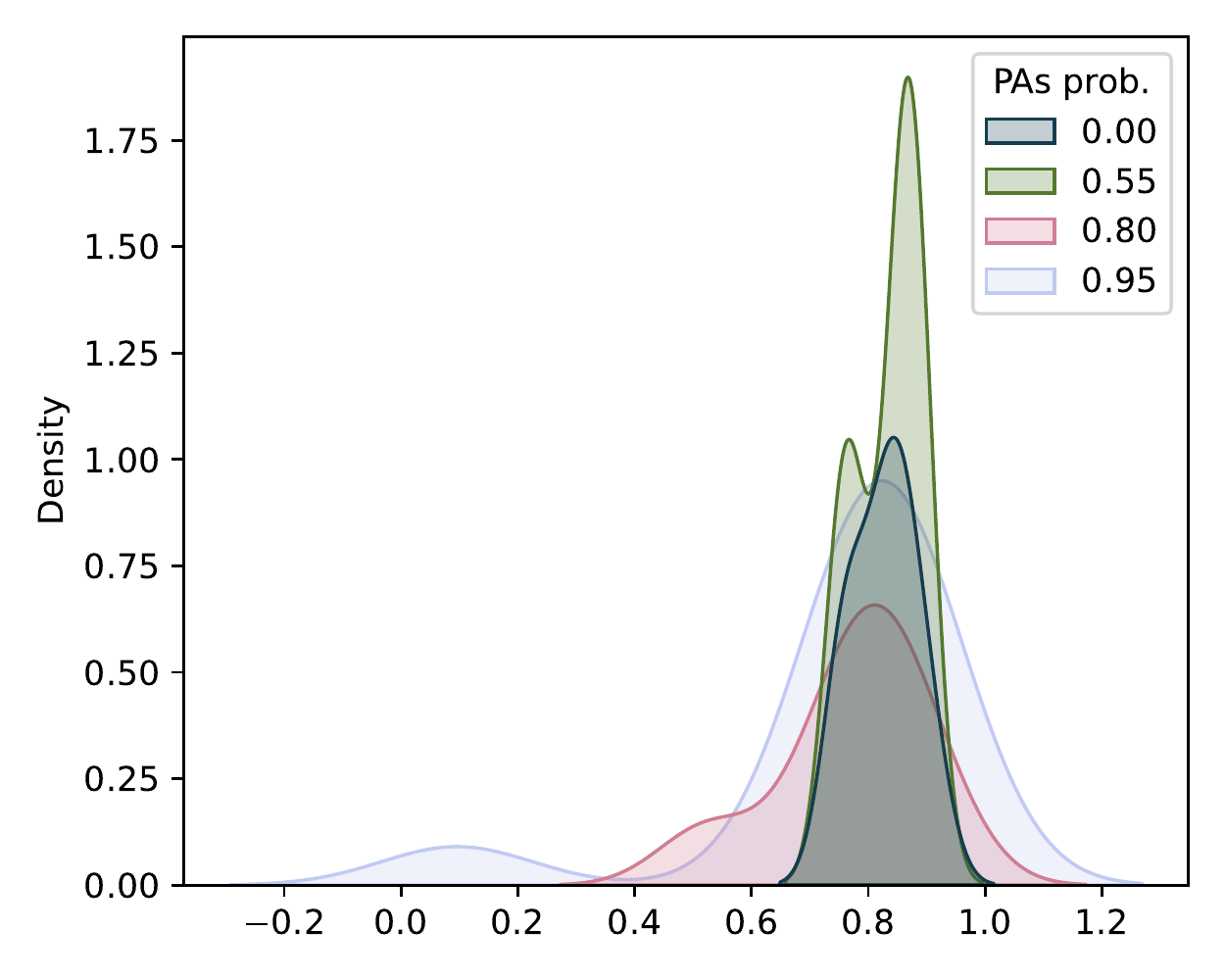}
 \caption{\label{fig:ccm-rel-kde}\reliability}
\end{subfigure}
\begin{subfigure}{.48\linewidth}
  \includegraphics[width=\textwidth]{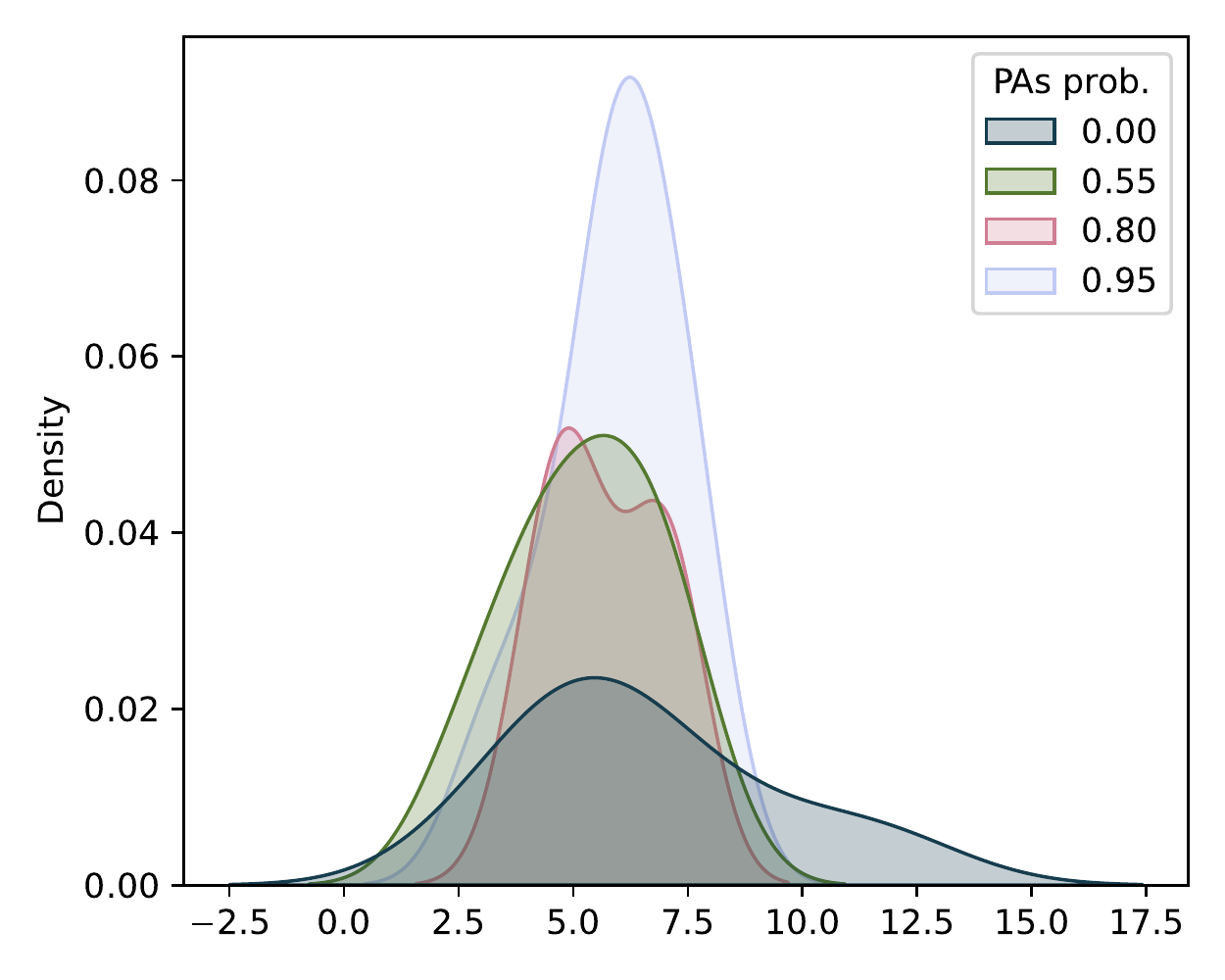}
 \caption{\label{fig:ccm-changes-kde}\achanges}
\end{subfigure}
\begin{subfigure}{.48\linewidth}
  \includegraphics[width=\textwidth]{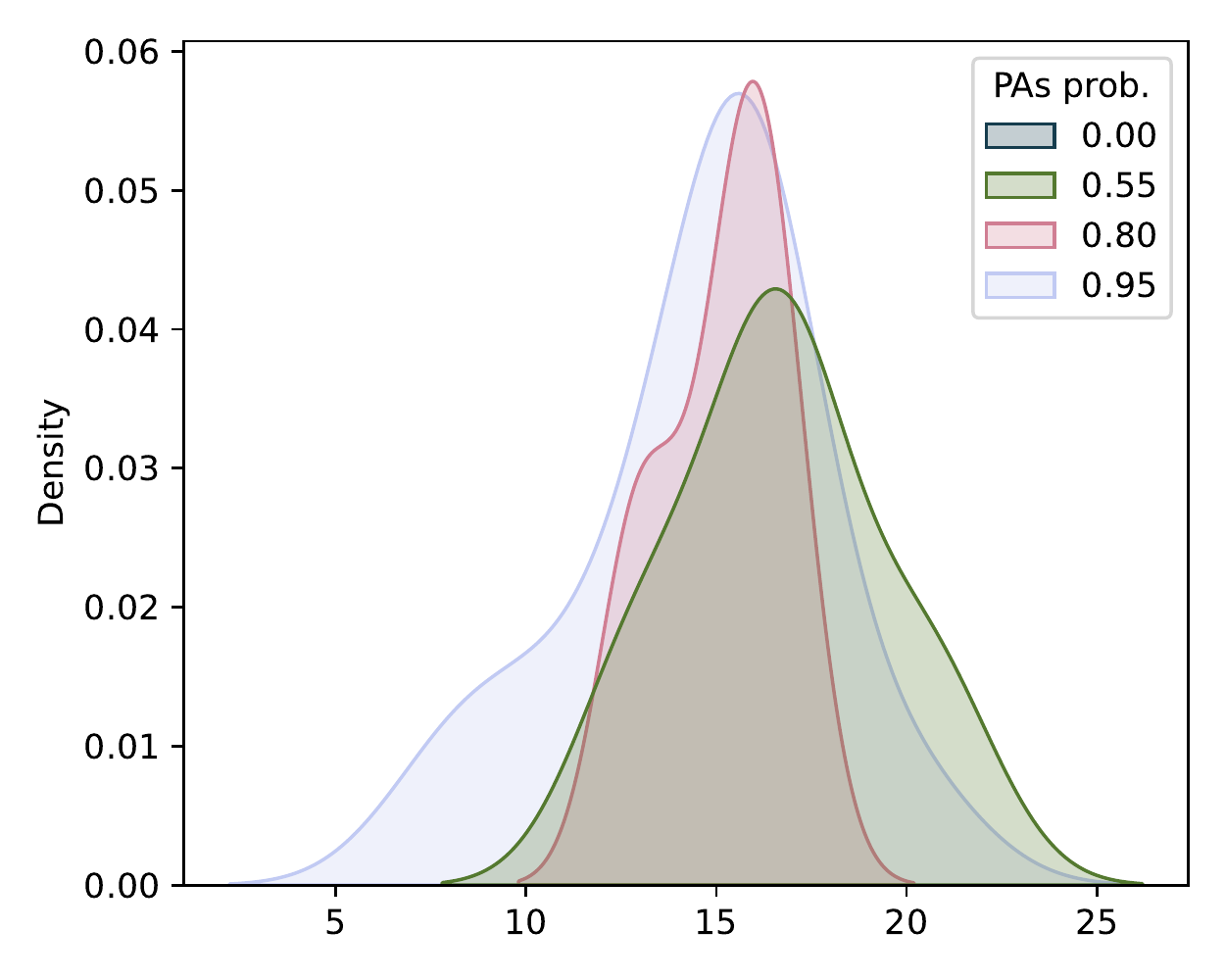}
 \caption{\label{fig:ccm-pas-kde}\pas}
\end{subfigure}
 \caption{\label{fig:ccm-kdes}The KDE plots of the \ccm case study while varying the Performance Antipattern fuzziness probabilities. The $\probpas=0.00$ means performance antipatterns were ignored as objectives. Each plot is referring to the objective in the label.}
\end{figure}

\noindent In order to answer this research question, we varied the values of the fuzziness threshold of the performance antipatterns detection within \{0.50, 0.80, 0.95\} for the two case studies. 
\figref{fig:ttbs-kdes} and \figref{fig:ccm-kdes} depict the kernel density estimate (KDE) plots showing each possible combination among objectives for \ttbs and \ccm respectively. Each plot depicts the KDE of the relative objectives, \eg \figref{fig:ttbs-perfq-kde} shows the \perfq KDE for the \ttbs case study.

\paragraph{Train Ticket Booking Service}

For the \ttbs case study, we have noticed larger variability of \perfq when performance antipatterns are ignored, see the flattest curve in \figref{fig:ttbs-perfq-kde}. In addition, \perfq is narrower to the mean ($\approx 0.2$) when performance antipatterns are involved in the fitness function, which means less variability in terms of performance in the model alternatives. 
With regard to the \reliability (\figref{fig:ttbs-rel-kde}), it seems to be more stable without performance antipattern detection. Moreover, the performance antipattern detection helps including solutions with higher reliability values than the case without them. \figref{fig:ttbs-changes-kde} shows that the lower the fuzziness the more stable the \achanges values, which means less variability in the model alternatives discovered by the search.
Finally, the $0.95$ fuzziness reduces the variability of the \pas objective (\figref{fig:ttbs-pas-kde}). Thus, the more deterministic, the higher the probability of discovering true positive performance antipatterns.

\paragraph{CoCoME}

We notice that Pareto frontiers obtained while ignoring performance antipatterns in the fitness function showed larger variability than the ones obtained while considering them. This is depicted in \figref{fig:ccm-perfq-kde} where \perfq shows negative values and the curve is flatter than the other cases. 
For \ccm we notice that the higher the performance antipattern \probpas, the higher \perfq, which becomes similar to a normal distribution with mean falling on $0.3$ for a $\probpas=0.95$ of performance fuzziness. In the case of the lowest fuzziness value, \perfq assumed the highest value in our experiments.
With regard to \achanges (\figref{fig:ccm-changes-kde}), it increases when \pas are ignored. Moreover, the higher the \probpas, the more stable \achanges, which means less variability in the model alternatives.
Again, due to the high value of \reliability for the initial model, \ccm shows most of the \reliability values around $0.9$ (\figref{fig:ccm-rel-kde}). 

\paragraph{Discussion}

Our analysis shows that in most of the cases the higher \probpas, the closer to the mean is the distribution of \perfq, which means less variability for \perfq. Therefore, it seems better to use a more deterministic antipattern detection (\ie higher values of \probpas). However, a deterministic detection has the drawback of relying on fixed thresholds that must be computed in advance for each model alternative. The trade-off between better quality solution and the effort to bind thresholds is likely domain-dependent and worth to be more investigated.

\medskip

\answerRQ{On the basis of our experimentation, we can state that performance antipattern fuzzy detection does not help to improve the quality of Pareto frontiers.} 

\subsubsection{RQ1.3}

\RQ{RQ1.3}{To what extent does the architectural distance contribute to find better alternatives?}

In order to answer this research question, we run the same problem configurations by varying the baseline refactoring factor value. In particular, we decided to activate (\brf) and deactivate (\nobrf) the baseline refactoring factor to study how it contributes to the generation of Pareto frontiers.

\paragraph{Train Ticket Booking Service}

\begin{figure}
\centering
\begin{subfigure}{.9\linewidth}
 \includegraphics[width=0.9\textwidth]{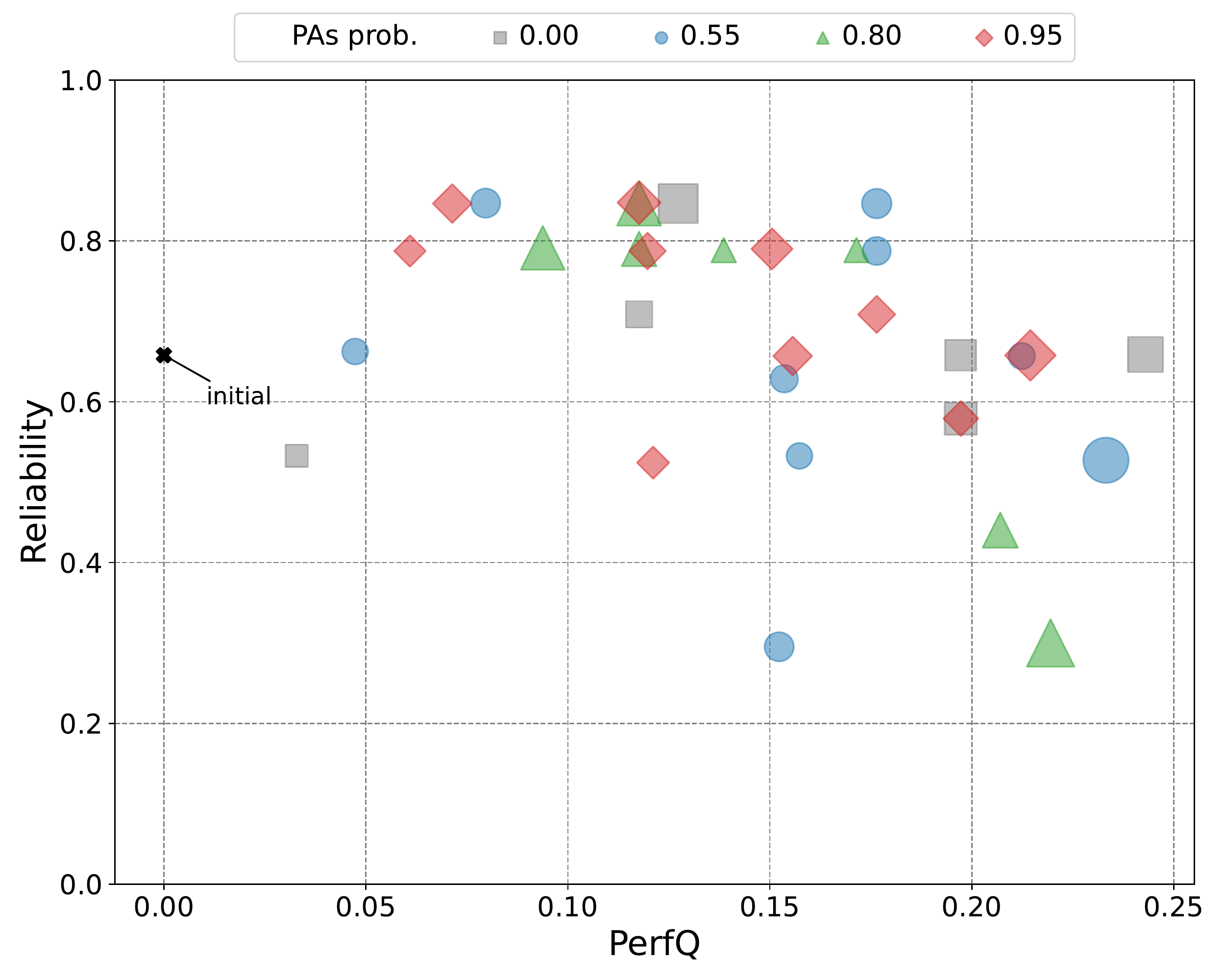}
\caption{\label{fig:ttbs-2dscatter-brf}\brf.} 
\end{subfigure}
\begin{subfigure}{.9\linewidth}
 \includegraphics[width=0.9\textwidth]{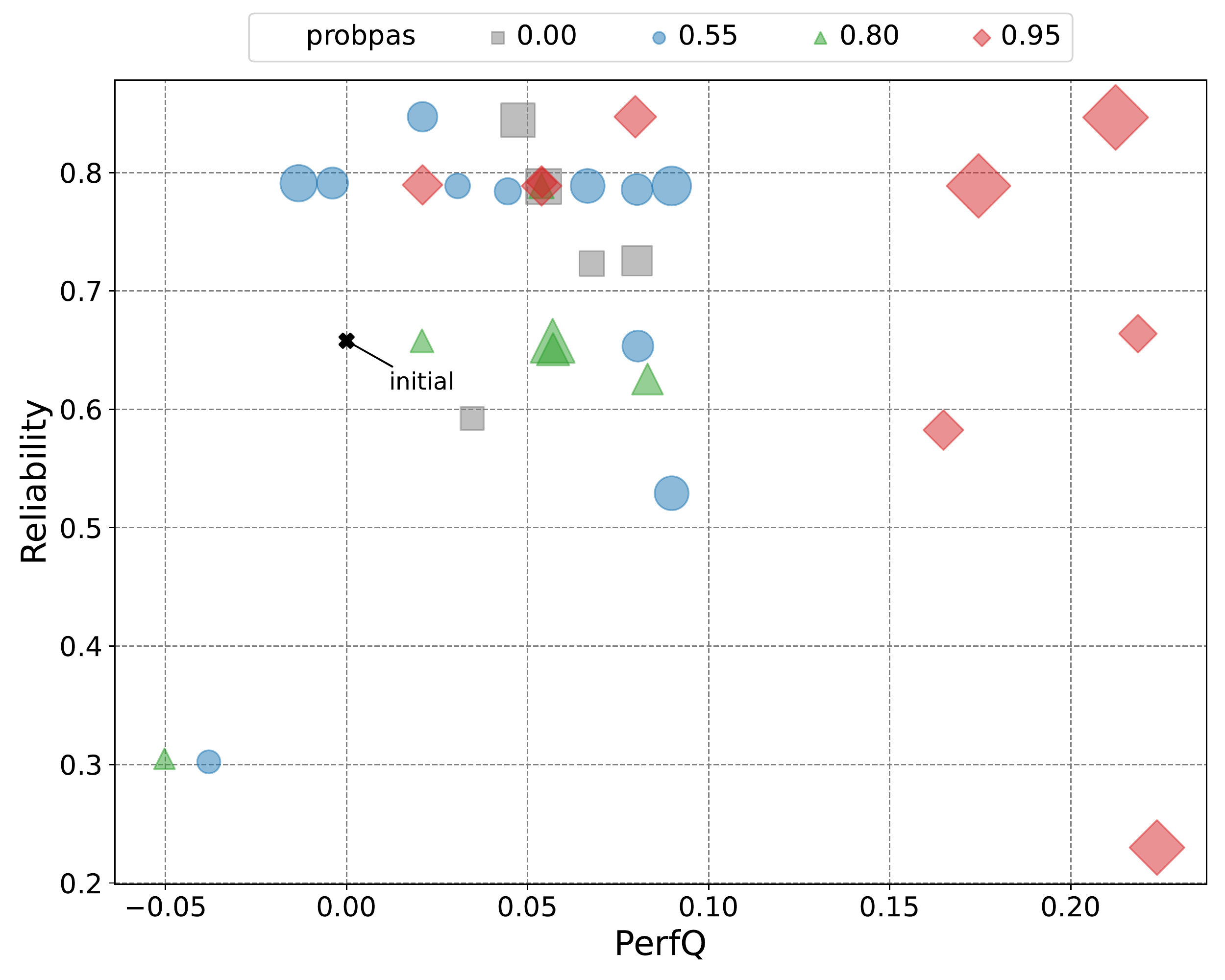}
\caption{\label{fig:ttbs-2dscatter-nobrf}\nobrf.} 
\end{subfigure}
 \caption{\label{fig:ttbs-2dscatter-brf-nobrf}The scatter plot of Train Ticket Booking Service Pareto frontiers while varying the fuzziness after $72$ genetic evolutions with \brf, and \nobrf configurations.} 
\end{figure}

\figref{fig:ttbs-2dscatter-brf} and \figref{fig:ttbs-2dscatter-nobrf} show Pareto frontiers obtained with \brf and \nobrf configurations, respectively.
We can see that results with \nobrf are narrower to the initial solution (\ie the black marker in figure) than the case where \brf is activated. \nobrf seems to penalize performance antipatterns with higher fuzziness, in fact $\probpas=0.95$ generates the best alternatives in terms of \perfq and \reliability (see the topmost right corner in \figref{fig:ttbs-2dscatter-nobrf}). However, the highest \perfq in the case of \nobrf is lower than the one in the case of \brf. Hence, \brf helps the search finding better solutions in terms of \perfq for the \ttbs case. Also, the \nobrf configuration shows, in a few cases, a detriment of the initial performance and reliability (see the left bottom-most corner) that it never happened when the \brf is active.

\paragraph{CoCoME}

\begin{figure}
\centering
\begin{subfigure}{.9\linewidth}
 \includegraphics[width=0.9\textwidth]{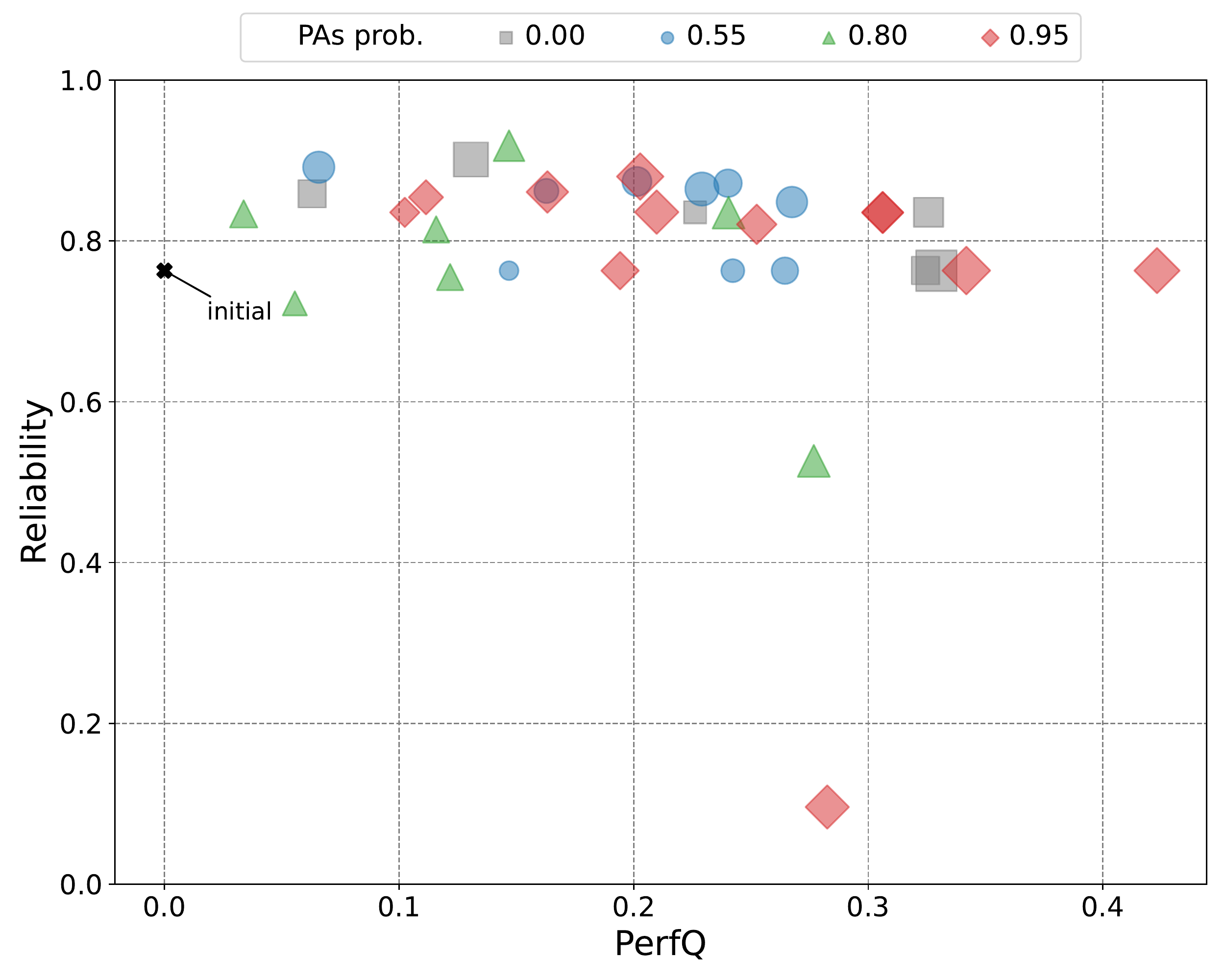}
 \caption{\label{fig:ccm-2dscatter-brf} \brf}
\end{subfigure}
\begin{subfigure}{.9\linewidth}
 \includegraphics[width=0.9\textwidth]{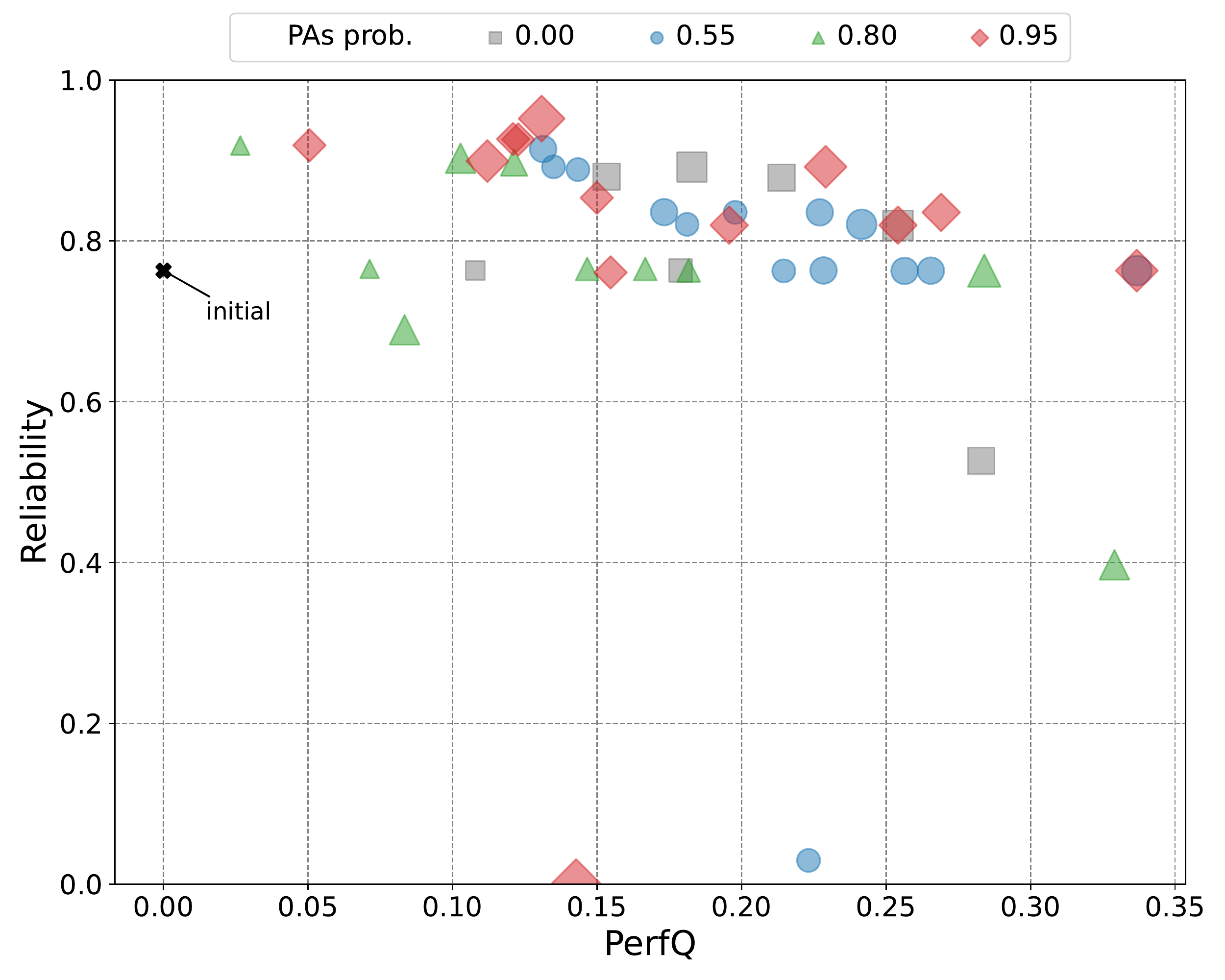}
 \caption{\label{fig:ccm-2dscatter-nobrf} \nobrf}
\end{subfigure}
 \caption{\label{fig:ccm-2dscatter-brf-nobrf}The scatter plot of \ccm Pareto frontiers while varying the fuzziness after $72$ genetic evolutions with \brf and \nobrf configurations.}
\end{figure}

\figref{fig:ccm-2dscatter-nobrf} shows Pareto frontiers obtained with \nobrf configuration. By comparing this plot with the one shown in \figref{fig:ccm-2dscatter-brf}, we can see that  the \brf exclusion generates more densely populated frontiers than the other case. Furthermore, no extreme differences arise between the executions with \brf and \nobrf configurations. In both cases \perfq and \reliability fall within the same region of the plot, where alternatives with \brf reached better \perfq (see $\perfq > 0.4$ in \figref{fig:ccm-2dscatter-brf}). With regard to the \reliability, we can see that \nobrf configuration found few model alternatives showing lower values.

\paragraph{Discussion}
Based on our analysis, the \emph{baseline refactoring factor} helps generating better alternatives in terms of objectives. We noticed that the \reliability is penalized with \nobrf configurations. Also, the \brf deactivation penalized \perfq in few cases.
A deeper investigation is required on how \brf might affect the computed Pareto frontiers quality. For example, we can introduce more complex cost models, \eg COCOMO~\cite{boehm2009software}, to improve its estimation. However, we preferred having a more straightforward cost estimation to avoid burdening the search algorithm with additional computational costs.

\answerRQ{Based on our results, we can state that \brf helps better estimating \achanges of refactoring actions, which generates Pareto frontiers showing higher quality (or at least it does not worsen the Pareto frontier quality).}
 \subsection{RQ2}

\RQ{RQ2}{Is it possible to increase reliability without performance degradation?}

\begin{figure}[htbp]
    \centering
    \begin{subfigure}{\linewidth}
        \centering
        \includegraphics[width=0.9\linewidth]{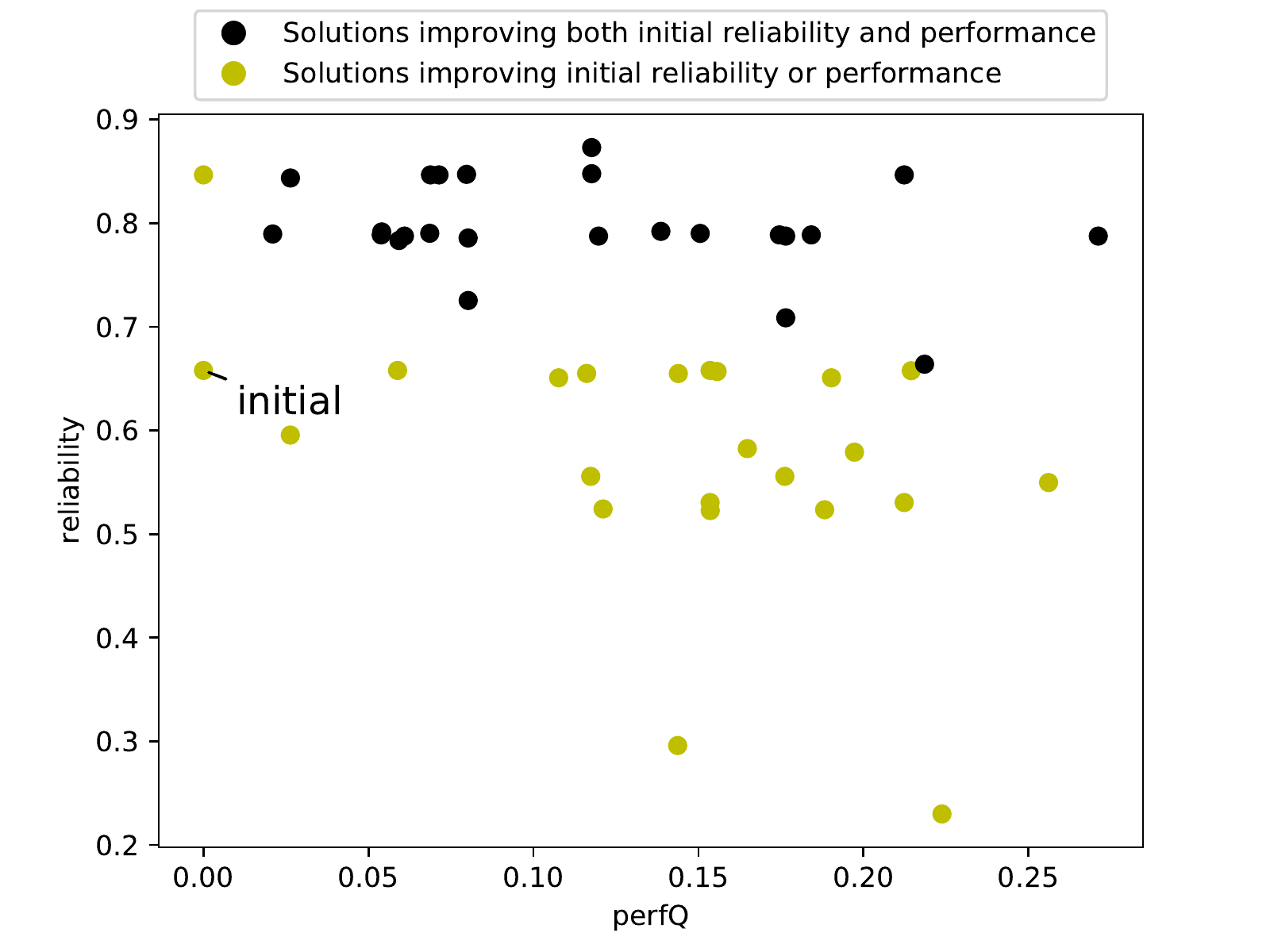}
        \caption{Train Ticket Booking Service}
        \label{fig:rq3_tt}
    \end{subfigure}
    \hfill
    \begin{subfigure}{\linewidth}
        \centering
        \includegraphics[width=0.9\linewidth]{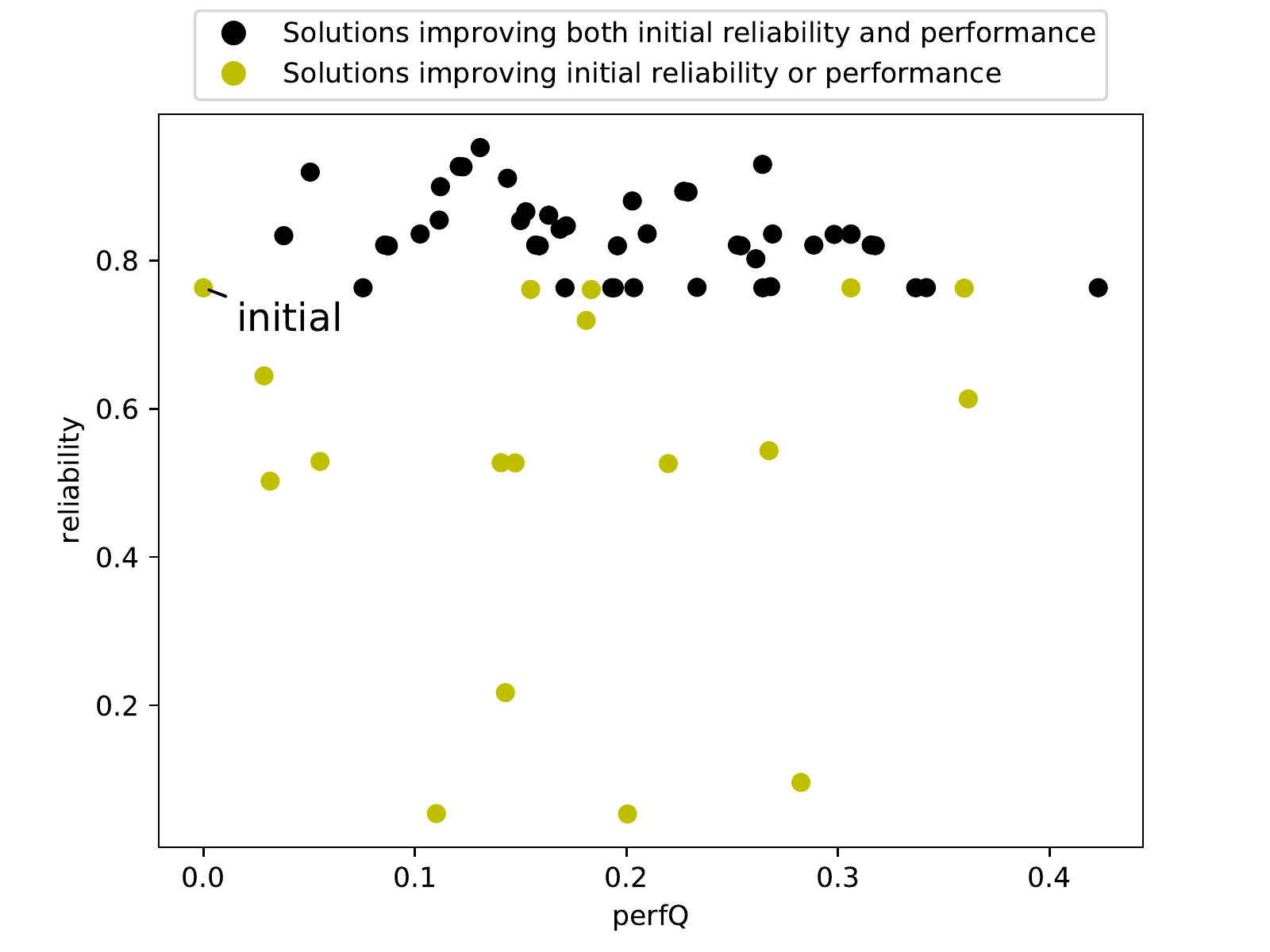}
        \caption{\ccm}
        \label{fig:rq3_cocome}
    \end{subfigure}
    \caption{Solutions of the Pareto frontiers displayed according to their reliability and performance.}
    \label{fig:rq3}
\end{figure}

We answer RQ2 by looking for model alternatives, within the computed Pareto frontiers (\computedP), that improve both initial reliability and performance. 

\figref{fig:rq3} shows the results obtained on the \computedP of \ttbs and \ccm. The dark dots represent the alternatives we are looking for, \ie those improving both \reliability and \perfq. Instead, the bright dots represent the model alternatives that improve one of the two non-functional aspects.

\paragraph{Train Ticket Booking Service}
\figref{fig:rq3_tt} shows that in \ttbs we obtained 54\% of the model alternatives improving \reliability and \perfq. Thus, there is a portion (\ie 46\%) presenting a detriment of the \reliability but an improvement in terms of performance. This is confirmed by looking at the model alternatives within the \referenceP: 18 over 26 alternatives are those taken from the examined Pareto. In this case, model alternatives that guarantee an improvement can be very important for a designer, as we find a performance upgrade of up to 27\% and a reliability increase of up to 32\%.

\paragraph{\ccm} The case of \ccm, in \figref{fig:rq3_cocome}, strengthens the observations made for \ttbs. In this case, the majority (\ie 74\%) of the model alternatives improve both \perfq and \reliability of the initial model. This is confirmed by the number of improving alternatives in the \referenceP: 38 out of 48. We got an improvement of the reliability up to 24\%, which is smaller than \ttbs but likely affected by the fact that, in this case, the starting model has higher initial reliability (\ie 0.75). Instead, the performance improvement is higher, \ie up to 42\%.

\paragraph{Discussion} The set of model alternatives, which have been found while answering to RQ1, are characterized by a neat improvement of two quality attributes: \reliability and \perfq. This result could be fundamental for designers, as they could do further analysis or use the model as a starting point in subsequent stages of the development process. 

\answerRQ{Our experimentation shows that, our approach can find design alternatives characterized by a significant improvement of both reliability and performance.}
 \subsection{RQ3}

\RQ{RQ3}{What type of refactoring actions are more likely to lead to better solutions?}

With this research question, we investigate whether some refactoring actions are more likely to be selected than others in the Pareto optimal front during the optimization process.
This could potentially lead to more general insights on the effectiveness of specific types of refactoring actions to improve the considered objectives.

\paragraph{Train Ticket Booking Service}
\tabref{tab:ref_types_tt} reports the share of refactoring types for Train Ticket Booking Service.
Each row represents a configuration (\ie an experiment) with a different combination of \brf, \maxeval, and \probpas. The rightmost four columns represent the refactoring action types that we have considered in our approach.
The last row shows the percentages computed over all the configurations.

It is evident that the genetic algorithms prefer to select certain types of refactorings.
\emph{MO2C} and \emph{Clon} are clearly more likely to be selected, with a slight preference for \emph{Clon} in most configurations and, consequently, on average across all configurations.
These refactorings are inherently very beneficial for the performance: cloning a component will frequently split the utilization in half, and moving an operation to a new component will not only reserve a node for a single operation, but will also relieve the original component of the load related to that operation.
Also, they are unlikely to disrupt the reliability objective, since the new nodes will have the same probability of failure as the ones they are cloned from.
Conversely, the \emph{ReDe} refactoring may be advantageous for performance and reliability only when the component to be redeployed is sharing the current node with many other components, and this is not the case in the initial model.
This is most probably the reason why the \emph{ReDe} refactoring is considerably less likely to be selected, and there is even a configuration in which it was not selected in any Pareto solution (\brf: yes, \maxeval: 72, \probpas: 0.95).

\begin{table}[htbp]
    \centering
    \footnotesize
\begin{tabular}{lllrrrr}
\toprule
brf & maxeval & probpas &  Clon &  MO2N &  MO2C &  ReDe \\
\midrule
 no &      72 &    0.00 & 31.77 & 42.71 & 12.50 & 13.02 \\
 no &      72 &    0.55 & 39.58 & 47.40 &  2.60 & 10.42 \\
 no &      72 &    0.80 & 31.25 & 53.12 &  9.90 &  5.73 \\
 no &      72 &    0.95 & 34.38 & 28.12 & 18.23 & 19.27 \\
 no &      82 &    0.00 & 56.25 & 27.08 &  2.60 & 14.06 \\
 no &      82 &    0.55 & 36.98 & 39.06 & 17.71 &  6.25 \\
 no &      82 &    0.80 & 23.96 & 51.04 & 11.98 & 13.02 \\
 no &      82 &    0.95 & 42.71 & 30.73 & 23.44 &  3.12 \\
 no &     102 &    0.00 & 42.71 & 30.73 & 16.15 & 10.42 \\
 no &     102 &    0.55 & 35.94 & 27.08 & 17.71 & 19.27 \\
 no &     102 &    0.80 & 40.10 & 30.21 & 14.58 & 15.10 \\
 no &     102 &    0.95 & 25.00 & 58.85 & 10.94 &  5.21 \\
yes &      72 &    0.00 & 40.10 & 30.21 & 16.15 & 13.54 \\
yes &      72 &    0.55 & 37.50 & 36.98 & 14.06 & 11.46 \\
yes &      72 &    0.80 & 42.71 & 19.27 & 16.15 & 21.88 \\
yes &      72 &    0.95 & 49.48 & 37.50 & 13.02 &  0.00 \\
yes &      82 &    0.00 & 19.79 & 57.81 & 10.42 & 11.98 \\
yes &      82 &    0.55 & 39.06 & 36.98 & 22.40 &  1.56 \\
yes &      82 &    0.80 & 27.60 & 40.62 & 13.54 & 18.23 \\
yes &      82 &    0.95 & 43.75 & 34.90 & 20.31 &  1.04 \\
yes &     102 &    0.00 & 43.75 & 35.94 & 16.67 &  3.65 \\
yes &     102 &    0.55 & 41.15 & 25.00 &  9.38 & 24.48 \\
yes &     102 &    0.80 & 35.42 & 40.10 & 10.42 & 14.06 \\
yes &     102 &    0.95 & 54.17 & 22.92 & 16.67 &  6.25 \\
\midrule
\multicolumn{3}{l}{Total} & 38.13 & 36.85 & 14.06 & 10.96 \\
\bottomrule
\end{tabular}
    \caption{Share of refactoring types in Train Ticket.}
    \label{tab:ref_types_tt}
\end{table}

\paragraph{\ccm}
Analogously, we report the share of refactoring actions for \ccm in \tabref{tab:ref_types_ccm}.
The overall preferences in the selection of refactorings seem to be similar to the Train Ticket Booking Service case.
However, we can notice an even stronger preference for the \emph{Clon} refactoring.
Since this refactoring largely decreases the utilization of nodes, it may be reasonable to conclude that, in the initial \ccm model, some nodes with high utilization are preventing the performance to improve.
While the \emph{ReDe} refactoring is still the less selected one, there are no configurations in which at least one refactoring of this type does not contribute to Pareto solutions.
However, in 13 configurations over a total of 24, the \emph{ReDe} refactoring has a share below 10\%.

\begin{table}[htbp]
    \centering
    \footnotesize
\begin{tabular}{lllrrrr}
\toprule
brf & maxeval & probpas &  Clon &  MO2N &  MO2C &  ReDe \\
\midrule
 no &      72 &    0.00 & 30.21 & 37.50 & 19.79 & 12.50 \\
 no &      72 &    0.55 & 54.69 & 24.48 & 12.50 &  8.33 \\
 no &      72 &    0.80 & 42.19 & 32.81 & 18.75 &  6.25 \\
 no &      72 &    0.95 & 45.83 & 37.50 &  9.90 &  6.77 \\
 no &      82 &    0.00 & 43.23 & 25.52 & 17.19 & 14.06 \\
 no &      82 &    0.55 & 48.96 & 27.60 & 12.50 & 10.94 \\
 no &      82 &    0.80 & 37.50 & 41.67 & 10.42 & 10.42 \\
 no &      82 &    0.95 & 53.12 & 28.12 &  5.73 & 13.02 \\
 no &     102 &    0.00 & 20.83 & 36.46 & 17.71 & 25.00 \\
 no &     102 &    0.55 & 44.27 & 28.12 & 20.31 &  7.29 \\
 no &     102 &    0.80 & 55.73 & 27.08 &  1.04 & 16.15 \\
 no &     102 &    0.95 & 56.25 & 28.65 & 13.54 &  1.56 \\
yes &      72 &    0.00 & 41.15 & 29.17 & 23.96 &  5.73 \\
yes &      72 &    0.55 & 38.02 & 32.81 & 20.83 &  8.33 \\
yes &      72 &    0.80 & 35.94 & 42.71 & 13.02 &  8.33 \\
yes &      72 &    0.95 & 61.98 & 23.44 & 10.94 &  3.65 \\
yes &      82 &    0.00 & 51.56 & 30.73 & 14.06 &  3.65 \\
yes &      82 &    0.55 & 41.67 & 33.85 & 18.75 &  5.73 \\
yes &      82 &    0.80 & 38.54 & 40.62 & 12.50 &  8.33 \\
yes &      82 &    0.95 & 44.27 & 26.56 & 16.67 & 12.50 \\
yes &     102 &    0.00 & 43.75 & 20.31 & 17.19 & 18.75 \\
yes &     102 &    0.55 & 66.67 &  6.25 & 20.31 &  6.77 \\
yes &     102 &    0.80 & 59.90 & 19.27 &  8.33 & 12.50 \\
yes &     102 &    0.95 & 61.46 & 16.67 &  8.33 & 13.54 \\
\midrule
\multicolumn{3}{l}{Total} & 46.57 & 29.08 & 14.34 & 10.00 \\
\bottomrule
\end{tabular}
    \caption{Share of refactoring types in \ccm.}
    \label{tab:ref_types_ccm}
\end{table}

\begin{figure}[htbp]
    \centering
    \begin{subfigure}[]{\linewidth}
        \centering
        \includegraphics[width=.8\linewidth]{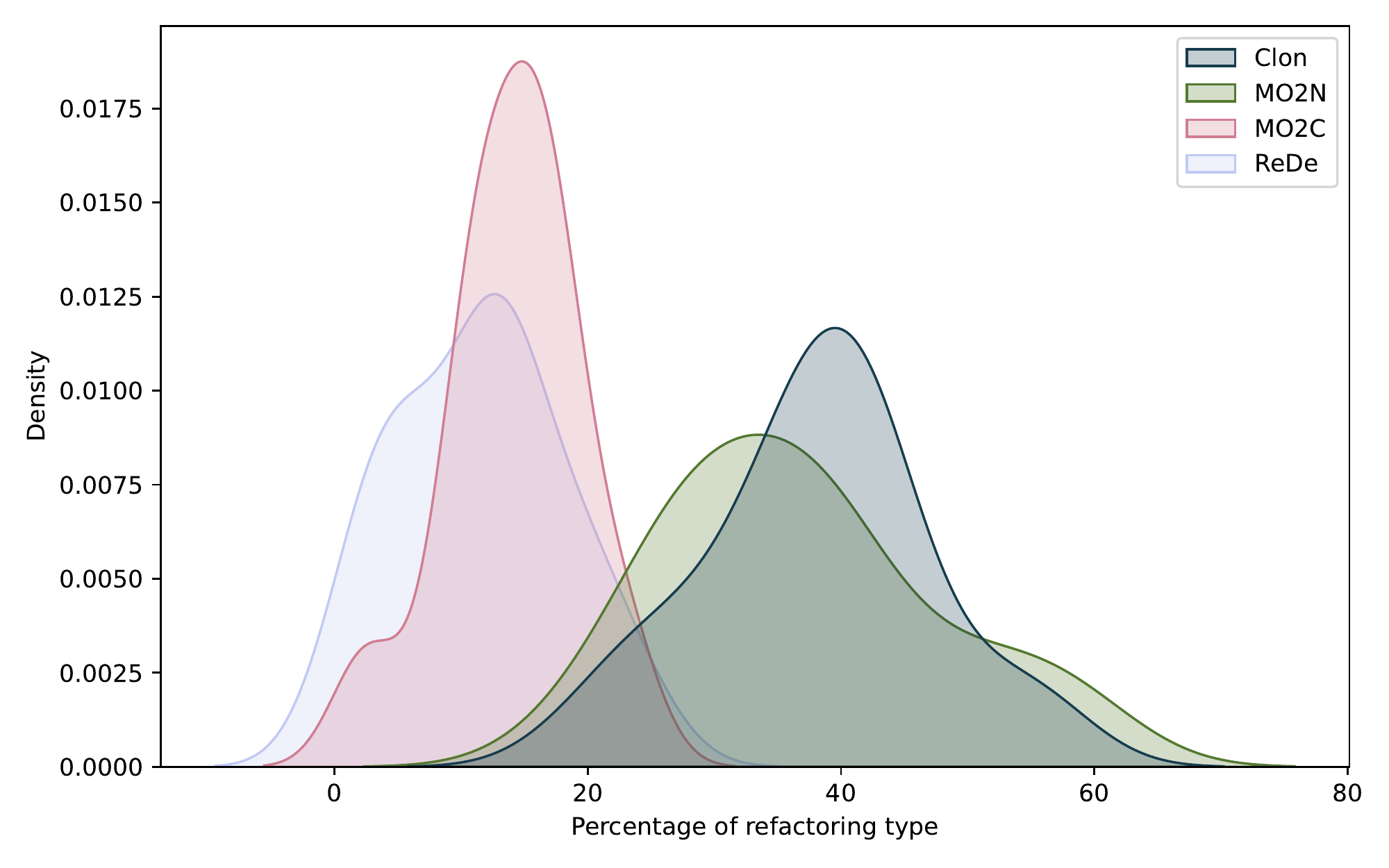}
        \caption{Train Ticket}
        \label{fig:ref_actions_dist_tt}
    \end{subfigure}
    \hfill
    \begin{subfigure}[]{\linewidth}
        \centering
        \includegraphics[width=.8\linewidth]{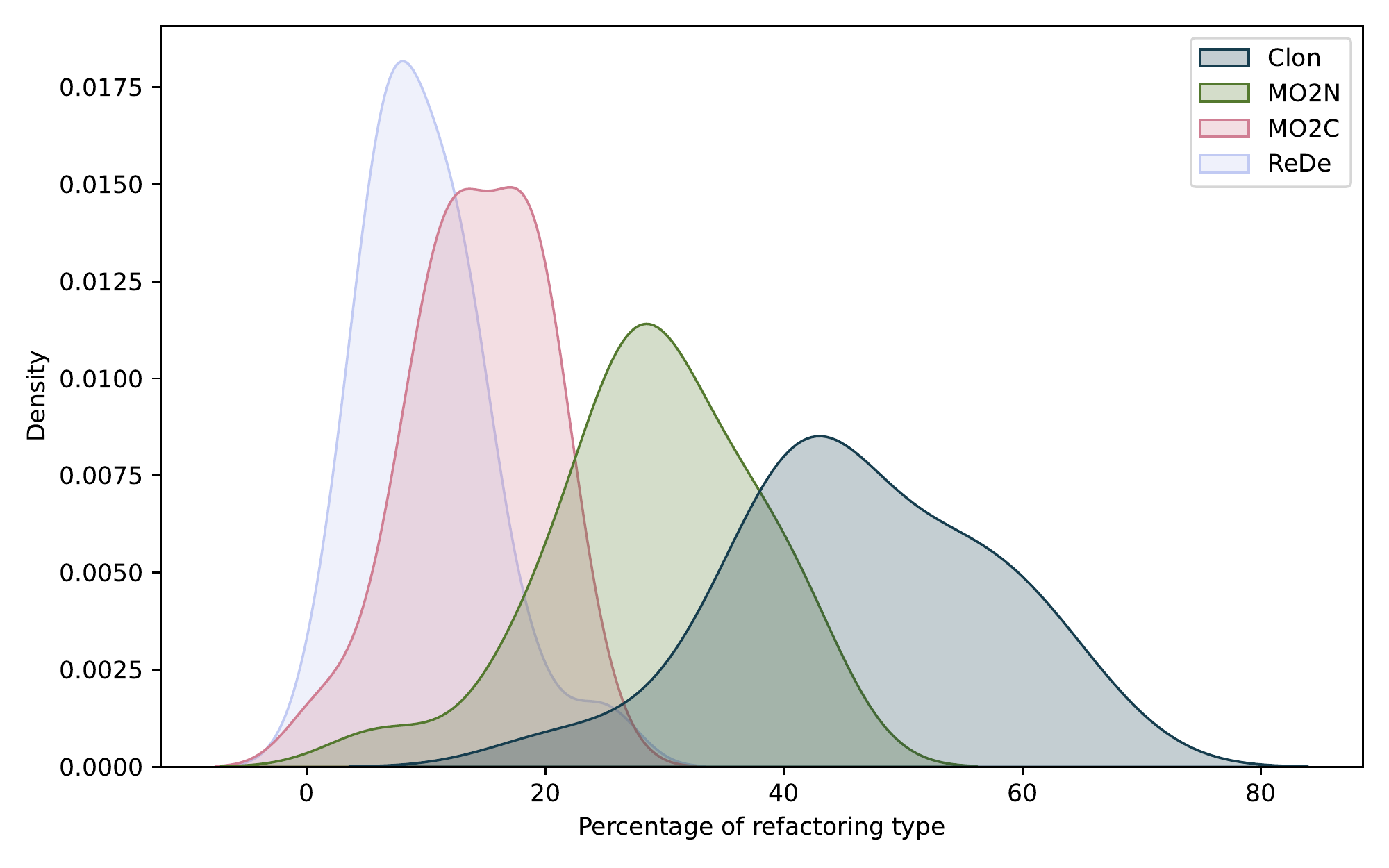}
        \caption{\ccm}
        \label{fig:ref_actions_dist_ccm}
    \end{subfigure}
    \caption{Distributions of refactoring types among different configurations.}
    \label{fig:ref_actions_dist}
\end{figure}

\paragraph{Discussion}
In both case studies, we can observe a common trend on preferring some refactoring types over other ones.
In order to confirm that the trend is consistent, we show in \figref{fig:ref_actions_dist} the density distributions of the shares of refactoring types across the different configurations.
The order in which the distributions are shifted along the x-axis is the same in both cases, and their overlapping is somehow similar.
This indicates that, on average, the refactoring types are selected with the same order of preference. 
We can also notice that, while in \ccm the variability decreases together with the average percentage, in Train Ticket Booking Service the situation is less clear.
A greater variability indicates that there are more chances that a change in the configuration will lead to a change in the selection preference of refactoring types, as it can be observed for \emph{Clon} and \emph{MO2N}.
On the other hand, a narrow distribution means that configuration changes have little effect on the selection choice, as it happens for \emph{MO2C} and \emph{ReDe}.
However, the refactorings that are more likely to be selected (\ie \emph{Clon} and \emph{MO2N}) exhibit larger variability in both case studies, thus meaning that these refactorings are also the most variable ones from one configuration to another.
This may indicate that, even if these two refactorings dominate, on average, the composition of solutions, the Pareto frontiers obtained by different configurations tend to be quite diverse.

Another aspect to consider is the influence of \brf on the choice of refactoring actions.
While \brf clearly has a direct impact on the \achanges objective, it looks like its presence is not enough to impose a different order of preference among the refactoring types.
On the one hand, it could be expected that the \emph{Clon} refactoring will be the most preferred because of its low \brf ($1.23$), but on the other hand the \emph{MO2N} refactoring, that is consistently in the second place, has the highest value of \brf.

In an attempt to understand if there is a stronger relation between refactoring types and the objectives, we have also performed a multiple regression analysis.
We tried to predict \perfq, \reliability, and \achanges using the refactoring types as predictors.
The coefficients of determination ($r^2$) we obtained for each objective and for both case studies are very low.
This means that the refactoring types are not suitable to explain most of the variability we observe in the objectives.
Such a result might be the indication that, at least for the two case studies we considered, we are not able to derive general refactoring strategies to improve the objectives without going through the optimization process.

\medskip

\answerRQ{From our experimentation, we were able to establish an order of preference among refactoring types that is consistent in both case studies.}
 
 \section{Threats to validity}\label{sec:t2v}

The validity of our study can be affected by different threats described by the Wohlin \etal classification~\cite{wohlin2012experimentation}. In the following, we detail each category by discussing the causes and motivations for each threat.

\paragraph{Construct validity}
The way we have designed our problem and our experimentation might be affected by \emph{Construct validity} threats. In particular, the role played by the architectural distance objective on the combination of refactoring actions might affect the selection of refactoring actions. However, we have studied the influence of our \brf in building \computedP in two different case studies, and it has coherently shown the ability to improve the overall quality of the non-dominated solutions in both cases. We will further investigate to what extent \brf could improve the overall quality with more accurate cost estimation, such as COCOMO~\cite{boehm2009software}, which might have as drawback the increase of the execution time for \brf estimation.

Another important aspect that might threaten our experimentation concerns the parameters of the initial UML model. For example, \ccm showed higher initial reliability that might affect the search. However, in our experiments, it seems that \ttbs and \ccm initial configurations did not threaten the optimization process. We will further investigate how different initial UML model parameters could change the optimization results. We remark that changing a single model parameter means starting the optimization process on a different point of the solution space that might produce completely different results.

\paragraph{Internal validity}
Our optimization approach might be affected by \emph{internal validity} threats. There are high degrees of freedom on our settings. For example, the variations of genetic configurations, such as the $P_{crossover}$ probability, may produce \computedP with different quality solutions. Also, the problem configuration variations may also change our results. The degrees of freedom in our experimentation generate unfeasible brute force investigation of each suitable combination. For this reason, we limit the variability to subsets of problem configurations, as shown in \tabref{tab:config_params}. We also mitigate this threat by involving two different case studies derived from the literature, thus reducing biases in their construction.

A fruitful investigation will be on the length of the sequence of refactoring actions. At this stage, we fixed the length to four actions. It will be interesting to investigate how the length of the sequence affects results. At a glance, the longer the sequence, the farther the solutions can go from the initial one, and it means that having a long sequence of refactoring actions might be unfeasible because it generates different model alternatives.

\paragraph{External validity}

Our results might be affected by \emph{external validity} threats, as their generalization might be limited to some of the assumptions behind our approach.

In the first place, a threat might be represented by the use of a single modeling notation.
We cannot generalize our results to other modeling notations, which could imply using a different portfolio of refactoring actions. The syntax and semantics of the modeling notation determine the amount and nature of refactoring actions that can be performed. 
However, we have adopted UML, which is the de facto standard in the software modeling domain. In general terms, this threat can be mitigated by porting the whole approach on a different modeling notation, but this is out of this paper scope. 

Another threat might be found in the fact that we have validated our approach on two case studies.
While the two case studies were selected from the available literature, they might not represent all the possible challenges that our approach could face in practice.
Nonetheless, our results could presumably hold in all the cases in which the modeling assumptions described in \secref{ref:assumptions} are met.
Specifically, the performance antipattern detection and the refactoring actions are designed to rely on information coming from static, dynamic, and deployment views of the system.
Without such information, even if in most cases the refactoring actions would still be applicable, they would not be as effective.

Finally, this study is limited to the use of a single algorithm.
Therefore, our results are influenced by the ability of \nsga of exploring the solution space, given the objectives of our approach.
While comparing the effectiveness of genetic algorithms in this context is out of the scope of this paper, we started investigating this issue~\cite{CORTELLESSA2021106568,SEAA2022}, and we will continue in future work.

\paragraph{Conclusion validity}
Our results might be affected by \emph{Conclusion validity} threats, since our considerations might change with deeply-tuned parameters for the \nsga. Also, parameter configurations might threaten our conclusion. We did not perform an extensive tuning phase for the latter due to the long duration of each run, while we used common parameters for the \nsga, which should mitigate these threats.
We can also soften this threat by employing other generic algorithms to generalize our results. Each algorithm will require its tuning phase, which is a clear drawback in execution time. 

Another aspect that might affect our results is the estimation of the reference Pareto frontier (\referenceP). \referenceP is used for extracting the quality indicators as described in \secref{sec:results}. We soften this threat by building the \referenceP overall our \computedP for each case study. Therefore, the reference Pareto should optimistically contain all non-dominated solutions across all configurations.

 \paragraph{Takeaways}\label{sec:takeaways}
Model-based multi-objective refactoring optimization presents a variety of challenges  that may jeopardize the validity of results.

Genetic algorithms contain a number of configuration options, to start.
Every parameter assignment may have an effect on the outcomes quality.
Indeed, there is opportunity for research direction here, since it would be impractical to evaluate every parameter combination.
There have been studies on determining the (almost) ideal configuration of genetic algorithms in diverse contexts.
We employed the standard genetic algorithm setup, such as the crossover probability~\citep{DBLP:journals/ese/ArcuriF13}. However, it would be interesting to see which study applies to our situation as well.
We plan to examine how different configurations affect the outcomes quality in future work.

The initial model setup is another factor taken into account.
Studies that mix running data (such as traces) and model artifacts already exist to address this problem.
There are plenty of shortcomings with these studies.
We recently investigated the potential of model-based performance predictions when models are fed with running application data~\citep{DBLP:journals/jss/CortellessaPET22}.
We discovered that if models take into account the confounding factors affecting application performance, such as network latency, they can anticipate the performance of the running application.

Moreover, the modeling notation affects how expressive the technique is.
For instance, the use of a domain specific language to speed up design time could impair models expressiveness.
Therefore, we chose to utilize UML, even though its broad general-purpose character is one of its disadvantages.
With regard to the modeling and annotation practices in industry, the effort dedicated to these activities can largely vary depending on the field where industries work. As an example, automotive industries have adopted (since many decades) model-driven engineering approaches for designing their embedded software systems.
For instance, \citet{DBLP:journals/tse/AmellerFGMABCCF21} provide an interesting study on the adoption in industrial contexts of modeling for sake of non-functional analysis.

Finally, regarding the applicability of the approach, it is difficult to establish a category of systems for which our approach would be better suited. Indeed, the only constraint that we require for its applicability is the usage of UML with the DAM~\cite{BernardiMP11} and MARTE~\cite{MARTE} profiles. Obviously, such approach should be applied in systems where performance and reliability requirements have high priority. For example: distributed systems where reliable connections and timely response are main critical issues; embedded domains (e.g., automotive) where resources with limited hardware capability must guarantee high reliability. Centralized systems represent a further category of systems that may be subject to stringent performance requirement because, for example, a single host machine and its hardware resources must manage a complex software system.
 \section{Related Work}\label{sec:related}

In the last decade, software model multi-objective optimization studies have been introduced to optimize various quality attributes (e.g., reliability, and energy \cite{Martens:2010bn,5949650,DBLP:conf/qosa/MeedeniyaBAG10,10.1007/978-3-642-13821-8_8}) with different degrees of freedom in the model modification (\eg service selection~\cite{DBLP:conf/IEEEscc/RosenbergMLMBD10,Cardellini:2009:QRA:1595696.1595718}). A systematic literature review on model optimization can be found in~\cite{Aleti:2013gp}. We consider here, as related work, those approaches that directly involve multi-objective evolutionary algorithms, and the ones that exploit LQN as performance modelling notation \cite{DBLP:conf/wosp/WoodsidePPSIM05,DBLP:conf/qest/LiAZCP17,DBLP:conf/cascon/AltamimiP07,Koziolek:2011cg}.

We split this section in two subsections, namely \emph{Software Architecture optimization} and \emph{Layered Queueing Network approaches}. The partition is not strict, as it might happen that some studies fall in both conceptual areas. In order to prevent duplication, we chose to describe these studies in only one specific area. 

\subsection{Software Architecture optimization}

Menasce \etal have presented a framework for architectural design and quality optimization~\cite{DBLP:conf/wosp/MenasceEGMS10}, where architectural patterns are used to support the search process (e.g., load balancing, fault tolerance). Two limitations affects the approach: the architecture has to be designed in a tool-related notation and not in a standard modelling language (as we do in this paper), and it uses equation-based analytical models for performance indices that could be too simple to capture architectural details and resource contention.

Aleti \etal~\cite{DBLP:conf/mompes/AletiBGM09} have presented an approach for modeling and analyzing AADL architectures~\cite{DBLP:books/daglib/0030032}. They have also introduced a tool aimed at optimizing different quality attributes while varying the architecture deployment and the component redundancy. Our work relies on UML models and considers more complex refactoring actions, as well as different target attributes for the fitness function. Besides, we investigate the role of performance antipatterns in the context of many-objective software model refactoring optimization.

A recent work compares the ability of two different multi-objective optimization approaches to improve non-functional attributes~\cite{NI2021106565}, where randomized search rules have been applied to improve the software model. The study of Ni~\etal is based on a specific modelling notation (\ie Palladio Component Model) and it has implicitly shown that the multi-objective optimization problem at model level is still an open challenge. They applied architectural tactics, which in general do not represent structured refactoring actions, to find optimal solutions. Conversely, we applied refactoring actions that change the structure of the initial model by preserving the original behavior. Another difference is the modelling notation, as we use UML with the goal of experimenting on a standard notation instead of a custom DSL.

Some authors of this paper have previously studied the sensitivity of multi-objective software model refactoring to configuration characteristics~\cite{CORTELLESSA2021106568}, where models are defined in \AE milia, which is a performance-oriented ADL. They compared two genetic algorithms in terms of Pareto frontiers quality. In this paper, we change the modelling notation from \AE milia to UML, and we add the reliability as a new objective. Both approaches provide a refactoring engine, however, in this paper, the refactoring engine offers more complex refactoring actions since UML is more expressive than \AE milia.

\citet{DBLP:journals/scp/EtemaadiC15} presented an approach aimed at improving architecture quality attributes through genetic algorithms.
The multi-objective optimization considers component-based architectures described through domain specific language (DSL), \ie AQOSA IR~\citep{DBLP:conf/cec/LiEEC11}.
The architecture evaluations can be obtained by means of several notation, such as Queueing Network and Fault Tree.
The genetic algorithm consider variation of designs (\eg number of hardware nodes) as objectives of the fitness function.
The main difference between our approach and the one of \citeauthor{DBLP:journals/scp/EtemaadiC15} is based on the types of the fitness function objectives. 
Yet, we used UML as the modeling notation instead of a DSL, and the LQN as the performance model. 

\subsection{Layered Queueing Network approaches}

\citeauthor{Koziolek:2011cg} have presented PerOpteryx~\cite{Koziolek:2011cg}, \ie a performance-oriented multi-objective optimization problem. In PerOpteryx the optimization process is guided by tactics referring to component reallocation, faster hardware, and more hardware. The latter ones do not represent structured refactoring actions, as we intend in this paper. PerOpteryx supports architectures specified in Palladio Component Model~\cite{Becker:2009cl} and produces, through model transformation, a LQN model for performance analysis. 

\citeauthor{10.1145/3132498.3132509} have presented SQuAT~\cite{10.1145/3132498.3132509}, which is an extensible platform aimed at including flexibility in the definition of an architecture optimization problem. SQuAT supports models conforming to Palladio Component Model language, exploits LQN for performance evaluation, and PerOpteryx tactics for architectural changes.
A main difference of our approach with PerOpteryx and SQuAt is that we use the UML modelling notation. 
We moved a step ahead with respect PerOpteryx and SQuAT. 
Beyond the modeling notation, we introduced more complex refactoring actions, and we use different objectives, \eg performance antipatterns.

Model-to-model (M2M) transformations from UML to LQN notations have been presented in~\cite{DBLP:conf/wosp/WoodsidePPSIM05,DBLP:conf/qest/LiAZCP17,DBLP:conf/cascon/AltamimiP07,altamimi_performance_2016}. 
For example, \citet{DBLP:conf/qest/LiAZCP17} presented a tool, namely Tulsa, aimed at enabling performance analysis of data intensive applications.
\citeauthor{DBLP:conf/qest/LiAZCP17} augmented UML models with the DICE profile, which allows expressing data intensive application domain specification.
Also, they introduced a model-to-model transformation aimed at allowing a performance analysis through Layered Queueing Network.
In contrast with these approaches, we present a novel M2M transformation mapping that employs UML Sequence Diagrams as the behavioral view of software architectures, instead of UML Activity Diagrams. UML Sequence Diagrams have two benefits: they are adopted more frequently than UML Activity Diagrams for software design~\cite{erickson2007can}, and they explicitly define method calls, while UML Activity Diagrams usually focus on workflows and processes. Therefore, our approach supports a more detailed behavioral representation in terms of time intervals between method calls.
 \section{Conclusions}\label{sec:conclusion}

In this work, we have used \nsga to optimize UML models with respect to performance and reliability properties, as well as the number of detected performance antipatterns and the architectural distance.
We focused our study on the impact that performance antipatterns may have on the quality of optimal refactoring solutions. We studied the composition of refactoring actions, and how the architectural distance metric can help the approach to compute Pareto frontiers.

From our experimentation, we gathered interesting insights about the quality of the generated solutions and the role of performance antipatterns as an objective of the algorithm.
In this regard, we showed that, by including the detection of performance antipatterns in the optimization process, we are able to obtain better solutions in terms of performance and reliability.
Moreover, we also showed that, the more we increase the probability of detecting a performance antipattern using the fuzziness threshold, the better the quality of the refactoring solutions. In addition, we noticed that the \emph{baseline refactoring factor} generally helps discovering better model alternatives.
Another important aspect of our study was to ensure that our approach did not worsen the reliability of the initial model.
In this respect, our experiments showed that we were in fact able to increase the reliability of model alternatives, with respect to the initial model, in the majority of cases.

As future work\label{sec:future-work}, we intend to tackle the threats to validity discussed before. In particular, we intend to investigate the influence of settings (\ie experiment and algorithm configurations) on the quality of Pareto frontiers. For example, we will investigate the impact of more dense populations in our analysis, in terms of computational time and quality of the computed Pareto frontiers (\computedP). 
Also, we are interested in the role played by \achanges, and specifically in studying the effect of estimating the \emph{baseline refactoring factor} through more complex cost model, such as COCOMO-II~\cite{boehm2009software}, on the combination of refactoring actions.
A fruitful investigation will be on the length of the sequence of refactoring actions, which is currently fixed to four refactoring actions, and we intend to extend the refactoring actions portfolio, for example, by including fault tolerance refactoring actions~\cite{DBLP:journals/infsof/CortellessaET20}.
We also intend to extend the reliability model to also take into account error propagation~\cite{DBLP:conf/cbse/CortellessaG07}.
We will involve other genetic algorithms in our process to study the contribution of different optimization techniques within the software model refactoring.

We also planned to study how modeling outcomes could be verified and estimated on real-systems.
As a first step to address this long-term study, we combined runtime traces (\ie traces from a running system) and modeling outcomes~\citep{DBLP:journals/jss/CortellessaPET22} and we found out that software models can help improve performance of software systems.

Another interesting aspect to investigate could be whether the refactoring actions proposed in the Pareto frontiers make sense form the point of view of the designer and within the established software development practices.
Therefore, we plan on using visualization techniques to conduct a detailed analysis of the solutions resulting from the optimization process.
Visualizing refactoring solutions also opens to a human-in-the-loop process, in which the designer could interactively drive the optimization towards acceptable solutions.

\section*{Acknowledgements}
\noindent Daniele Di Pompeo is supported by the Centre of EXcellence on Connected, Geo-Localized and Cybersecure Vehicle (EX-Emerge), funded by the Italian Government under CIPE resolution n. 70/2017 (Aug. 7, 2017).

\noindent Michele Tucci is supported by the OP RDE project No. CZ.02.2.69/\-0.0/\-0.0/\-18\_053/\-0016976 ``International mobility of research, technical and administrative staff at the Charles University''.

\bibliographystyle{elsarticle-num} 
\bibliography{biblio}

\begin{thebibliography}{10}
\expandafter\ifx\csname url\endcsname\relax
  \def\url#1{\texttt{#1}}\fi
\expandafter\ifx\csname urlprefix\endcsname\relax\def\urlprefix{URL }\fi
\expandafter\ifx\csname href\endcsname\relax
  \def\href#1#2{#2} \def\path#1{#1}\fi

\bibitem{fowler2018refactoring}
M.~Fowler, Refactoring: improving the design of existing code, Addison-Wesley
  Professional, 2018.

\bibitem{Bavota:2014kr}
G.~Bavota, M.~D. Penta, R.~Oliveto,
  \href{https://doi.org/10.1007/978-3-642-45398-4_4}{Search based software
  maintenance: Methods and tools}, in: T.~Mens, A.~Serebrenik, A.~Cleve (Eds.),
  Evolving Software Systems, Springer, 2014, pp. 103--137.
\newblock \href {https://doi.org/10.1007/978-3-642-45398-4_4}
  {\path{doi:10.1007/978-3-642-45398-4_4}}.
\newline\urlprefix\url{https://doi.org/10.1007/978-3-642-45398-4_4}

\bibitem{Kessentini:2012cb}
M.~Kessentini, H.~A. Sahraoui, M.~Boukadoum, O.~Benomar,
  \href{https://doi.org/10.1007/s10270-010-0175-7}{Search-based model
  transformation by example}, Softw. Syst. Model. 11~(2) (2012) 209--226.
\newblock \href {https://doi.org/10.1007/s10270-010-0175-7}
  {\path{doi:10.1007/s10270-010-0175-7}}.
\newline\urlprefix\url{https://doi.org/10.1007/s10270-010-0175-7}

\bibitem{Mariani:2017jd}
T.~Mariani, S.~R. Vergilio,
  \href{https://doi.org/10.1016/j.infsof.2016.11.009}{A systematic review on
  search-based refactoring}, Inf. Softw. Technol. 83 (2017) 14--34.
\newblock \href {https://doi.org/10.1016/j.infsof.2016.11.009}
  {\path{doi:10.1016/j.infsof.2016.11.009}}.
\newline\urlprefix\url{https://doi.org/10.1016/j.infsof.2016.11.009}

\bibitem{Ouni:2015db}
A.~Ouni, R.~G. Kula, M.~Kessentini, K.~Inoue,
  \href{https://doi.org/10.1145/2739480.2754724}{Web service antipatterns
  detection using genetic programming}, in: S.~Silva, A.~I.
  Esparcia{-}Alc{\'{a}}zar (Eds.), Proceedings of the Genetic and Evolutionary
  Computation Conference, {GECCO} 2015, Madrid, Spain, July 11-15, 2015, {ACM},
  2015, pp. 1351--1358.
\newblock \href {https://doi.org/10.1145/2739480.2754724}
  {\path{doi:10.1145/2739480.2754724}}.
\newline\urlprefix\url{https://doi.org/10.1145/2739480.2754724}

\bibitem{Ouni:2017db}
A.~Ouni, M.~Kessentini, K.~Inoue, M.~{\'{O}}. Cinn{\'{e}}ide,
  \href{https://doi.org/10.1109/TSC.2015.2502595}{Search-based web service
  antipatterns detection}, {IEEE} Trans. Serv. Comput. 10~(4) (2017) 603--617.
\newblock \href {https://doi.org/10.1109/TSC.2015.2502595}
  {\path{doi:10.1109/TSC.2015.2502595}}.
\newline\urlprefix\url{https://doi.org/10.1109/TSC.2015.2502595}

\bibitem{Ramirez:2018uz}
A.~Ram{\'{\i}}rez, J.~R. Romero, S.~Ventura,
  \href{https://doi.org/10.1016/j.jss.2018.12.015}{A survey of many-objective
  optimisation in search-based software engineering}, J. Syst. Softw. 149
  (2019) 382--395.
\newblock \href {https://doi.org/10.1016/j.jss.2018.12.015}
  {\path{doi:10.1016/j.jss.2018.12.015}}.
\newline\urlprefix\url{https://doi.org/10.1016/j.jss.2018.12.015}

\bibitem{Ray:2014ip}
M.~Ray, D.~P. Mohapatra,
  \href{https://doi.org/10.1007/s11334-014-0234-2}{Multi-objective test
  prioritization via a genetic algorithm}, Innov. Syst. Softw. Eng. 10~(4)
  (2014) 261--270.
\newblock \href {https://doi.org/10.1007/s11334-014-0234-2}
  {\path{doi:10.1007/s11334-014-0234-2}}.
\newline\urlprefix\url{https://doi.org/10.1007/s11334-014-0234-2}

\bibitem{DBLP:conf/mompes/AletiBGM09}
A.~Aleti, S.~Bj{\"{o}}rnander, L.~Grunske, I.~Meedeniya,
  \href{https://doi.org/10.1109/MOMPES.2009.5069138}{Archeopterix: An
  extendable tool for architecture optimization of {AADL} models}, in: {ICSE}
  2009 Workshop on Model-Based Methodologies for Pervasive and Embedded
  Software, {MOMPES} 2009, May 16, 2009, Vancouver, Canada, {IEEE} Computer
  Society, 2009, pp. 61--71.
\newblock \href {https://doi.org/10.1109/MOMPES.2009.5069138}
  {\path{doi:10.1109/MOMPES.2009.5069138}}.
\newline\urlprefix\url{https://doi.org/10.1109/MOMPES.2009.5069138}

\bibitem{Aleti:2013gp}
A.~Aleti, B.~Buhnova, L.~Grunske, A.~Koziolek, I.~Meedeniya,
  \href{https://doi.org/10.1109/TSE.2012.64}{Software architecture optimization
  methods: {A} systematic literature review}, {IEEE} Trans. Software Eng.
  39~(5) (2013) 658--683.
\newblock \href {https://doi.org/10.1109/TSE.2012.64}
  {\path{doi:10.1109/TSE.2012.64}}.
\newline\urlprefix\url{https://doi.org/10.1109/TSE.2012.64}

\bibitem{Martens:2010bn}
A.~Martens, H.~Koziolek, S.~Becker, R.~H. Reussner,
  \href{https://doi.org/10.1145/1712605.1712624}{Automatically improve software
  architecture models for performance, reliability, and cost using evolutionary
  algorithms}, in: A.~Adamson, A.~B. Bondi, C.~Juiz, M.~S. Squillante (Eds.),
  Proceedings of the first joint {WOSP/SIPEW} International Conference on
  Performance Engineering, San Jose, California, USA, January 28-30, 2010,
  {ACM}, 2010, pp. 105--116.
\newblock \href {https://doi.org/10.1145/1712605.1712624}
  {\path{doi:10.1145/1712605.1712624}}.
\newline\urlprefix\url{https://doi.org/10.1145/1712605.1712624}

\bibitem{CORTELLESSA2021106568}
V.~Cortellessa, D.~{Di Pompeo},
  \href{https://doi.org/10.1016/j.infsof.2021.106568}{Analyzing the sensitivity
  of multi-objective software architecture refactoring to configuration
  characteristics}, Inf. Softw. Technol. 135 (2021) 106568.
\newblock \href {https://doi.org/10.1016/j.infsof.2021.106568}
  {\path{doi:10.1016/j.infsof.2021.106568}}.
\newline\urlprefix\url{https://doi.org/10.1016/j.infsof.2021.106568}

\bibitem{NI2021106565}
Y.~Ni, X.~Du, P.~Ye, L.~L. Minku, X.~Yao, M.~Harman, R.~Xiao,
  \href{https://doi.org/10.1016/j.infsof.2021.106565}{Multi-objective software
  performance optimisation at the architecture level using randomised search
  rules}, Inf. Softw. Technol. 135 (2021) 106565.
\newblock \href {https://doi.org/10.1016/j.infsof.2021.106565}
  {\path{doi:10.1016/j.infsof.2021.106565}}.
\newline\urlprefix\url{https://doi.org/10.1016/j.infsof.2021.106565}

\bibitem{Koziolek:2011cg}
A.~Koziolek, H.~Koziolek, R.~H. Reussner,
  \href{https://doi.org/10.1145/2000259.2000267}{Peropteryx: automated
  application of tactics in multi-objective software architecture
  optimization}, in: I.~Crnkovic, J.~A. Stafford, D.~C. Petriu, J.~Happe,
  P.~Inverardi (Eds.), 7th International Conference on the Quality of Software
  Architectures, QoSA 2011 and 2nd International Symposium on Architecting
  Critical Systems, {ISARCS} 2011. Boulder, CO, USA, June 20-24, 2011,
  Proceedings, {ACM}, 2011, pp. 33--42.
\newblock \href {https://doi.org/10.1145/2000259.2000267}
  {\path{doi:10.1145/2000259.2000267}}.
\newline\urlprefix\url{https://doi.org/10.1145/2000259.2000267}

\bibitem{Deb:2002ut}
K.~Deb, S.~Agrawal, A.~Pratap, T.~Meyarivan,
  \href{https://doi.org/10.1109/4235.996017}{A fast and elitist multiobjective
  genetic algorithm: {NSGA-II}}, {IEEE} Trans. Evol. Comput. 6~(2) (2002)
  182--197.
\newblock \href {https://doi.org/10.1109/4235.996017}
  {\path{doi:10.1109/4235.996017}}.
\newline\urlprefix\url{https://doi.org/10.1109/4235.996017}

\bibitem{DBLP:journals/tse/NeilsonWPM95}
J.~E. Neilson, C.~M. Woodside, D.~C. Petriu, S.~Majumdar,
  \href{https://doi.org/10.1109/32.464543}{Software bootlenecking in
  client-server systems and rendezvous networks}, {IEEE} Trans. Software Eng.
  21~(9) (1995) 776--782.
\newblock \href {https://doi.org/10.1109/32.464543}
  {\path{doi:10.1109/32.464543}}.
\newline\urlprefix\url{https://doi.org/10.1109/32.464543}

\bibitem{CortellessaSC02}
V.~Cortellessa, H.~Singh, B.~Cukic,
  \href{https://doi.org/10.1145/584369.584415}{Early reliability assessment of
  {UML} based software models}, in: Third International Workshop on Software
  and Performance, WOSP@ISSTA 2002, July 24-26, 2002, Rome, Italy, {ACM}, 2002,
  pp. 302--309.
\newblock \href {https://doi.org/10.1145/584369.584415}
  {\path{doi:10.1145/584369.584415}}.
\newline\urlprefix\url{https://doi.org/10.1145/584369.584415}

\bibitem{DBLP:journals/infsof/ArcelliCP18}
D.~Arcelli, V.~Cortellessa, D.~{Di Pompeo},
  \href{https://doi.org/10.1016/j.infsof.2017.09.006}{Performance-driven
  software model refactoring}, Inf. Softw. Technol. 95 (2018) 366--397.
\newblock \href {https://doi.org/10.1016/j.infsof.2017.09.006}
  {\path{doi:10.1016/j.infsof.2017.09.006}}.
\newline\urlprefix\url{https://doi.org/10.1016/j.infsof.2017.09.006}

\bibitem{Arcelli:2018vo}
D.~Arcelli, V.~Cortellessa, M.~D'Emidio, D.~{Di Pompeo},
  \href{https://doi.org/10.1109/ICSA.2018.00020}{{EASIER:} an evolutionary
  approach for multi-objective software architecture refactoring}, in: {IEEE}
  International Conference on Software Architecture, {ICSA} 2018, Seattle, WA,
  USA, April 30 - May 4, 2018, {IEEE} Computer Society, 2018, pp. 105--114.
\newblock \href {https://doi.org/10.1109/ICSA.2018.00020}
  {\path{doi:10.1109/ICSA.2018.00020}}.
\newline\urlprefix\url{https://doi.org/10.1109/ICSA.2018.00020}

\bibitem{DBLP:conf/wosp/SmithW00}
C.~U. Smith, L.~G. Williams,
  \href{https://doi.org/10.1145/350391.350420}{Software performance
  antipatterns}, in: Second International Workshop on Software and Performance,
  {WOSP} 2000, Ottawa, Canada, September 17-20, 2000, {ACM}, 2000, pp.
  127--136.
\newblock \href {https://doi.org/10.1145/350391.350420}
  {\path{doi:10.1145/350391.350420}}.
\newline\urlprefix\url{https://doi.org/10.1145/350391.350420}

\bibitem{DBLP:conf/cmg/SmithW01a}
C.~U. Smith, L.~G. Williams, Software performance antipatterns; common
  performance problems and their solutions, in: 27th International Computer
  Measurement Group Conference, Anaheim, CA, USA, December 2-7, 2001, Computer
  Measurement Group, 2001, pp. 797--806.

\bibitem{Smith:2003wv}
C.~U. Smith, L.~G. Williams, {More New Software Performance Antipatterns: Even
  More Ways to Shoot Yourself in the Foot}, in: 29th International Computer
  Measurement Group Conference, 2003, pp. 717--725.

\bibitem{MARTE}
O.~M. Group, \href{http://www.omg.org/omgmarte/}{{A UML profile for MARTE:
  modeling and analysis of real-time embedded systems}}, Object Management
  Group (2008).
\newline\urlprefix\url{http://www.omg.org/omgmarte/}

\bibitem{BernardiMP11}
S.~Bernardi, J.~Merseguer, D.~C. Petriu,
  \href{https://doi.org/10.1007/s10270-009-0128-1}{A dependability profile
  within {MARTE}}, Softw. Syst. Model. 10~(3) (2011) 313--336.
\newblock \href {https://doi.org/10.1007/s10270-009-0128-1}
  {\path{doi:10.1007/s10270-009-0128-1}}.
\newline\urlprefix\url{https://doi.org/10.1007/s10270-009-0128-1}

\bibitem{SEAA2021}
V.~Cortellessa, D.~{Di Pompeo}, V.~Stoico, M.~Tucci,
  \href{https://doi.org/10.1109/SEAA53835.2021.00036}{On the impact of
  performance antipatterns in multi-objective software model refactoring
  optimization}, in: M.~T. Baldassarre, G.~Scanniello, A.~Skavhaug (Eds.), 47th
  Euromicro Conference on Software Engineering and Advanced Applications,
  {SEAA} 2021, Palermo, Italy, September 1-3, 2021, {IEEE}, 2021, pp. 224--233.
\newblock \href {https://doi.org/10.1109/SEAA53835.2021.00036}
  {\path{doi:10.1109/SEAA53835.2021.00036}}.
\newline\urlprefix\url{https://doi.org/10.1109/SEAA53835.2021.00036}

\bibitem{DBLP:conf/staf/Pompeo0CE19}
D.~{Di Pompeo}, M.~Tucci, A.~Celi, R.~Eramo,
  \href{http://ceur-ws.org/Vol-2405/06_paper.pdf}{A microservice reference case
  study for design-runtime interaction in {MDE}}, in: A.~Bagnato,
  H.~Bruneli{\`{e}}re, L.~Burgue{\~{n}}o, R.~Eramo, A.~G{\'{o}}mez (Eds.),
  {STAF} 2019 Co-Located Events Joint Proceedings: 1st Junior Researcher
  Community Event, 2nd International Workshop on Model-Driven Engineering for
  Design-Runtime Interaction in Complex Systems, and 1st Research Project
  Showcase Workshop co-located with Software Technologies: Applications and
  Foundations {(STAF} 2019), Eindhoven, The Netherlands, July 15 - 19, 2019,
  Vol. 2405 of {CEUR} Workshop Proceedings, CEUR-WS.org, 2019, pp. 23--32.
\newline\urlprefix\url{http://ceur-ws.org/Vol-2405/06_paper.pdf}

\bibitem{DBLP:journals/tse/ZhouPXSJLD21}
X.~Zhou, X.~Peng, T.~Xie, J.~Sun, C.~Ji, W.~Li, D.~Ding,
  \href{https://doi.org/10.1109/TSE.2018.2887384}{Fault analysis and debugging
  of microservice systems: Industrial survey, benchmark system, and empirical
  study}, {IEEE} Trans. Software Eng. 47~(2) (2021) 243--260.
\newblock \href {https://doi.org/10.1109/TSE.2018.2887384}
  {\path{doi:10.1109/TSE.2018.2887384}}.
\newline\urlprefix\url{https://doi.org/10.1109/TSE.2018.2887384}

\bibitem{Herold2008}
S.~Herold, H.~Klus, Y.~Welsch, C.~Deiters, A.~Rausch, R.~Reussner, K.~Krogmann,
  H.~Koziolek, R.~Mirandola, B.~Hummel, M.~Meisinger, C.~Pfaller,
  \href{https://doi.org/10.1007/978-3-540-85289-6_3}{Cocome - the common
  component modeling example}, in: The Common Component Modeling Example:
  Comparing Software Component Models, Vol. 5153 of LNCS, Springer Berlin
  Heidelberg, Berlin, Heidelberg, 2008, pp. 16--53.
\newblock \href {https://doi.org/10.1007/978-3-540-85289-6_3}
  {\path{doi:10.1007/978-3-540-85289-6_3}}.
\newline\urlprefix\url{https://doi.org/10.1007/978-3-540-85289-6_3}

\bibitem{Mansoor:2015bm}
U.~Mansoor, M.~Kessentini, M.~Wimmer, K.~Deb,
  \href{https://doi.org/10.1007/s11219-015-9284-4}{Multi-view refactoring of
  class and activity diagrams using a multi-objective evolutionary algorithm},
  Softw. Qual. J. 25~(2) (2017) 473--501.
\newblock \href {https://doi.org/10.1007/s11219-015-9284-4}
  {\path{doi:10.1007/s11219-015-9284-4}}.
\newline\urlprefix\url{https://doi.org/10.1007/s11219-015-9284-4}

\bibitem{DBLP:conf/qest/LiAZCP17}
C.~Li, T.~Altamimi, M.~H. Zargari, G.~Casale, D.~C. Petriu,
  \href{https://doi.org/10.1007/978-3-319-66335-7_18}{Tulsa: {A} tool for
  transforming {UML} to layered queueing networks for performance analysis of
  data intensive applications}, in: N.~Bertrand, L.~Bortolussi (Eds.),
  Quantitative Evaluation of Systems - 14th International Conference, {QEST}
  2017, Berlin, Germany, September 5-7, 2017, Proceedings, Vol. 10503 of
  Lecture Notes in Computer Science, Springer, 2017, pp. 295--299.
\newblock \href {https://doi.org/10.1007/978-3-319-66335-7_18}
  {\path{doi:10.1007/978-3-319-66335-7_18}}.
\newline\urlprefix\url{https://doi.org/10.1007/978-3-319-66335-7_18}

\bibitem{DBLP:books/sp/03/SmithW03}
C.~U. Smith, L.~G. Williams,
  \href{https://doi.org/10.1007/0-306-48738-1\_16}{Software performance
  engineering}, in: L.~Lavagno, G.~Martin, B.~Selic (Eds.), {UML} for Real -
  Design of Embedded Real-Time Systems, Kluwer, 2003, pp. 343--365.
\newblock \href {https://doi.org/10.1007/0-306-48738-1\_16}
  {\path{doi:10.1007/0-306-48738-1\_16}}.
\newline\urlprefix\url{https://doi.org/10.1007/0-306-48738-1\_16}

\bibitem{DBLP:journals/sosym/CortellessaMT14}
V.~Cortellessa, A.~D. Marco, C.~Trubiani,
  \href{https://doi.org/10.1007/s10270-012-0246-z}{An approach for modeling and
  detecting software performance antipatterns based on first-order logics},
  Softw. Syst. Model. 13~(1) (2014) 391--432.
\newblock \href {https://doi.org/10.1007/s10270-012-0246-z}
  {\path{doi:10.1007/s10270-012-0246-z}}.
\newline\urlprefix\url{https://doi.org/10.1007/s10270-012-0246-z}

\bibitem{DBLP:conf/fase/ArcelliCT15}
D.~Arcelli, V.~Cortellessa, C.~Trubiani,
  \href{https://doi.org/10.1007/978-3-662-46675-9_10}{Performance-based
  software model refactoring in fuzzy contexts}, in: A.~Egyed, I.~Schaefer
  (Eds.), Fundamental Approaches to Software Engineering - 18th International
  Conference, {FASE} 2015, Held as Part of the European Joint Conferences on
  Theory and Practice of Software, {ETAPS} 2015, London, UK, April 11-18, 2015.
  Proceedings, Vol. 9033 of Lecture Notes in Computer Science, Springer, 2015,
  pp. 149--164.
\newblock \href {https://doi.org/10.1007/978-3-662-46675-9_10}
  {\path{doi:10.1007/978-3-662-46675-9_10}}.
\newline\urlprefix\url{https://doi.org/10.1007/978-3-662-46675-9_10}

\bibitem{DBLP:conf/wcre/ArcelliCP19}
D.~Arcelli, V.~Cortellessa, D.~D. Pompeo,
  \href{https://doi.org/10.1109/SANER.2019.8667967}{Automating performance
  antipattern detection and software refactoring in {UML} models}, in: X.~Wang,
  D.~Lo, E.~Shihab (Eds.), 26th {IEEE} International Conference on Software
  Analysis, Evolution and Reengineering, {SANER} 2019, Hangzhou, China,
  February 24-27, 2019, {IEEE}, 2019, pp. 639--643.
\newblock \href {https://doi.org/10.1109/SANER.2019.8667967}
  {\path{doi:10.1109/SANER.2019.8667967}}.
\newline\urlprefix\url{https://doi.org/10.1109/SANER.2019.8667967}

\bibitem{DBLP:conf/kbse/ArcelliCP18}
D.~Arcelli, V.~Cortellessa, D.~{Di Pompeo},
  \href{https://doi.org/10.1145/3242163.3242167}{A metamodel for the
  specification and verification of model refactoring actions}, in: A.~Ouni,
  M.~Kessentini, M.~{\'{O}}. Cinn{\'{e}}ide (Eds.), Proceedings of the 2nd
  International Workshop on Refactoring, IWoR@ASE 2018, Montpellier, France,
  September 4, 2018, IWoR@ACM, 2018, pp. 14--21.
\newblock \href {https://doi.org/10.1145/3242163.3242167}
  {\path{doi:10.1145/3242163.3242167}}.
\newline\urlprefix\url{https://doi.org/10.1145/3242163.3242167}

\bibitem{boehm2009software}
B.~W. Boehm, C.~Abts, A.~W. Brown, S.~Chulani, B.~K. Clark, E.~Horowitz,
  R.~Madachy, D.~J. Reifer, B.~Steece, Software cost estimation with COCOMO II,
  Prentice Hall Press, 2009.

\bibitem{trendowicz2013software}
A.~Trendowicz, Software Cost Estimation, Benchmarking, and Risk Assessment: The
  Software Decision-Makers' Guide to Predictable Software Development, Springer
  Science \& Business Media, 2013.

\bibitem{DBLP:conf/gecco/NebroDV15}
A.~J. Nebro, J.~J. Durillo, M.~Vergne,
  \href{https://doi.org/10.1145/2739482.2768462}{Redesigning the jmetal
  multi-objective optimization framework}, in: S.~Silva, A.~I.
  Esparcia{-}Alc{\'{a}}zar (Eds.), Genetic and Evolutionary Computation
  Conference, {GECCO} 2015, Madrid, Spain, July 11-15, 2015, Companion Material
  Proceedings, {ACM}, 2015, pp. 1093--1100.
\newblock \href {https://doi.org/10.1145/2739482.2768462}
  {\path{doi:10.1145/2739482.2768462}}.
\newline\urlprefix\url{https://doi.org/10.1145/2739482.2768462}

\bibitem{Ali_Arcaini_Pradhan_Safdar_Yue_2020}
S.~Ali, P.~Arcaini, D.~Pradhan, S.~A. Safdar, T.~Yue,
  \href{https://doi.org/10.1145/3375636}{Quality indicators in search-based
  software engineering: An empirical evaluation}, {ACM} Trans. Softw. Eng.
  Methodol. 29~(2) (2020) 10:1--10:29.
\newblock \href {https://doi.org/10.1145/3375636} {\path{doi:10.1145/3375636}}.
\newline\urlprefix\url{https://doi.org/10.1145/3375636}

\bibitem{DBLP:conf/cec/ZhouJZST06}
A.~Zhou, Y.~Jin, Q.~Zhang, B.~Sendhoff, E.~P.~K. Tsang,
  \href{https://doi.org/10.1109/CEC.2006.1688406}{Combining model-based and
  genetics-based offspring generation for multi-objective optimization using a
  convergence criterion}, in: {IEEE} International Conference on Evolutionary
  Computation, {CEC} 2006, part of {WCCI} 2006, Vancouver, BC, Canada, 16-21
  July 2006, {IEEE}, 2006, pp. 892--899.
\newblock \href {https://doi.org/10.1109/CEC.2006.1688406}
  {\path{doi:10.1109/CEC.2006.1688406}}.
\newline\urlprefix\url{https://doi.org/10.1109/CEC.2006.1688406}

\bibitem{DBLP:conf/cec/IshibuchiMN16}
H.~Ishibuchi, H.~Masuda, Y.~Nojima,
  \href{https://doi.org/10.1109/CEC.2016.7743912}{Sensitivity of performance
  evaluation results by inverted generational distance to reference points},
  in: {IEEE} Congress on Evolutionary Computation, {CEC} 2016, Vancouver, BC,
  Canada, July 24-29, 2016, {IEEE}, 2016, pp. 1107--1114.
\newblock \href {https://doi.org/10.1109/CEC.2016.7743912}
  {\path{doi:10.1109/CEC.2016.7743912}}.
\newline\urlprefix\url{https://doi.org/10.1109/CEC.2016.7743912}

\bibitem{DBLP:journals/tec/ZitzlerT99}
E.~Zitzler, L.~Thiele,
  \href{https://doi.org/10.1109/4235.797969}{Multiobjective evolutionary
  algorithms: a comparative case study and the strength pareto approach},
  {IEEE} Trans. Evol. Comput. 3~(4) (1999) 257--271.
\newblock \href {https://doi.org/10.1109/4235.797969}
  {\path{doi:10.1109/4235.797969}}.
\newline\urlprefix\url{https://doi.org/10.1109/4235.797969}

\bibitem{DBLP:journals/tec/ZitzlerTLFF03}
E.~Zitzler, L.~Thiele, M.~Laumanns, C.~M. Fonseca, V.~G. da~Fonseca,
  \href{https://doi.org/10.1109/TEVC.2003.810758}{Performance assessment of
  multiobjective optimizers: an analysis and review}, {IEEE} Trans. Evol.
  Comput. 7~(2) (2003) 117--132.
\newblock \href {https://doi.org/10.1109/TEVC.2003.810758}
  {\path{doi:10.1109/TEVC.2003.810758}}.
\newline\urlprefix\url{https://doi.org/10.1109/TEVC.2003.810758}

\bibitem{DBLP:journals/ese/ArcuriF13}
A.~Arcuri, G.~Fraser,
  \href{https://doi.org/10.1007/s10664-013-9249-9}{Parameter tuning or default
  values? an empirical investigation in search-based software engineering},
  Empir. Softw. Eng. 18~(3) (2013) 594--623.
\newblock \href {https://doi.org/10.1007/s10664-013-9249-9}
  {\path{doi:10.1007/s10664-013-9249-9}}.
\newline\urlprefix\url{https://doi.org/10.1007/s10664-013-9249-9}

\bibitem{Arcuri_Fraser_2011}
A.~Arcuri, G.~Fraser, \href{https://doi.org/10.1007/978-3-642-23716-4_6}{On
  parameter tuning in search based software engineering}, in: M.~B. Cohen,
  M.~{\'{O}}. Cinn{\'{e}}ide (Eds.), Search Based Software Engineering - Third
  International Symposium, {SSBSE} 2011, Szeged, Hungary, September 10-12,
  2011. Proceedings, Vol. 6956 of Lecture Notes in Computer Science, Springer,
  2011, pp. 33--47.
\newblock \href {https://doi.org/10.1007/978-3-642-23716-4_6}
  {\path{doi:10.1007/978-3-642-23716-4_6}}.
\newline\urlprefix\url{https://doi.org/10.1007/978-3-642-23716-4_6}

\bibitem{wohlin2012experimentation}
C.~Wohlin, P.~Runeson, M.~H{\"{o}}st, M.~C. Ohlsson, B.~Regnell,
  \href{https://doi.org/10.1007/978-3-642-29044-2}{Experimentation in Software
  Engineering}, Springer, 2012.
\newblock \href {https://doi.org/10.1007/978-3-642-29044-2}
  {\path{doi:10.1007/978-3-642-29044-2}}.
\newline\urlprefix\url{https://doi.org/10.1007/978-3-642-29044-2}

\bibitem{SEAA2022}
D.~{Di Pompeo}, M.~Tucci,
  \href{https://doi.org/10.1109/SEAA56994.2022.00070}{Search budget in
  multi-objective refactoring optimization: a model-based empirical study}, in:
  48th Euromicro Conference on Software Engineering and Advanced Applications,
  {SEAA} 2022, {IEEE}, 2022, pp. 406--413, to appear.
\newblock \href {https://doi.org/10.1109/SEAA56994.2022.00070}
  {\path{doi:10.1109/SEAA56994.2022.00070}}.
\newline\urlprefix\url{https://doi.org/10.1109/SEAA56994.2022.00070}

\bibitem{DBLP:journals/jss/CortellessaPET22}
V.~Cortellessa, D.~{Di Pompeo}, R.~Eramo, M.~Tucci,
  \href{https://doi.org/10.1016/j.jss.2021.111084}{A model-driven approach for
  continuous performance engineering in microservice-based systems}, J. Syst.
  Softw. 183 (2022) 111084.
\newblock \href {https://doi.org/10.1016/j.jss.2021.111084}
  {\path{doi:10.1016/j.jss.2021.111084}}.
\newline\urlprefix\url{https://doi.org/10.1016/j.jss.2021.111084}

\bibitem{DBLP:journals/tse/AmellerFGMABCCF21}
D.~Ameller, X.~Franch, C.~G{\'{o}}mez, S.~Mart{\'{\i}}nez{-}Fern{\'{a}}ndez,
  J.~Ara{\'{u}}jo, S.~Biffl, J.~Cabot, V.~Cortellessa, D.~M. Fern{\'{a}}ndez,
  A.~Moreira, H.~Muccini, A.~Vallecillo, M.~Wimmer, V.~Amaral, W.~B{\"{o}}hm,
  H.~Bruneli{\`{e}}re, L.~Burgue{\~{n}}o, M.~Goul{\~{a}}o, S.~Teufl,
  L.~Berardinelli, \href{https://doi.org/10.1109/TSE.2019.2904476}{Dealing with
  non-functional requirements in model-driven development: {A} survey}, {IEEE}
  Trans. Software Eng. 47~(4) (2021) 818--835.
\newblock \href {https://doi.org/10.1109/TSE.2019.2904476}
  {\path{doi:10.1109/TSE.2019.2904476}}.
\newline\urlprefix\url{https://doi.org/10.1109/TSE.2019.2904476}

\bibitem{5949650}
R.~Li, R.~Etemaadi, M.~T.~M. Emmerich, M.~R.~V. Chaudron,
  \href{https://doi.org/10.1109/CEC.2011.5949650}{An evolutionary
  multiobjective optimization approach to component-based software architecture
  design}, in: Proceedings of the {IEEE} Congress on Evolutionary Computation,
  {CEC} 2011, New Orleans, LA, USA, 5-8 June, 2011, {IEEE}, 2011, pp. 432--439.
\newblock \href {https://doi.org/10.1109/CEC.2011.5949650}
  {\path{doi:10.1109/CEC.2011.5949650}}.
\newline\urlprefix\url{https://doi.org/10.1109/CEC.2011.5949650}

\bibitem{DBLP:conf/qosa/MeedeniyaBAG10}
I.~Meedeniya, B.~Buhnova, A.~Aleti, L.~Grunske,
  \href{https://doi.org/10.1007/978-3-642-13821-8_6}{Architecture-driven
  reliability and energy optimization for complex embedded systems}, in: G.~T.
  Heineman, J.~Kofron, F.~Plasil (Eds.), Research into Practice - Reality and
  Gaps, 6th International Conference on the Quality of Software Architectures,
  QoSA 2010, Prague, Czech Republic, June 23 - 25, 2010. Proceedings, Vol. 6093
  of Lecture Notes in Computer Science, Springer, 2010, pp. 52--67.
\newblock \href {https://doi.org/10.1007/978-3-642-13821-8_6}
  {\path{doi:10.1007/978-3-642-13821-8_6}}.
\newline\urlprefix\url{https://doi.org/10.1007/978-3-642-13821-8_6}

\bibitem{10.1007/978-3-642-13821-8_8}
A.~Martens, D.~Ardagna, H.~Koziolek, R.~Mirandola, R.~H. Reussner,
  \href{https://doi.org/10.1007/978-3-642-13821-8_8}{A hybrid approach for
  multi-attribute qos optimisation in component based software systems}, in:
  G.~T. Heineman, J.~Kofron, F.~Plasil (Eds.), Research into Practice - Reality
  and Gaps, 6th International Conference on the Quality of Software
  Architectures, QoSA 2010, Prague, Czech Republic, June 23 - 25, 2010.
  Proceedings, Vol. 6093 of Lecture Notes in Computer Science, Springer, 2010,
  pp. 84--101.
\newblock \href {https://doi.org/10.1007/978-3-642-13821-8_8}
  {\path{doi:10.1007/978-3-642-13821-8_8}}.
\newline\urlprefix\url{https://doi.org/10.1007/978-3-642-13821-8_8}

\bibitem{DBLP:conf/IEEEscc/RosenbergMLMBD10}
F.~Rosenberg, M.~B. M{\"{u}}ller, P.~Leitner, A.~Michlmayr, A.~Bouguettaya,
  S.~Dustdar, \href{https://doi.org/10.1109/SCC.2010.58}{Metaheuristic
  optimization of large-scale qos-aware service compositions}, in: 2010 {IEEE}
  International Conference on Services Computing, {SCC} 2010, Miami, Florida,
  USA, July 5-10, 2010, {IEEE} Computer Society, 2010, pp. 97--104.
\newblock \href {https://doi.org/10.1109/SCC.2010.58}
  {\path{doi:10.1109/SCC.2010.58}}.
\newline\urlprefix\url{https://doi.org/10.1109/SCC.2010.58}

\bibitem{Cardellini:2009:QRA:1595696.1595718}
V.~Cardellini, E.~Casalicchio, V.~Grassi, F.~L. Presti, R.~Mirandola,
  \href{https://doi.org/10.1145/1595696.1595718}{Qos-driven runtime adaptation
  of service oriented architectures}, in: H.~van Vliet, V.~Issarny (Eds.),
  Proceedings of the 7th joint meeting of the European Software Engineering
  Conference and the {ACM} {SIGSOFT} International Symposium on Foundations of
  Software Engineering, 2009, Amsterdam, The Netherlands, August 24-28, 2009,
  {ACM}, 2009, pp. 131--140.
\newblock \href {https://doi.org/10.1145/1595696.1595718}
  {\path{doi:10.1145/1595696.1595718}}.
\newline\urlprefix\url{https://doi.org/10.1145/1595696.1595718}

\bibitem{DBLP:conf/wosp/WoodsidePPSIM05}
C.~M. Woodside, D.~C. Petriu, D.~B. Petriu, H.~Shen, T.~Israr, J.~Merseguer,
  \href{https://doi.org/10.1145/1071021.1071022}{Performance by unified model
  analysis {(PUMA)}}, in: Proceedings of the Fifth International Workshop on
  Software and Performance, {WOSP} 2005, Palma, Illes Balears, Spain, July
  12-14, 2005, {ACM}, 2005, pp. 1--12.
\newblock \href {https://doi.org/10.1145/1071021.1071022}
  {\path{doi:10.1145/1071021.1071022}}.
\newline\urlprefix\url{https://doi.org/10.1145/1071021.1071022}

\bibitem{DBLP:conf/cascon/AltamimiP07}
T.~Altamimi, D.~C. Petriu,
  \href{http://dl.acm.org/citation.cfm?id=3172810}{Incremental change
  propagation from {UML} software models to {LQN} performance models}, in:
  M.~Mindel, K.~A. Lyons, J.~Wigglesworth (Eds.), Proceedings of the 27th
  Annual International Conference on Computer Science and Software Engineering,
  {CASCON} 2017, Markham, Ontario, Canada, November 6-8, 2017, {IBM} / {ACM},
  2017, pp. 120--131.
\newline\urlprefix\url{http://dl.acm.org/citation.cfm?id=3172810}

\bibitem{DBLP:conf/wosp/MenasceEGMS10}
D.~A. Menasc{\'{e}}, J.~M. Ewing, H.~Gomaa, S.~Malek, J.~P. Sousa,
  \href{https://doi.org/10.1145/1712605.1712612}{A framework for utility-based
  service oriented design in {SASSY}}, in: A.~Adamson, A.~B. Bondi, C.~Juiz,
  M.~S. Squillante (Eds.), Proceedings of the first joint {WOSP/SIPEW}
  International Conference on Performance Engineering, San Jose, California,
  USA, January 28-30, 2010, {ACM}, 2010, pp. 27--36.
\newblock \href {https://doi.org/10.1145/1712605.1712612}
  {\path{doi:10.1145/1712605.1712612}}.
\newline\urlprefix\url{https://doi.org/10.1145/1712605.1712612}

\bibitem{DBLP:books/daglib/0030032}
P.~H. Feiler, D.~P. Gluch,
  \href{http://www.pearsoned.co.uk/bookshop/detail.asp?item=100000000518651}{Model-Based
  Engineering with {AADL} - An Introduction to the {SAE} Architecture Analysis
  and Design Language}, {SEI} series in software engineering, Addison-Wesley,
  2012.
\newline\urlprefix\url{http://www.pearsoned.co.uk/bookshop/detail.asp?item=100000000518651}

\bibitem{DBLP:journals/scp/EtemaadiC15}
R.~Etemaadi, M.~R.~V. Chaudron,
  \href{https://doi.org/10.1016/j.scico.2014.06.012}{New degrees of freedom in
  metaheuristic optimization of component-based systems architecture:
  Architecture topology and load balancing}, Sci. Comput. Program. 97 (2015)
  366--380.
\newblock \href {https://doi.org/10.1016/j.scico.2014.06.012}
  {\path{doi:10.1016/j.scico.2014.06.012}}.
\newline\urlprefix\url{https://doi.org/10.1016/j.scico.2014.06.012}

\bibitem{DBLP:conf/cec/LiEEC11}
R.~Li, R.~Etemaadi, M.~T.~M. Emmerich, M.~R.~V. Chaudron,
  \href{https://doi.org/10.1109/CEC.2011.5949650}{An evolutionary
  multiobjective optimization approach to component-based software architecture
  design}, in: Proceedings of the {IEEE} Congress on Evolutionary Computation,
  {CEC} 2011, New Orleans, LA, USA, 5-8 June, 2011, {IEEE}, 2011, pp. 432--439.
\newblock \href {https://doi.org/10.1109/CEC.2011.5949650}
  {\path{doi:10.1109/CEC.2011.5949650}}.
\newline\urlprefix\url{https://doi.org/10.1109/CEC.2011.5949650}

\bibitem{Becker:2009cl}
S.~Becker, H.~Koziolek, R.~H. Reussner,
  \href{https://doi.org/10.1016/j.jss.2008.03.066}{The palladio component model
  for model-driven performance prediction}, J. Syst. Softw. 82~(1) (2009)
  3--22.
\newblock \href {https://doi.org/10.1016/j.jss.2008.03.066}
  {\path{doi:10.1016/j.jss.2008.03.066}}.
\newline\urlprefix\url{https://doi.org/10.1016/j.jss.2008.03.066}

\bibitem{10.1145/3132498.3132509}
A.~Rago, S.~A. Vidal, J.~A. Diaz{-}Pace, S.~Frank, A.~van Hoorn,
  \href{https://doi.org/10.1145/3132498.3132509}{Distributed quality-attribute
  optimization of software architectures}, in: Proceedings of the 11th
  Brazilian Symposium on Software Components, Architectures and Reuse, {SBCARS}
  2017, Fortaleza, CE, Brazil, September 18 - 19, 2017, {ACM}, 2017, pp.
  7:1--7:10.
\newblock \href {https://doi.org/10.1145/3132498.3132509}
  {\path{doi:10.1145/3132498.3132509}}.
\newline\urlprefix\url{https://doi.org/10.1145/3132498.3132509}

\bibitem{altamimi_performance_2016}
T.~Altamimi, M.~H. Zargari, D.~C. Petriu,
  \href{http://dl.acm.org/citation.cfm?id=3049899}{Performance analysis
  roundtrip: automatic generation of performance models and results feedback
  using cross-model trace links}, in: M.~Mindel, B.~Jones, H.~A. M{\"{u}}ller,
  V.~Onut (Eds.), Proceedings of the 26th Annual International Conference on
  Computer Science and Software Engineering, {CASCON} 2016, Toronto, Ontario,
  Canada, October 31 - November 2, 2016, {IBM} / {ACM}, 2016, pp. 208--217.
\newline\urlprefix\url{http://dl.acm.org/citation.cfm?id=3049899}

\bibitem{erickson2007can}
J.~Erickson, K.~Siau, \href{http://ceur-ws.org/Vol-365/paper9.pdf}{Can {UML} be
  simplified? practitioner use of {UML} in separate domains}, in: E.~Proper,
  T.~A. Halpin, J.~Krogstie (Eds.), Proceedings of the 12th International
  Workshop on Exploring Modeling Methods for Systems Analysis and Design,
  {EMMSAD} 2008, held in conjunction with the 19th Conference on Advanced
  Information Systems (CAiSE 2007), Trondheim, Norway, 11-15 June, 2007, Vol.
  365 of {CEUR} Workshop Proceedings, CEUR-WS.org, 2007, pp. 81--90.
\newline\urlprefix\url{http://ceur-ws.org/Vol-365/paper9.pdf}

\bibitem{DBLP:journals/infsof/CortellessaET20}
V.~Cortellessa, R.~Eramo, M.~Tucci,
  \href{https://doi.org/10.1016/j.infsof.2020.106362}{From software
  architecture to analysis models and back: Model-driven refactoring aimed at
  availability improvement}, Inf. Softw. Technol. 127 (2020) 106362.
\newblock \href {https://doi.org/10.1016/j.infsof.2020.106362}
  {\path{doi:10.1016/j.infsof.2020.106362}}.
\newline\urlprefix\url{https://doi.org/10.1016/j.infsof.2020.106362}

\bibitem{DBLP:conf/cbse/CortellessaG07}
V.~Cortellessa, V.~Grassi,
  \href{https://doi.org/10.1007/978-3-540-73551-9_10}{A modeling approach to
  analyze the impact of error propagation on reliability of component-based
  systems}, in: H.~W. Schmidt, I.~Crnkovic, G.~T. Heineman, J.~A. Stafford
  (Eds.), Component-Based Software Engineering, 10th International Symposium,
  {CBSE} 2007, Medford, MA, USA, July 9-11, 2007, Proceedings, Vol. 4608 of
  Lecture Notes in Computer Science, Springer, 2007, pp. 140--156.
\newblock \href {https://doi.org/10.1007/978-3-540-73551-9_10}
  {\path{doi:10.1007/978-3-540-73551-9_10}}.
\newline\urlprefix\url{https://doi.org/10.1007/978-3-540-73551-9_10}

\end{thebibliography}

\end{document}